\documentclass[11pt,letterpaper]{article}
\pdfoutput=1

\usepackage{jcappub}

\usepackage{color}
\usepackage{graphicx}
\usepackage[space]{grffile}

\usepackage{verbatim}
\usepackage{amsmath}
\usepackage{amssymb}
\usepackage[caption=false]{subfig}
\usepackage{url}
\usepackage{bbold}
\usepackage{slashed}
\usepackage{epstopdf}
\usepackage{xspace}

\usepackage{multirow}
\usepackage{threeparttable}
\usepackage{paralist}

\newcommand{\keV}{\text{keV}}
\newcommand{\MeV}{\text{MeV}}
\newcommand{\GeV}{\text{GeV}}

\newcommand{\vev}[1]{\langle #1 \rangle}

\DeclareRobustCommand{\Sec}[1]{Sec.\,\ref{#1}}
\DeclareRobustCommand{\Secs}[2]{Secs.\,\ref{#1} and \ref{#2}}
\DeclareRobustCommand{\App}[1]{App.\,\ref{#1}}
\DeclareRobustCommand{\Tab}[1]{Table\,\ref{#1}}

\DeclareRobustCommand{\Fig}[1]{Fig.\,\ref{#1}}
\DeclareRobustCommand{\Figs}[2]{Figs.\,\ref{#1} and \ref{#2}}
\DeclareRobustCommand{\Eq}[1]{Eq.\,(\ref{#1})}
\DeclareRobustCommand{\Eqs}[2]{Eqs.\,(\ref{#1}) and (\ref{#2})}
\DeclareRobustCommand{\Ref}[1]{Ref.\,\cite{#1}}
\DeclareRobustCommand{\Refs}[1]{Refs.\,\cite{#1}}

\newcommand{\be}{\begin{equation}}
\newcommand{\ee}{\end{equation}}
\newcommand{\bea}{\begin{eqnarray}}
\newcommand{\eea}{\end{eqnarray}}
\newcommand{\bi}{\begin{itemize}}
\newcommand{\ei}{\end{itemize}}

\usepackage{xspace}

\newcommand{\A}{\psi_A}
\newcommand{\Abar}{\overline{\psi}_A}
\newcommand{\mA}{m_A}
\newcommand{\OmegaA}{\Omega_A}

\newcommand{\B}{\psi_B}
\newcommand{\Bbar}{\overline{\psi}_B}
\newcommand{\mB}{m_B}
\newcommand{\OmegaB}{\Omega_B}

\begin{document}

\title{(In)direct Detection of \\ Boosted Dark Matter}

\author[a]{Kaustubh Agashe,}
\author[a]{Yanou Cui,}

\author[b]{Lina Necib,}
\author[b]{and Jesse Thaler}

\affiliation[a]{Maryland Center for Fundamental Physics, University of Maryland, \\
 College Park, MD 20742, USA}

\affiliation[b]{Center for Theoretical Physics, Massachusetts Institute of Technology, \\
Cambridge, MA 02139, USA}

\emailAdd{kagashe@umd.edu}
\emailAdd{cuiyo@umd.edu}
\emailAdd{lnecib@mit.edu}
\emailAdd{jthaler@mit.edu}

 \abstract{We initiate the study of novel thermal dark matter (DM) scenarios where present-day annihilation of DM in the galactic center produces boosted stable particles in the dark sector.  These stable particles are typically a subdominant DM component, but because they are produced with a large Lorentz boost in this process, they can be detected in large volume terrestrial experiments via neutral-current-like interactions with electrons or nuclei.  This novel DM signal thus combines the production mechanism associated with indirect detection experiments (i.e.\ galactic DM annihilation) with the detection mechanism associated with direct detection experiments (i.e.\ DM scattering off terrestrial targets).  Such processes are generically present in multi-component DM scenarios or those with non-minimal DM stabilization symmetries.  As a proof of concept, we present a model of two-component thermal relic DM, where the dominant heavy DM species has no tree-level interactions with the standard model and thus largely evades direct and indirect DM bounds. Instead, its thermal relic abundance is set by annihilation into a subdominant lighter DM species, and the latter can be detected in the boosted channel via the same annihilation process occurring today. Especially for dark sector masses in the 10 MeV--10 GeV range, the most promising signals are electron scattering events pointing toward the galactic center.  These can be detected in experiments designed for neutrino physics or proton decay, in particular Super-K and its upgrade Hyper-K, as well as the PINGU/MICA extensions of IceCube.  This boosted DM phenomenon highlights the distinctive signatures possible from non-minimal dark sectors.}

\arxivnumber{1405.7370}

\preprint{MIT-CTP {4538}~~ UMD-PP-014-005}

\maketitle

\section{Introduction}
\label{sec:introduction}

A preponderance of gravitational evidence points to the existence of dark matter (DM) \cite{Zwicky:1933gu,Begeman:1991iy,Bertone:2010zza}.  Under the compelling assumption that DM is composed of one or more species of massive particles,  DM particles in our Milky Way halo today are expected to be non-relativistic, with velocities $v_{\rm DM,0} \simeq \mathcal{O}(10^{-3})$. Because of this small expected velocity, DM indirect detection experiments are designed to look for nearly-at-rest annihilation or decay of DM, and DM direct detection experiments are designed to probe small nuclear recoil energies on the order of $\frac{\mu^2}{m_N}v_{\rm DM,0}^2$ ($\mu$ is the reduced mass of the DM-nucleus system, $m_N$ is the nucleus mass). In addition, these conventional detection strategies are based on the popular (and well-motivated) assumption that DM is a weakly-interacting massive particle (WIMP) whose thermal relic abundance is set by its direct couplings to the standard model (SM).

In this paper, we explore a novel possibility that a small population of DM (produced non-thermally by late-time processes) is in fact relativistic, which we call ``boosted DM''.  As a concrete example, consider two species of DM, $\A$ and $\B$ (which need not be fermions), with masses $\mA > \mB$.  Species $\A$ constitutes the dominant DM component, with no direct couplings to the SM.  Instead, its thermal relic abundance is set by the annihilation process\footnote{To our knowledge, the first use of $\A \Abar \to \B \Bbar$ to set the relic abundance of $\A$ appears in the assisted freeze-out scenario \cite{Belanger:2011ww}.  As an interesting side note, we will find that assisted freeze-out of $\A$ can lead to a novel ``balanced freeze-out'' behavior for $\B$.  In \App{app:ABrelicstory}, we show that the relic abundance can scale like $\Omega_B \propto 1/\sqrt{\sigma_B}$ (unlike $\Omega_B \propto 1/\sigma_B$ for standard freeze-out).  In this paper, of course, we are more interested in the boosted $\B$ population, not the thermal relic $\B$ population.}
\be
\label{eq:AAtoBB}
\A \Abar \to \B \Bbar.
\ee
At the present day, non-relativistic $\A$ particles undergo the same annihilation process in the galactic halo today, producing relativistic final state $\B$ particles, with Lorentz factor $\gamma = \mA / \mB$.  These boosted DM particles can then be detected via their interactions with SM matter at large volume terrestrial experiments that are designed for detecting neutrinos and/or proton decay, such as Super-K/Hyper-K \cite{Fukuda:2002uc, Abe:2011ts}, IceCube/PINGU/MICA \cite{Ahrens:2002dv, Aartsen:2014oha, MICA}, KM3NeT \cite{Katz:2006wv}, and ANTARES \cite{Collaboration:2011nsa}, as well as recent proposals based on liquid Argon such as LAr TPC and GLACIER \cite{Bueno:2007um,Badertscher:2010sy}, and liquid scintillator experiments like JUNO \cite{PhysRevD.88.013008,Li:2014qca}. In such experiments, boosted DM can scatter via the neutral-current-like process
\be
\B X \to \B X^{(\prime)},
\ee
similar to high energy neutrinos.  This boosted DM phenomenon is generic in multi-component DM scenarios and in single-component DM models with non-minimal stabilization symmetries), where boosted DM can be produced in DM conversion $\psi_i \psi_j \to \psi_k \psi_\ell$ \cite{DEramo:2010ep,SungCheon:2008ts,Belanger:2011ww}, semi-annihilation $\psi_i \psi_j \to \psi_k \phi$ (where $\phi$ is a non-DM state) \cite{DEramo:2010ep,Hambye:2008bq,Hambye:2009fg,Arina:2009uq,Belanger:2012vp}, $3 \rightarrow 2$ self-annihilation \cite{Carlson:1992fn,deLaix:1995vi,Hochberg:2014dra}, or decay transition $\psi_i \to \psi_j + \phi$.

In order to be detectable, of course, boosted DM must have an appreciable cross section to scatter off SM targets.  Based on \Eq{eq:AAtoBB} alone and given our assumption that $\A$ is isolated from the SM, one might think that $\B$ could also have negligible SM interactions.  In that case, however, the dark sector would generally have a very different temperature from the SM sector, with the temperature difference depending on details related to reheating, couplings to the inflaton, and entropy releases in the early universe \cite{Berezhiani:1995am, Berezhiani:2000gw, Ciarcelluti:2010zz, Feng:2008mu}. So if we want to preserve the most attractive feature of the WIMP paradigm---namely, that the thermal relic abundance of $\A$ is determined by its annihilation cross section, insensitive to other details---then $\B$ must have efficient enough interactions with the SM to keep $\A$ in thermal equilibrium at least until $\A \Abar \to \B \Bbar$ freezes out. Such $\B$-SM couplings then offer a hope for detecting the dark sector even if the major DM component $\A$ has no direct SM couplings.

\begin{figure}[t]
  \centerline{\includegraphics[scale=0.6]{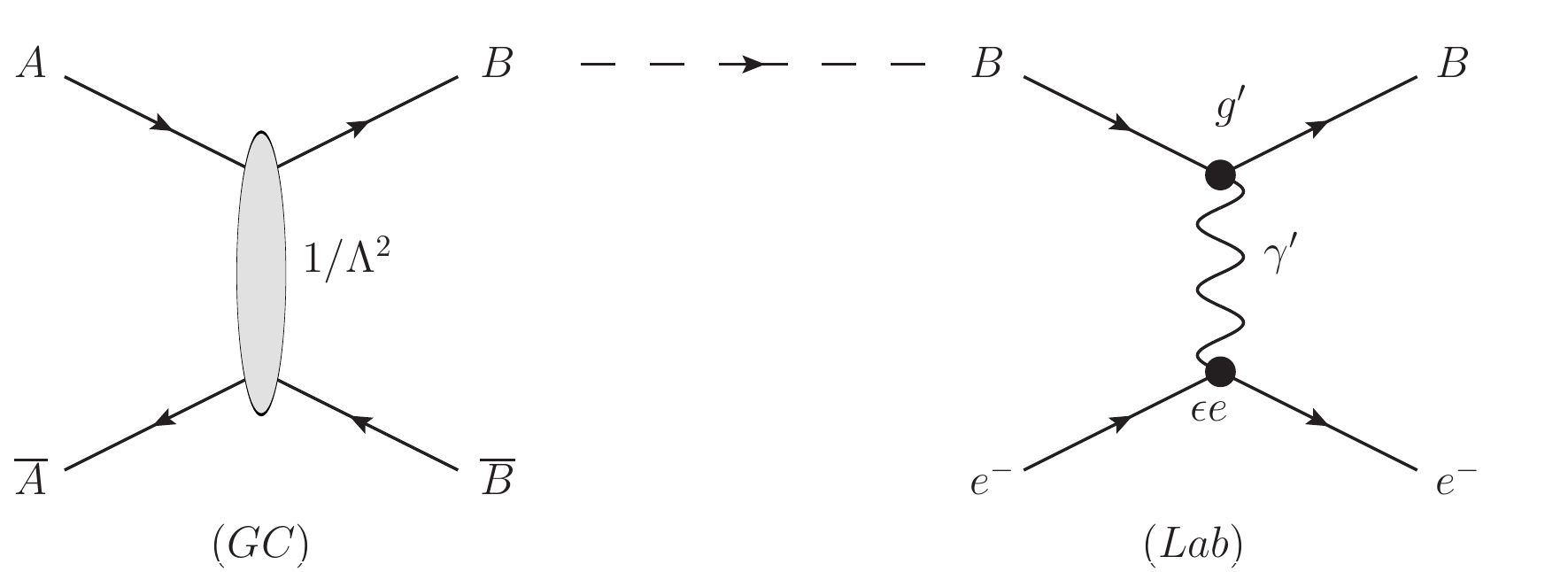}} 
  \caption{(Left) Production of boosted $\B$ particles through $\A$ annihilation in the galactic center:  $\A\Abar\rightarrow \B\Bbar$.  This process would be considered ``indirect detection'' of  $\A$.  (Right) Scattering of $\B$ off terrestrial electron targets: $\B e^-\rightarrow\B e^-$.  This process would be considered ``direct detection'' of  $\B$.}
\label{fig:feynman}
\end{figure}

As a simple proof of concept, we present a two-component DM model of the above type, with $\A$/$\B$ now being specified as fermions. The dominant DM component $\A$ has no (tree-level) interactions with the SM, such that traditional DM searches are largely insensitive to it.  In contrast, the subdominant DM component $\B$ has significant interactions with the SM via a dark photon $\gamma'$ that is kinetically-mixed with the SM photon.   The two processes related to the (in)direct detection of the $\A$/$\B$ dark sector are illustrated in \Fig{fig:feynman}.  In the early universe, the process on the left, due to a contact interaction between $\A$ and $\B$, sets both the thermal relic abundance of $\A$ as well as the production rate of boosted $\B$ in the galactic halo today.  The resulting boosted $\B$ population has large scattering cross sections off nuclei and electrons via dark photon exchange, shown on the right of \Fig{fig:feynman}.  Assuming that $\B$ itself has a small thermal relic abundance (which is expected given a large SM scattering cross section), and is light enough to evade standard DM detection bounds, then (direct) detection of boosted $\B$ via (indirect) detection of $\A$ annihilation would offer the best non-gravitational probe of the dark sector.\footnote{\label{footnote:suncapture}Because $\A$ has no direct coupling to the SM, the $\A$ solar capture rate is suppressed.  By including a finite $\A$-SM coupling, one could also imagine boosted DM coming from annihilation in the sun. The possibility of detecting fast-moving DM emerging from the sun has been studied previously in the context of induced nucleon decay \cite{Huang:2013xfa}, though not with the large boost factors we envision here which enable detection via Cherenkov radiation. Note, however, that $\B$ particles are likely to become trapped in the sun due to energy loss effects (see \Sec{subsec:earth}), limiting solar capture as a viable signal channel.}

Beyond just the intrinsic novelty of the boosted DM signal, there are other reasons to take this kind of DM scenario seriously.  First, having the dominant DM component $\A$ annihilate into light stable $\B$ particles (i.e.~assisted freeze-out \cite{Belanger:2011ww}) is a novel way to ``seclude''  DM from the SM while still maintaining the successes of the thermal freeze-out paradigm of WIMP-type DM.\footnote{For variations such as annihilating to dark radiation or to dark states that decay back to the SM, see for instance \Refs{Pospelov:2007mp,ArkaniHamed:2008qn,Ackerman:mha,Nomura:2008ru,Mardon:2009gw,AB_CMB}.}   Such a feature enables this model to satisfy the increasingly severe constraints from DM detection experiments.  A key lesson from secluded DM scenarios \cite{Pospelov:2007mp} is that it is often easier to detect the ``friends'' of DM (in this case $\B$) rather than the dominant DM component itself \cite{Bjorken:2009mm}.  Second, our study here can be seen as exploring the diversity of phenomenological possibilities present (in general) in multi-component DM scenarios.  Non-minimal dark sectors are quite reasonable, especially considering the non-minimality of the SM (with protons and electrons stabilized by separate $B$- and $L$-number symmetries).  Earlier work along these lines includes, for instance, the possibility of a mirror DM sector \cite{Hodges:1993yb, Mohapatra:2000qx,Berezhiani:2000gw, Foot:2001hc}.  Recently, multi-component DM scenarios have drawn rising interest motivated by anomalies in DM detection experiments \cite{Fairbairn:2008fb,Zurek:2008qg, Profumo:2009tb} and possible new astrophysical phenomena such as a ``dark disk'' \cite{Fan:2013yva}.  Boosted DM provides yet another example of how the expected kinematics, phenomenology, and search strategies for multi-component DM can be very different from single-component DM. 

The outline of the rest of this paper is as follows.  In \Sec{sec:preliminaries}, we present the above model in more detail.  In \Sec{sec:production}, we describe the annihilation processes of both $\A$ and $\B$, which sets their thermal relic abundances and the rate of boosted DM production today, and we discuss the detection mechanisms for boosted DM in \Sec{sec:detection}.  We assess the discovery prospects at present and future experiments in \Sec{sec:experiments}, where we find that Super-K should already be sensitive to boosted DM by looking for single-ring electron events from the galactic center (GC).  We summarize the relevant constraints on this particular model in \Sec{sec:constraints}, and we conclude in \Sec{sec:conclusions} with a discussion of other DM scenarios with similar phenomenology.  More details are relegated to the appendices.

\section{Two Component Dark Matter}
\label{sec:preliminaries}

Consider two species of fermion DM $\A$ and $\B$ with Dirac masses $\mA > \mB$, which interact via a contact operator\footnote{Via a Fierz rearrangement, we can rewrite this operator as
$$
-\frac{1}{ 4 \Lambda^2} \Bigl( \Abar \A \Bbar \B + \Abar \gamma^\mu \A \Bbar \gamma_\mu \B + \frac{1}{2} \Abar \Sigma^{\mu \nu} \A \Bbar \Sigma_{\mu \nu} \B + \Abar \gamma^5 \A \Bbar \gamma^5 \B- \Abar \gamma^\mu \gamma^5 \A \Bbar \gamma_\mu \gamma^5 \B \Bigr),
$$
where $\Sigma^{\mu \nu}  = \frac{i}{2} [\gamma^\mu, \gamma^\nu]$.
}
\begin{equation}
\label{eq:AABBint}
\mathcal{L}_{\rm int} = \frac{1}{\Lambda^2} \Abar \B  \Bbar \A.
\end{equation}
This operator choice ensures an $s$-wave annihilation channel \cite{Cui:2010ud}, $\A \Abar \to \B \Bbar$ as in \Fig{fig:feynman}, which is important for having a sizable production rate of boosted $\B$ today. A UV completion for such operator is shown in \Fig{fig:AApp} in \App{app:ADirectDetection}. Other Lorentz structures are equally plausible (as long as they lead to $s$-wave annihilation).  

As an extreme limit, we assume that \Eq{eq:AABBint} is the sole (tree-level) interaction for $\A$ at low energies and that $\A$ is the dominant DM component in the universe today.  We assume that both $\A$ and $\B$ are exactly stable because of separate stabilizing symmetries (e.g.~a $\mathbb{Z}_2 \times \mathbb{Z}_2$). 

The subdominant species $\B$ is charged under a dark $U(1)'$ gauge group, with charge $+1$ for definiteness. This group is spontaneously broken, giving rise to a massive dark photon $\gamma'$ with the assumed mass hierarchy
\be
\mA > \mB > m_{\gamma'}.
\ee
We will take the gauge coupling $g'$ of the dark $U(1)'$ to be sufficiently large (yet perturbative) such that the process $\B \Bbar \to \gamma' \gamma'$ efficiently depletes $\B$ and gives rise to a small thermal relic abundance (see \Eq{eq:Babundanceestimate} below).  

Via kinetic mixing with the SM photon \cite{Holdom:1985ag,Okun:1982xi,Galison:1983pa} (strictly speaking, the hypercharge gauge boson),
\be
\mathcal{L} \supset -\frac{\epsilon}{2} F_{\mu\nu}' F^{\mu \nu},
\ee
$\gamma'$ acquires $\epsilon$-suppressed couplings to SM fields.  
In this way, we can get a potentially large cross section for $\B$ to scatter off terrestrial SM targets, in particular $\B e^- \to \B e^-$ from $\gamma'$ exchange (with large $g'$ and suitable $\epsilon$) as in 
\Fig{fig:feynman}.  In principle, we would need to account for the possibility of a dark Higgs boson $H'$ in the spectrum, but for simplicity, we assume that such a state is irrelevant to the physics we consider here, perhaps due to a Stuckelberg mechanism for the $U(1)'$ \cite{Stueckelberg:1900zz,Kors:2005uz} or negligible couplings of $H'$ to matter fields.

The parameter space of this model is defined by six parameters
\be
\{m_A,  m_B,  m_{\gamma'},  \Lambda,  g',  \epsilon\}.
\ee
Throughout this paper, we will adjust $\Lambda$ to yield the desired DM relic abundance of $\A$, assuming that any DM asymmetry is negligible.  Because the process $\B e^- \to \B e^-$ has homogeneous scaling with $g'$ and $\epsilon$, the dominant phenomenology depends on just the three mass parameters: $m_A$, $m_B$, and $m_{\gamma'}$.   To achieve a sufficiently large flux of boosted $\B$ particles, we need a large number density of $\A$ particles in the galactic halo.  For this reason, we will focus on somewhat low mass thermal DM, with typical scales:
\be
m_A \simeq \mathcal{O}(10~\GeV), \quad m_B \simeq \mathcal{O}(100~\MeV), \quad m_{\gamma'} \simeq \mathcal{O}(10~\MeV). \label{eq: mass scales}
\ee
Constraints on this scenario from standard DM detection methods are summarized later in \Sec{sec:constraints}.
This includes direct detection and CMB constraints on the thermal relic $\B$ population.
In addition, $\A$ can acquire couplings to $\gamma'$ through a $\B$-loop, thus yielding constraints from direct detection of $\A$, and we introduce a simple UV completion for \Eq{eq:AABBint} in \App{app:ADirectDetection} which allows us to compute this effect without having to worry about UV divergences.

There are a variety of possible extensions and modifications to this simple scenario.  One worth mentioning explicitly is that $\A$ and/or $\B$ could have small Majorana masses which lead to mass splittings within each multiplet (for $\psi_B$ this would appear after $U(1)'$ breaking) \cite{TuckerSmith:2001hy,Cui:2009xq}.  As discussed in \Refs{Finkbeiner:2009mi,Graham:2010ca,Pospelov:2013nea}, both components in an inelastic DM multiplet can be cosmologically stable, such that the current day annihilation is not suppressed.  These splittings, however, would typically soften the bounds on the non-relativistic component of $\A/\B$ from conventional direct detection experiments, since the scattering would be inelastic (either endothermic or exothermic).  This is one way to avoid the direct detection of bounds discussed in \Sec{sec:constraints}.

\section{Thermal Relic Abundances and Present-Day Annihilation}
\label{sec:production}

To find the relic density of $\A$/$\B$, we need to write down their coupled Boltzmann equations.  In \App{app:ABrelicstory}, we provide details about this Boltzmann system  (see also \Refs{Belanger:2011ww,Bhattacharya:2013hva,Modak:2013jya}), as well as analytic estimates for the freeze-out temperature and relic abundance in certain limits.  Here, we briefly summarize the essential results.  

The annihilation channel $\A\Abar \rightarrow \B\Bbar$ not only determines the thermal freeze-out of the dominant DM component $\A$ but also sets the present-day production rate for boosted $\B$ particles in Milky Way.  Considering just the operator from \Eq{eq:AABBint}, the thermally-averaged cross section in the $s$-wave limit is:  
\be
\langle\sigma_{A \overline{A} \rightarrow B \overline{B}} v\rangle_{v \rightarrow 0}=  \frac{1}{8 \pi \Lambda^4} \left(  m_A + m_B \right)^2     \sqrt{1 - \frac{m_B^2}{m_A^2}}   \label{eq:thermav}.
\ee
As discussed in \App{app:ABrelicstory}, the Boltzmann equation for $\A$ effectively decouples from $\B$ when $\langle \sigma_{B\bar{B}\rightarrow\gamma'\gamma'} v \rangle \gg \langle \sigma_{A \bar{A} \rightarrow B \bar{B}} v \rangle$.  In this limit, the relic density $\Omega_A$ takes the standard form expected of WIMP DM (assuming $s$-wave annihilation):  
\be \label{eq:omegaA}
\Omega_A\simeq 0.2 \left( \frac{5\times10^{-26}~\text{cm}^3/\text{s}}{\langle\sigma_{A \bar{A} \rightarrow B \bar{B}} v\rangle} \right). 
\ee
Notice that in order to get the observed DM relic abundance $\OmegaA \approx0.2$, the thermal annihilation cross section is around twice the ``standard'' thermal cross section $3\times10^{-26}~\text{cm}^3/\text{s}$ where a Majorana fermion DM with $\simeq 100$ GeV mass is assumed. The slight discrepancy is because our $\Omega_A$ is the sum of the abundances of both Dirac particles $\A$ and $\Abar$, and the $\A$ we are interested in has lower mass $\lesssim 20$ GeV (see, e.g., \Ref{Steigman:2012nb}).

In the limit that $m_B \ll m_A$, we have
\be
\langle\sigma_{A \overline{A} \rightarrow B \overline{B}} v\rangle \approx 5  \times 10^{-26} ~\text{cm}^3/\text{s} \left( \frac{m_A}{20 ~\GeV} \right) ^2  \left(\frac{250 ~ \GeV}{ \Lambda} \right)^4.
\ee
Note that $m_A \ll \Lambda$ for our benchmark mass $m_A = 20~\GeV$, so it is consistent to treat the annihilation of $\A$ as coming just from the effective operator in \Eq{eq:AABBint}.

The thermal relic abundance of $\B$ is more subtle. In the absence of $\A$, the relic abundance of $\B$ would be determined just by the annihilation process $\B\Bbar\rightarrow\gamma'\gamma'$, whose thermally-averaged cross section in the $s$-wave limit is  
\be
\label{eq:thermbv}
\langle\sigma_{B\overline{B} \rightarrow\gamma'\gamma'} v\rangle_{v\rightarrow0} = \frac{g'^4}{2 \pi} \frac{\left( m_B^2 -  m_{\gamma'}^2 \right)}{ (m_{\gamma'}^2 - 2 m_B^2)^2}  \sqrt{1 - \frac{m_{\gamma'}^2}{m_B^2}}.
\ee
However, the process $\A\Abar \rightarrow \B\Bbar$ is still active even after $\A$ freezes out with a nearly constant $\A$ abundance well above its equilibrium value, which can have impact on the relic abundance of $\B$.  Let $x_{f,B} = m_B/T_{f,B}$, $T_{f,B}$ being the temperature at $\B$ freeze-out.  As explained in \App{app:ABrelicstory}, when $\frac{\sigma_B}{\sigma_A}(\frac{m_B}{m_A})^2\gg(x_{f})^2$ (i.e.~large $g'$), a good approximation to the relic abundance $\OmegaB$ is
\be
\label{eq:Babundanceestimate}
\frac{\OmegaB}{\OmegaA} \simeq  \frac{m_B}{m_A}    \sqrt{\frac{\langle\sigma_{A\overline{A} \rightarrow B\overline{B}} v\rangle}{\langle\sigma_{B\overline{B}\rightarrow\gamma'\gamma'} v\rangle}}.
\ee
This $\Omega \propto 1/\sqrt{\sigma}$ behavior is very different from the usual DM abundance relation $\Omega \propto 1/ \sigma$.   It arises because in this limit, there is a balance between depletion from $\B$ annihilation and replenishment from $\A\Abar \rightarrow \B\Bbar$ conversion.  To our knowledge, this ``balanced freeze-out'' behavior has not been discussed before in the DM literature.

\begin{figure}[t]%
    \centering
    \subfloat[\label{fig:abundance:a}]{{\includegraphics[scale=0.5]{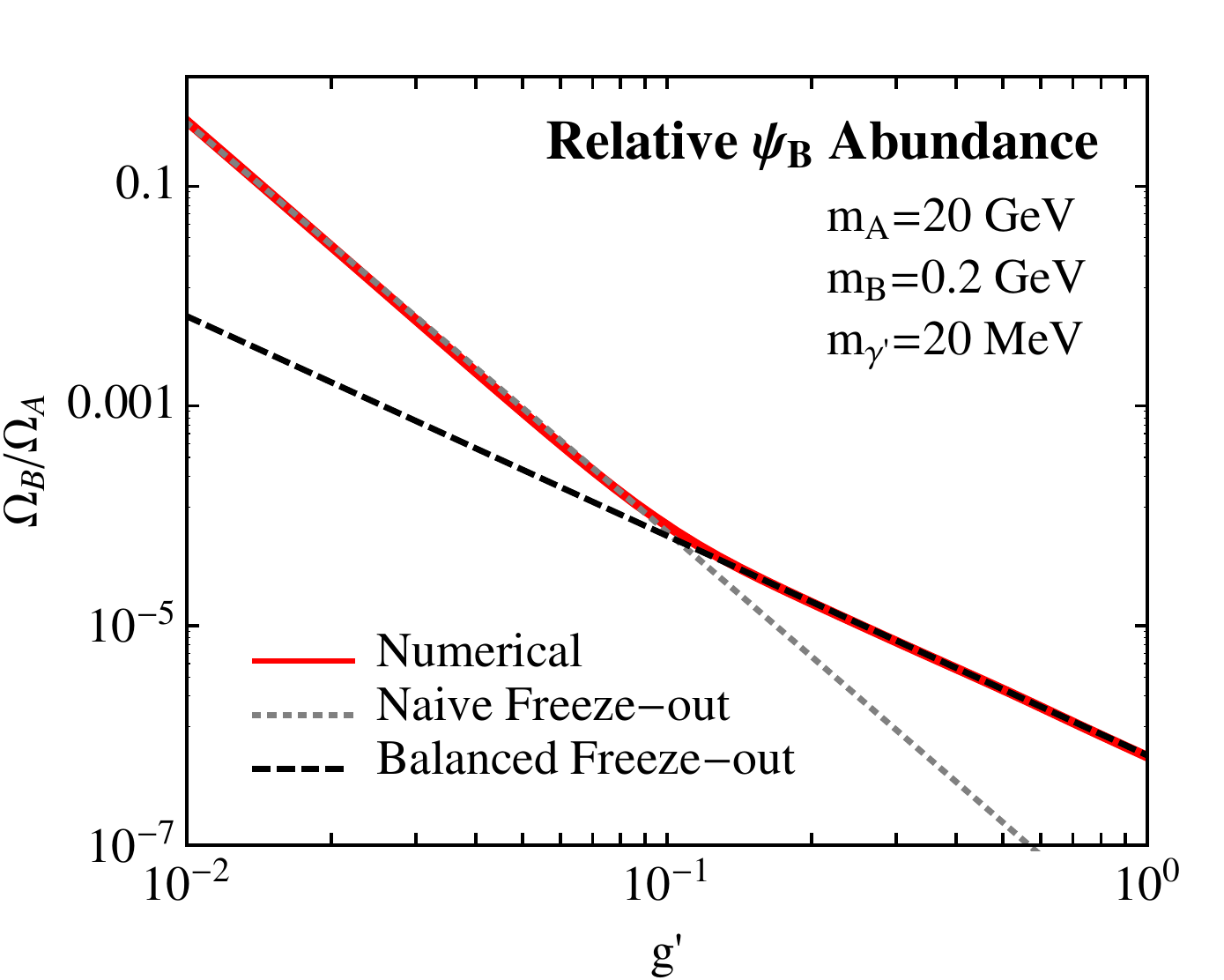} }}%
    \qquad
       \subfloat[\label{fig:abundance:b}]{{\includegraphics[scale=0.5]{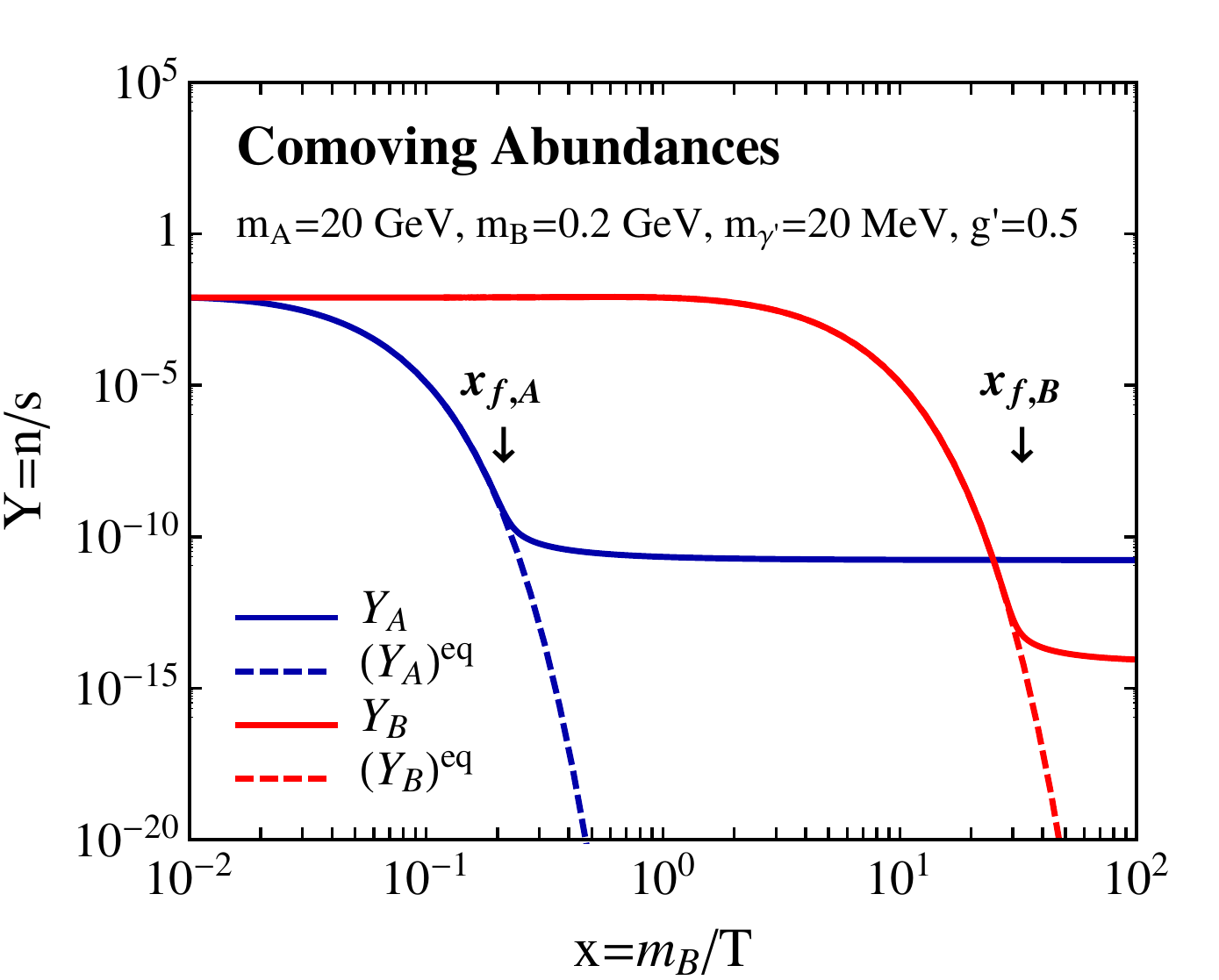} }} %
        \caption{(a)  Ratio of the abundances $\OmegaB/\OmegaA$ as a function of $g'$, fixing $m_A = 20~\GeV$, $m_B= 0.2~\GeV$, and $m_{\gamma'} = 20~\MeV$.  The solid line is the numerical solution of the Boltzmann equation in \Eq{eq:Boltzmann}, the dotted line is the analytic estimate from assuming independent thermal freeze-out of $\A$ and $\B$ (naive freeze-out), and the dashed line is the analytic estimate from \Eq{eq:Babundanceestimate} (balanced freeze-out).  (b) Evolution of the co-moving abundances $Y_A$ and $Y_B$ as a function of $x=m_B/T$ for the benchmark in \Eq{eq:keybenchmark}.  The solid lines show the actual densities per unit entropy, while the dashed lines are the equilibrium curves.}%
    \label{fig:abundance}%
\end{figure}

In \Fig{fig:abundance:a}, we show numerical results for $\OmegaB$ as a function of $g'$: for small $g'$, $\B$ freezes out in the standard way with $\OmegaB \propto 1/ \sigma_B$, while for large $g'$, $\OmegaB$ exhibits the $1/\sqrt{\sigma_B}$ scaling from balanced freeze-out.  Thus, as long as $g'$ is sufficiently large, then $\B$ will be a subdominant DM component as desired.  In \Fig{fig:abundance:b}, we show the full solution to the coupled Boltzmann equations for $\A$ and $\B$ (see \Eq{eq:Boltzmann}) for the following benchmark scenario:
\be
\label{eq:keybenchmark}
m_A = 20~\GeV, \quad m_B =200~\MeV, \quad m_{\gamma'}= 20~ \MeV, \quad g'=0.5, \quad\epsilon = 10^{-3},
\ee
where we have adjusted $\Lambda = 250~\GeV$ to yield the cross section $\langle \sigma_{A \overline{A} \rightarrow B \overline{B}} v \rangle = 5 \times10^{-26} \text{cm}^3/\text{s}$ 
 needed to achieve $\OmegaA  \simeq\Omega_{\rm DM}\approx0.2$.  For this benchmark, $\B$ has a much smaller abundance $\OmegaB \simeq 2.6 \times 10^{-6} \, \Omega_{\rm DM}$.  We have chosen the reference masses to be safe from existing constraints but visible with a reanalysis of existing Super-K data, and we have chosen the reference value of $g'$ to be comparable to hypercharge in the SM.  The values of $m_{\gamma'}$ and $\epsilon$ are also interesting for explaining the muon $g-2$ anomaly \cite{Fayet:2007ua,Pospelov:2008zw}.

This model, though simple, exhibits a novel $\B$ freeze-out behavior, and the ``balancing condition'' behind \Eq{eq:Babundanceestimate} may be interesting to study in other contexts. For much of parameter space of our interest in this paper, the $\OmegaB \propto 1/\sqrt{\sigma_B}$ scaling affects the CMB and direct detection constraints on $\B$.  As discussed in \Sec{sec:constraints}, this scaling implies that the constraints from CMB heating on $\B$ annihilation are largely independent of $g'$.  Similarly, unless there is some kind of inelastic splitting within the $\B$ multiplet, there is a firm direct detection bound on $m_B$ that is also largely independent of $g'$.   Note that the benchmark scenario in \Eq{eq:keybenchmark} indeed satisfies these bounds (see the star in \Fig{fig:significance}).

\section{Detecting Boosted Dark Matter}
\label{sec:detection}

With $\A$ being the dominant DM species, the annihilation process $\A \Abar \to \B \Bbar$ is active in the galactic halo today, producing boosted $\B$ particles. To compute the flux of $\B$ incident on the earth, we can recycle the standard formulas from indirect detection of WIMP DM.   Roughly speaking, the (in)direct detection of boosted $\B$ particles from $\A$ annihilation is analogous to the familiar process of indirect detection of neutrinos from WIMP annihilation.  For this reason, the natural experiments to detect boosted DM are those designed to detect astrophysical neutrinos.  As we will see, $\B$ typically needs to have stronger interactions with the SM than real neutrinos in order to give detectable signals in current/upcoming experiments.

We also want to comment that, due to the small mass and suppressed thermal abundance, the non-relativistic relic $\B$ particles can be difficult to detect through conventional direct and indirect DM searches, even with efficient interaction between $\B$ and SM states.   (See \Sec{sec:constraints} for existing bounds on $\B$.)  Therefore, detecting boosted $\B$ particles may be the only smoking gun from this two-component $\A$/$\B$ system.

\subsection{Flux of Boosted Dark Matter}

The flux of $\B$ from the GC is 
\be
\frac{d \Phi_{\text{GC}}}{d \Omega \, d E_B} = \frac{1}{4} \frac{r_{\text{Sun}}}{4 \pi} \left( \frac{\rho_{\text{local}}}{m_A}\right) ^2 J \,  \langle\sigma_{A\overline{A} \rightarrow B\overline{B}} v\rangle_{v \rightarrow 0} \frac{d N_B}{dE_B},
\ee
where $r_{\text{Sun}}= 8.33~\text{kpc}$ is the distance from the sun to the GC and $\rho_{\text{local}} = 0.3~\GeV/\text{cm}^3$ is the local DM density.  Since the $\A \Abar \to \B \Bbar$ annihilation process yields two mono-energetic boosted $\B$ particles with energy $m_A$, the differential energy spectrum is simply
\be
\frac{d N_B}{d E_B} = 2 \, \delta(E_B - m_A).
\ee
The quantity $J$ is a halo-shape-dependent dimensionless integral over the line of sight,
\be
J = \int_{\text{l.o.s}} \frac{d s }{r_{\text{Sun}}} \left( \frac{\rho(r(s,\theta))}{\rho_{\text{local}}} \right)^2, 
\ee 
where $s$ is the line-of-sight distance to the earth, the coordinate $r(s,\theta) = (r_{\text{Sun}}^2 + s^2 - 2 r_{\text{Sun}} s \cos{\theta})^{1/2} $ is centered on the GC, and $\theta$ is the angle between the line-of-sight direction and the earth/GC axis.  Assuming the NFW halo profile \cite{Navarro:1995iw}, we use the interpolation functions $J(\theta)$ provided in \Ref{Cirelli:2010xx} and integrate them over angular range of interest.  In particular, when trying to mitigate neutrino backgrounds in \Sec{subsec:background}, we will require the $\B e^- \to \B e^-$ process to give final state electrons within a cone of angle $\theta_C$ from the GC.

To illustrate the scaling of the flux, we integrate over a $10^\circ$ cone around the GC and obtain 
\be
\label{eq:PhiGC}
\Phi^{10^\circ}_{\text{GC}}= 9.9 \times 10^{-8}~\text{cm}^{-2} \text{s}^{-1} \left( \frac{\langle \sigma_{A\overline{A} \rightarrow B\overline{B}} v \rangle}{5 \times 10^{-26}~\text{cm}^3/\text{s}} \right)  \left( \frac{20~\GeV}{m_A} \right)^2.
\ee
For completeness, the flux over the whole sky is:
\be
\label{eq:PhiAllsky}
\Phi^{4 \pi}_{\text{GC}} = 4.0 \times 10^{-7}~\text{cm}^{-2} \text{s}^{-1} \left( \frac{\langle \sigma_{A\overline{A} \rightarrow B\overline{B}} v \rangle}{5 \times 10^{-26}~\text{cm}^3/\text{s}} \right)  \left( \frac{20~\GeV}{m_A} \right)^2.
\ee
These estimates are subject to uncertainties on the DM profile; for example, an Einasto profile would increase the flux by an $\mathcal{O}(1)$ factor \cite{Cirelli:2010xx}. 

Note that this GC flux estimate is the same as for any mono-energetic DM annihilation products.\footnote{Up to factors of 2 if the particles considered are Majorana or Dirac, and the number of particles created in the final state.} Therefore we can estimate the expected bound on the boosted DM-SM cross section by reinterpreting neutrino bounds on DM annihilation.  Looking at \Ref{Yuksel:2007ac}, the anticipated Super-K limit on 1-100 GeV DM annihilating in the Milky Way exclusively to monochromatic neutrinos is $10^{-21}-10^{-22} \simeq \text{cm}^3/\text{sec}$.  This is four to five orders of magnitude weaker than a typical thermal annihilation cross section ($\simeq 10^{-26} ~\text{cm}^3/\text{sec}$).  Assuming thermal relic $\A$ DM exclusively annihilates to boosted $\B$ particles, we can estimate the bound on the $\B$-SM cross section by scaling down the charged current neutrino scattering cross section ($10^{-38}~\text{cm}^2$, see \Eq{eq:CCxsec}) by the corresponding factor.  This gives an estimated bound of
\be
\sigma_{B \, \text{SM} \to B \, \text{SM}} \lesssim 10^{-33}-10^{-34}~\text{cm}^2,
\ee
which is consistent with the cross section derived later in \Eq{eq:typicalsignalxsec} for a benchmark model that is on the edge of detectability.\footnote{Our numbers are less consistent with Super-K bounds shown in conference proceedings in \Ref{Mijakowski:2012dva}, which are two orders of magnitude more constraining than expected from \Ref{Beacom:2006tt}.  However, the details of the Super-K analysis are not available for direct comparison.}

\subsection{Detection of Boosted Dark Matter}
\label{bDM_detect}

The flux of boosted $\B$ particles estimated from \Eq{eq:PhiGC} is rather small.\footnote{For comparison, the flux of non-relativistic relic $\B$ particles incident on earth is approximately 
$$
\Phi_{\rm local} \simeq \frac{\rho_{\rm local} v_{\rm 0}}{m_B} \frac{\OmegaB}{\Omega_{\rm DM}} = 2.25 \times 10^3~\text{cm}^{-2} \text{s}^{-1} \left( \frac{200~\MeV}{m_B} \right) \left( \frac{\OmegaB}{10^{-5}} \right).
$$ where $v_0 \simeq $ 220 km/sec. 
}  Therefore, in order to detect boosted $\B$, one needs a large volume, small background detector sensitive to the (quasi-)elastic scattering process
\be
\label{eq:BXtoBX}
\B X \to \B X',
\ee
where $X$ and $X'$ are SM states (possibly the same).  Because the $\gamma'$ is kinetically-mixed with the photon, $\B$ can scatter off any SM state $X$ with electromagnetic couplings via $t$-channel exchange of $\gamma'$.\footnote{There are also subdominant scatterings from weak charges as well.}  A large scattering cross section favors light $m_{\gamma'}$, large $\epsilon$, and large $g'$; the values of $m_\gamma'\gtrsim10~\MeV$ and $\epsilon \sim 10^{-3}$ in the benchmark in \Eq{eq:keybenchmark} are (marginally) consistent with current limits on dark photons \cite{Essig:2013lka}.  

Existing neutrino detectors such as Super-K, IceCube, and their upgrades can be employed to detect boosted DM via \Eq{eq:BXtoBX}.  The strategy is to detect Cherenkov light from the final state charged particles, so the energy of outgoing $X'$ must be above the Cherenkov threshold.  In terms of a Lorentz factor, the threshold is
\be
\text{Water: }  \gamma_\text{Cherenkov} = 1.51, \qquad \text{Ice: }    \gamma_\text{Cherenkov} = 1.55,
\ee
where there is typically a stricter analysis threshold $E^{\rm thresh}$ on $X'$ as well, depending on experimental specifics.  Furthermore, one needs to distinguish $\B$ scattering from the large background of neutrino scattering events, which we discuss more in \Sec{subsec:background}.

\begin{figure}[t]%
    \centering
    \subfloat[]{{\includegraphics[scale=0.55, trim = 0 -0.5cm 0 0]{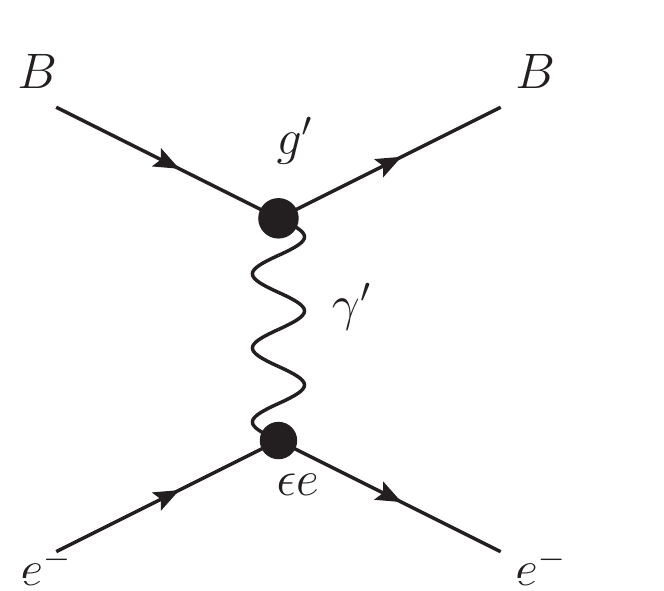} } \label{fig:feynmandetection:a}}%
    \qquad
       \subfloat[]{{\includegraphics[scale=0.55, trim = 0 -0.5cm 0 0]{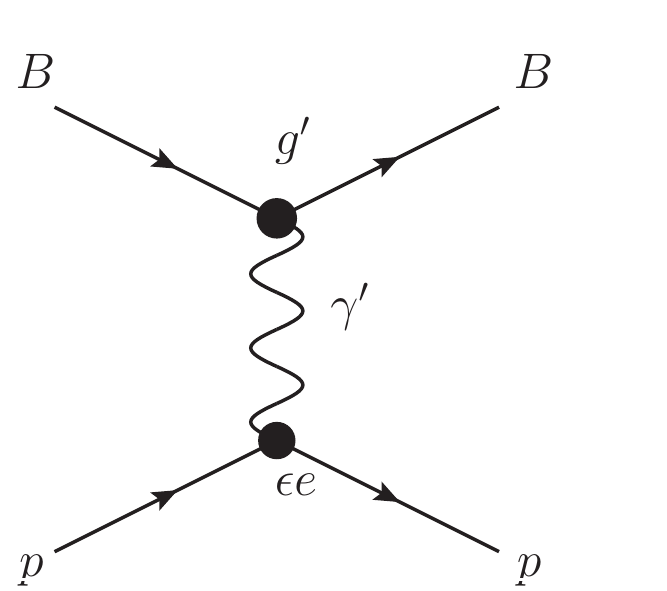} }}  \label{fig:feynmandetection:b}%
        \qquad
       \subfloat[]{{\includegraphics[scale=0.52]{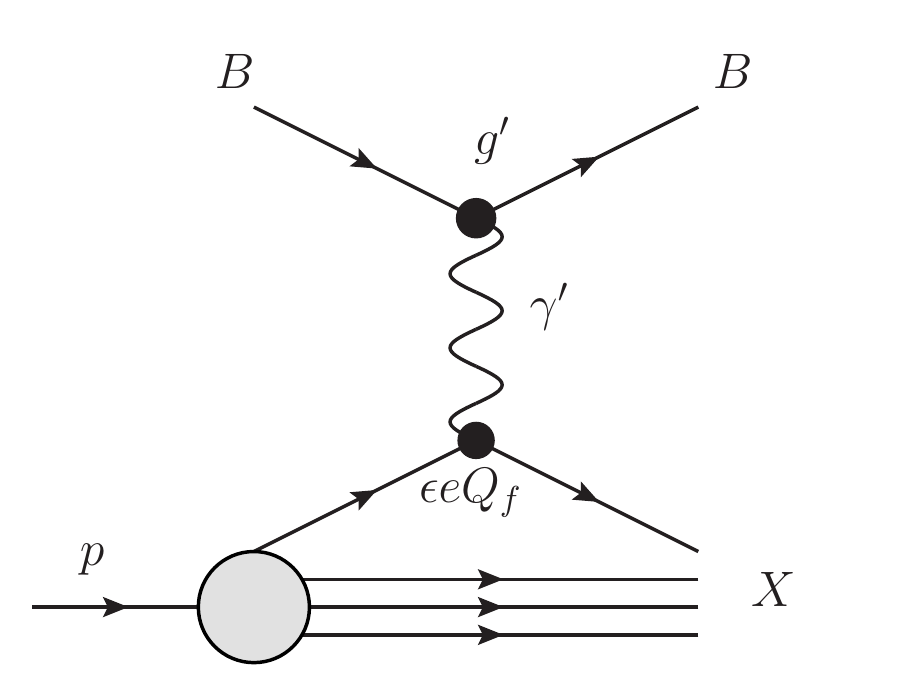} }}%
        \caption{Detection channels for boosted $\B$ in neutrino experiments.  (a) Elastic scattering on electrons. (b) Elastic scattering on protons (or nuclei). (c) Deep inelastic scattering on protons (or nuclei).  For Cherenkov experiments, we find that the most promising channel is electron scattering.}%
    \label{fig:feynmandetection}%
\end{figure}

As shown in \Fig{fig:feynmandetection}, there are three detection channels for boosted $\B$ at a neutrino detector: elastic scattering off electrons, elastic scattering off protons (or nuclei), and deep inelastic scattering (DIS) off protons (or nuclei).   As discussed in more detail in \App{app:detection}, although the total $\B$ scattering cross section off protons and nuclei can be sizable, the detectable signal strengths in these channels are suppressed relative to scattering off electrons.\footnote{\label{footnote:protonissue}The reason is that $\B$ scattering proceeds via $t$-channel exchange of the light mediator $\gamma'$, so the differential cross section peaks at small momentum transfers, while achieving Cherenkov radiation (or DIS scattering) requires large momentum transfers.  For elastic scattering, this logic favors electrons over protons in two different ways: an $\mathcal{O}(1~\GeV)$ $\B$ can more effectively transfer momentum to electrons compared to protons because of the heavier proton mass, and protons require a larger absolute momentum transfer to get above the Cherenkov threshold.  Compounding these issues, protons have an additional form-factor suppression, identifying proton tracks is more challenging than identifying electron tracks  \cite{PhysRevD.67.093001,Fechner:2009mq,Fechner:2009aa}, and the angular resolution for protons is worse than for electrons at these low energies \cite{Fechner:2009aa}.  We note that liquid Argon detectors are able to reconstruct hadronic final states using ionization instead of Cherenkov light, so they may be able to explore the (quasi-)elastic proton channels down to lower energies, even with smaller detector volumes \cite{Bueno:2007um,Badertscher:2010sy}.}  Thus, we focus on the elastic scattering off electrons
 \be
\B e^- \to \B e^-
\ee
as the most promising detection channel, though we present signal studies for the other channels in \App{app:detection}.  At detectors like Super-K, the signal would appear as single-ring electron events coming from the direction of the GC.

We start by discussing the kinematics of scattering off electrons (the same logic would hold for protons). In the rest frame of an electron target with mass $m_e$, the momenta of incoming and outgoing particles are:
\be
\begin{array}{rlrl}
\text{Incident~$\B$:} & p_1 = (E_B,\vec{p}\,), & \text{$\quad$ Scattered~$\B$:} &p_3 = (E_B',\vec{p}^{\, \prime}), \\
 \text{Initial~$e$:} &p_2= (m_e,0), & \text{Scattered~$e$:} &p_4= (E_e, \vec{q}\,). \label{eq:momentadefs}
\end{array}
\ee
For $\B$ coming from nearly-at-rest $\A$ annihilation,
\be
E_B = m_A.
\ee
The maximum scattered electron energy occurs when $\vec{p}$ and $\vec{p}^{\, \prime}$ are parallel:
\be
E_e^{\rm max} = m_e \frac{(E_B + m_e)^2 + E_B^2 - m_B^2}{(E_B + m_e)^2 - E_B^2 + m_B^2}. \label{eq:emax}
\ee
The minimum detectable energy is set by the analysis threshold (assumed to be above the Cherenkov threshold),
\be
E_e^{\rm min} = E_e^{\rm thresh} > \gamma_\text{Cherenkov} m_e. \label{eq:emin}
\ee
Of course, to have any viable phase space, $E_e^{\rm max} \geq E_e^{\rm min}$.  From \Eqs{eq:emax}{eq:emin}, we can also express the viable kinematic region in terms of boost factors $\gamma_e$ and $\gamma_B$ (taking $m_A \gg m_B \gg m_e$):
\be
\gamma_e^\text{min} = \frac{E_e^\text{thresh}}{m_e}, \qquad \gamma_e^\text{max} = 2 \gamma_B^2 -1, \qquad \gamma_B = \frac{E_B}{m_B} = \frac{m_A}{m_B}. \label{boost_relation}
\ee

The differential cross section for $\B$ elastic scattering off electrons is:  
\begin{equation}
\label{eq:diffBeBe}
\frac{d\sigma_{B e^- \rightarrow B e^-}}{dt }= \frac{1}{8 \pi}  \frac{ (\epsilon e g')^2}{(t - m_{\gamma'}^2)^2} \frac{8 E_B^2 m_e^2 + t(t+ 2s)}{\lambda(s,m_e^2, m_B^2)},
\end{equation}
where $\lambda (x,y,z)= x^2 + y^2 + z^2 - 2 x y - 2 xz - 2 yz$, $s = m_B^2 + m_e^2 + 2 E_B m_e$, $t = q^2 =  2 m_e (m_e - E_e)$, and one should make the replacement $E_B=m_A$ for our scenario. To give a numerical sense of the Cherenkov electron signal cross section, integrating \Eq{eq:diffBeBe} over the allowed kinematic region for the benchmark in \Eq{eq:keybenchmark} yields
\be
\label{eq:typicalsignalxsec}
\sigma_{B e^- \rightarrow B e^-}  = 1.2 \times 10^{-33}~\text{cm}^2 \left(\frac{\epsilon}{10^{-3}} \right)^2 \left(\frac{g'}{0.5} \right)^2 \left(\frac{20~\MeV}{m_{\gamma'}} \right)^2,
\ee
for an experimental threshold of $E_e^{\rm thresh} = 100~\MeV$.  
The approximate scaling is derived in the limit $m_e E_e^{\rm thresh} \ll m_{\gamma'}^2 \ll m_e E_e^{\rm max}$, where the dependance on $E_B$, $m_B$, and $E_e^{\rm thresh}$ is weaker than polynomial, which holds in the vicinity of the benchmark point but not in general. For completeness, the full cross section for $\B$-electron scattering without an energy threshold cut is
\be
\sigma_{B e^- \rightarrow B e^-}^\text{tot}  = 1.47 \times 10^{-33}~\text{cm}^2 \left(\frac{\epsilon}{10^{-3}} \right)^2 \left(\frac{g'}{0.5} \right)^2 \left(\frac{20~\MeV}{m_{\gamma'}} \right)^2. \label{eq:xsectot}
\ee
Since this cross section is rather high, we have to account for the possibility that $\B$ particles might be stopped as they pass through the earth.  In \Sec{subsec:earth}, we find that the attenuation of $\B$ particles is mild, so we will treat the earth as transparent to $\B$ particles in our analysis.

\begin{figure}[t]%
     \subfloat[]{\includegraphics[width= 7.1 cm]{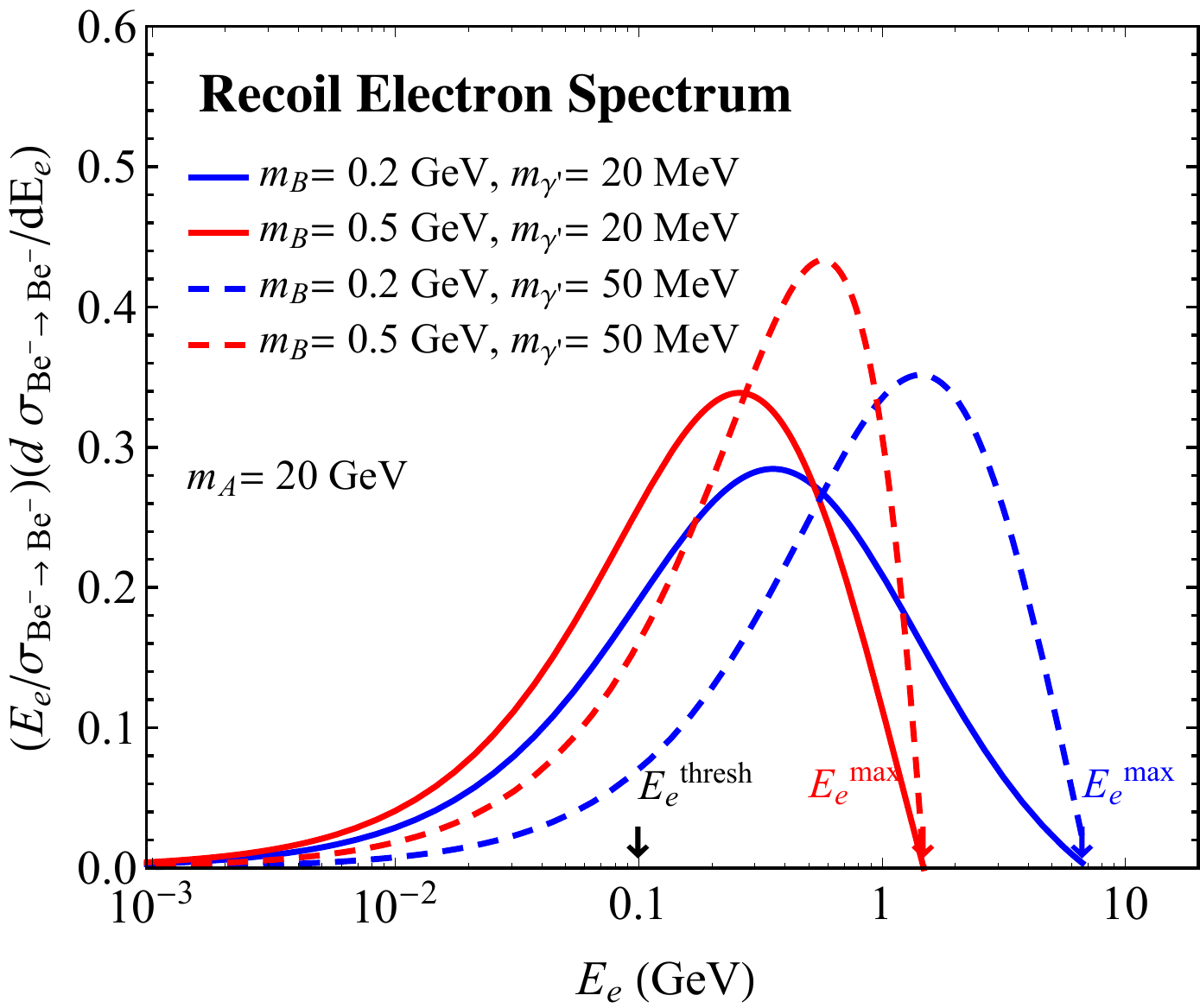}\label{fig:electronkinematics:a}}
    \qquad
       \subfloat[]{\includegraphics[width = 7.55 cm]{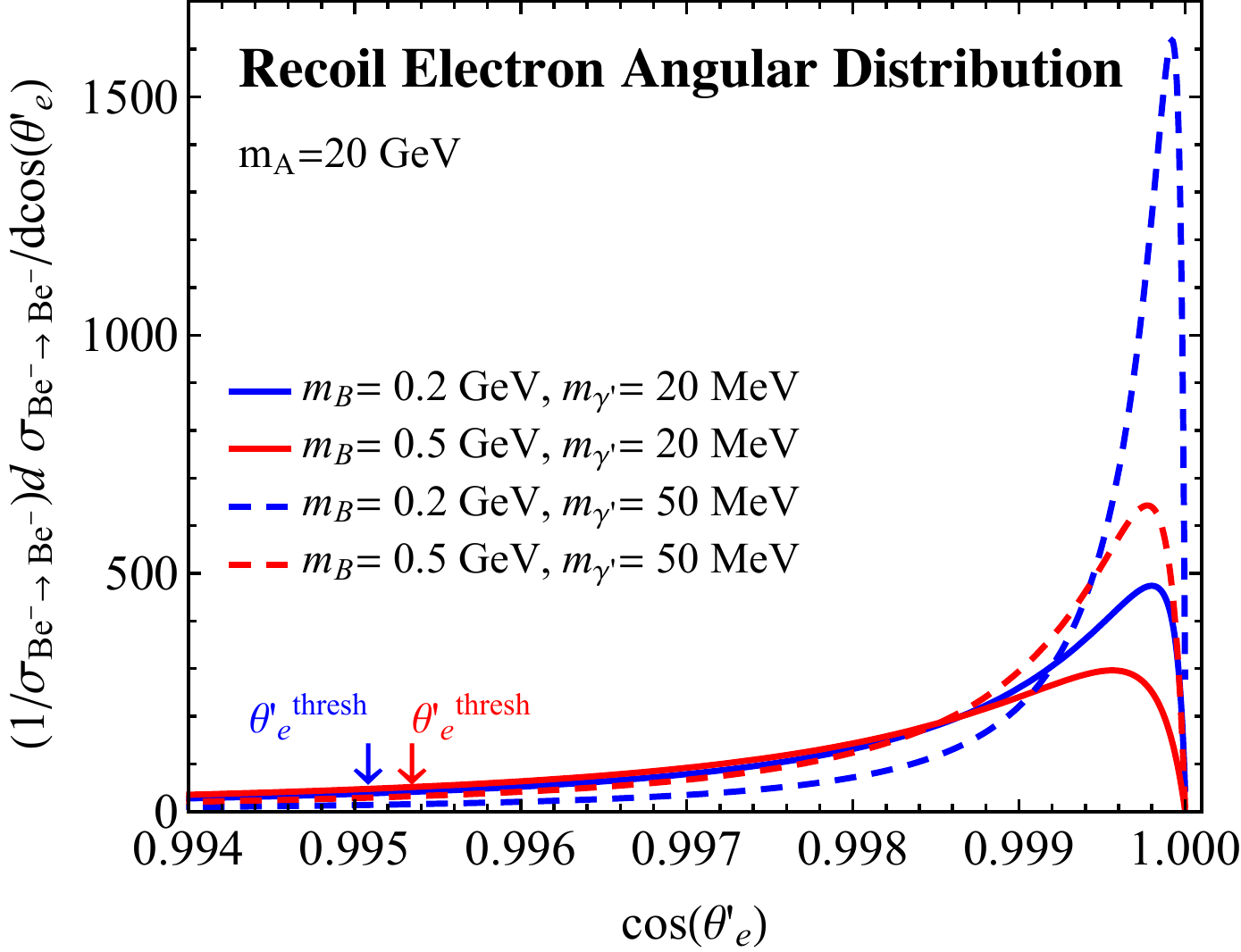}\label{fig:electronkinematics:b}} 
        \caption{(a) Normalized recoil electron spectrum for different benchmark scenarios.  Also indicated is the maximum scattered electron energy, given by \Eq{eq:emax} as well as the experimental threshold of Super-K in the Sub-GeV category (See \Eq{eq:subgev}). (b) Recoil electron angular distribution for the same signal benchmarks, assuming a $\B$ particle coming directly from the GC. The cutoff angle $\theta^{\prime \text{thresh}}_e$ is obtained by substituting the 100 MeV energy threshold into \Eq{eq:thetaprime}. }%
    \label{fig:electronkinematics}%
\end{figure}

 In \Fig{fig:electronkinematics:a}, we show the normalized, logarithmic electron spectrum for different benchmarks, including the one from \Eq{eq:keybenchmark}.  The electron energy $E_e$ peaks at relatively low values due to the $t$-channel $\gamma'$, as discussed further in footnote~\ref{footnote:protonissue}.   We note that the position of the peak depends both on $m_B$ and $m_{\gamma'}$, though the dominant effect of $m_{\gamma'}$ is to change the overall signal cross section (not visible in this normalized plot).  The angular distribution of the recoil electron is shown in \Fig{fig:electronkinematics:b}. The signal is very forward peaked, as expected from $m_B \gg m_e$. This is advantageous when looking for boosted DM from the GC, since the recoil electrons' direction is tightly correlated to that of the $\B$'s.  

\begin{figure}[t]%
	\centering
	\includegraphics[scale=0.7]{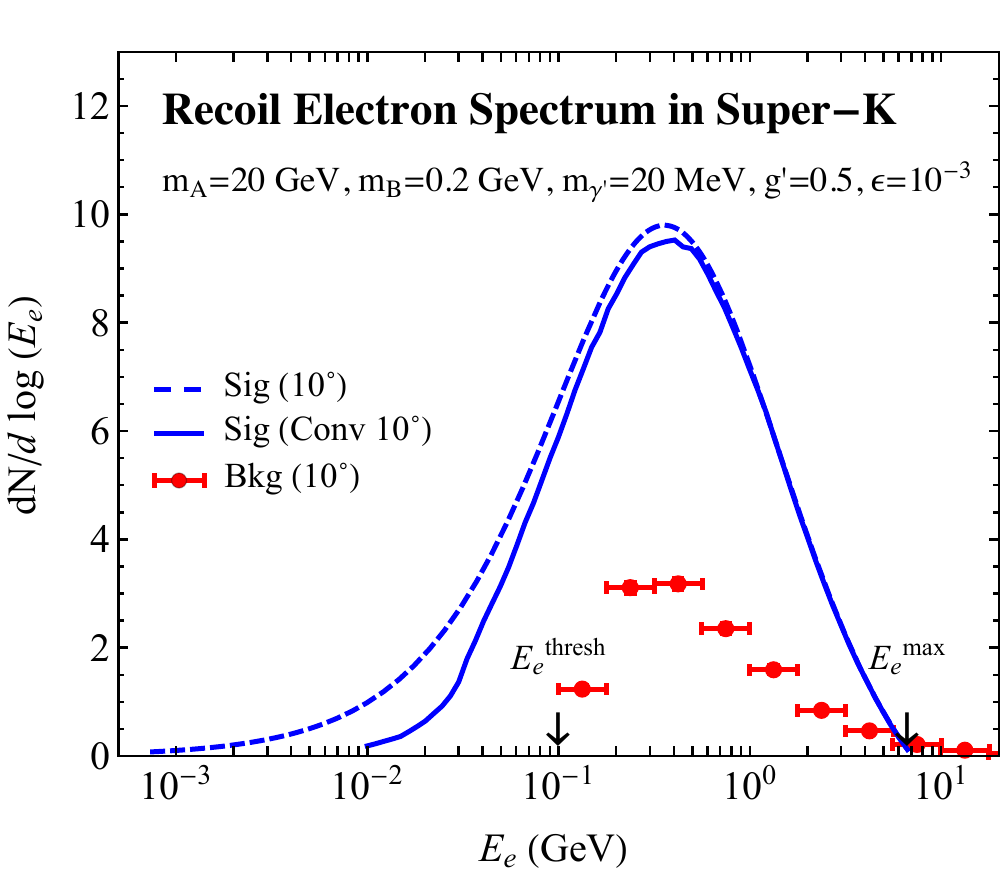} %
	\caption{Energy spectrum of signal and background events, normalized to the expected event yield over one year.  The blue dashed line corresponds to the naive formula in \Eq{eq:naivespectrum} for the number of signal events in a $10^\circ$ search cone.  The solid blue line is spectrum obtained from the convolution in \Eq{eq:signalconvolve}.   The background spectrum of CC $\nu_e$ and $\overline{\nu}_e$ events comes from Super-K \cite{Ashie:2005ik}, scaled by a factor $\pi (10^\circ)^2/(4 \pi)$ to account for the nominal $10^\circ$ search cone.  Note that data is available only for $E_e>100~\MeV$, which is the same experimental threshold given in \Eq{eq:subgev}.  Also indicated is the maximum scattered electron energy, given by \Eq{eq:emax}.}
	\label{fig:spectrum}%
\end{figure}

In \Fig{fig:spectrum}, we compare the energy profile of the signal to the observed background electron events at SK-I \cite{Ashie:2005ik}.  Using the benchmark model in \Eq{eq:keybenchmark}, we plot the (logarithmic) energy spectrum of the yearly signal event yield within a cone of $10^\circ$ around the GC:
\be
\label{eq:naivespectrum}
\frac{dN}{d\log E_e} = E_e \frac{dN}{dE_e} = \Delta T N_\text{target}  \Phi^{10^\circ}_\text{GC} E_e \frac{d \sigma_{B e^- \rightarrow B e^-}}{d E_e},
\ee
where $\Phi_\text{GC}^{10^\circ}$ is defined in \Eq{eq:PhiGC}, $\Delta T$ is a year, and $N_\text{target}$ is the number of targets (electrons) at Super-K.  Anticipating the analysis of \Sec{sec:signalrates}, we also plot a more realistic spectrum obtained by convolving the signal scattering cross section $\B e^- \rightarrow \B e^-$ with the shape of the DM halo.   This convolved spectrum matches nicely to the naive spectrum from \Eq{eq:naivespectrum}, as expected given the peaked nature of the angular spectrum in \Fig{fig:electronkinematics:b}, with signal losses at low energies arising because less energetic electrons can be more easily deflected outside the search cone.  Once the background from \Ref{Ashie:2005ik} is scaled by the appropriate factor of $(\pi (10^\circ)^2)/(4 \pi) \approx 8 \times 10^{-3}$, the signal for this benchmark is visible above the background, though the peak location is (accidentally) at a similar location.

\subsection{Backgrounds to Boosted Dark Matter}
\label{subsec:background}

The major background to the boosted DM signal comes from atmospheric neutrinos, which are produced through interactions of cosmic rays with protons and nuclei in the earth's atmosphere.  Atmospheric neutrino energy spectrum peaks around 1 GeV and follows a power law $E^{-2.7}$ at higher energies \cite{Amenomori:2008zzd}.  The scattering process $\B e^- \to \B e^-$ with an energetic outgoing electron faces a large background from charged-current (CC) electron-neutrino scattering $\nu_e n \rightarrow e^- p$ when the outgoing proton is not detected, as well as $\overline{\nu}_e p \rightarrow e^+ n$ since Cherenkov-based experiments cannot easily distinguish electrons from positrons.  For $\mathcal{O}(1~\GeV)$ neutrinos, the CC cross sections are \cite{Formaggio:2013kya} 
\begin{eqnarray}
\label{eq:CCxsec}
\sigma_{\rm CC}^{\nu_e} &\approx& 0.8 \times 10^{-38}~\text{cm}^2 \left(\frac{E_\nu}{\GeV} \right), \\
\sigma_{\rm CC}^{\overline{\nu}_e} &\approx& 0.3 \times 10^{-38}~\text{cm}^2 \left(\frac{E_\nu}{\GeV} \right).
\end{eqnarray}
While smaller than the expected signal cross section in \Eq{eq:typicalsignalxsec}, the atmospheric neutrino flux is much higher than the boosted $\B$ flux. The neutral current process $\nu_e e^- \rightarrow \nu_e e^-$ can also mimic the signal but it is subdominant to the CC interaction due to $m_e/m_p$ suppression \cite{Formaggio:2013kya}.

\begin{figure}[t]%
	\centering
	\includegraphics[scale=0.45]{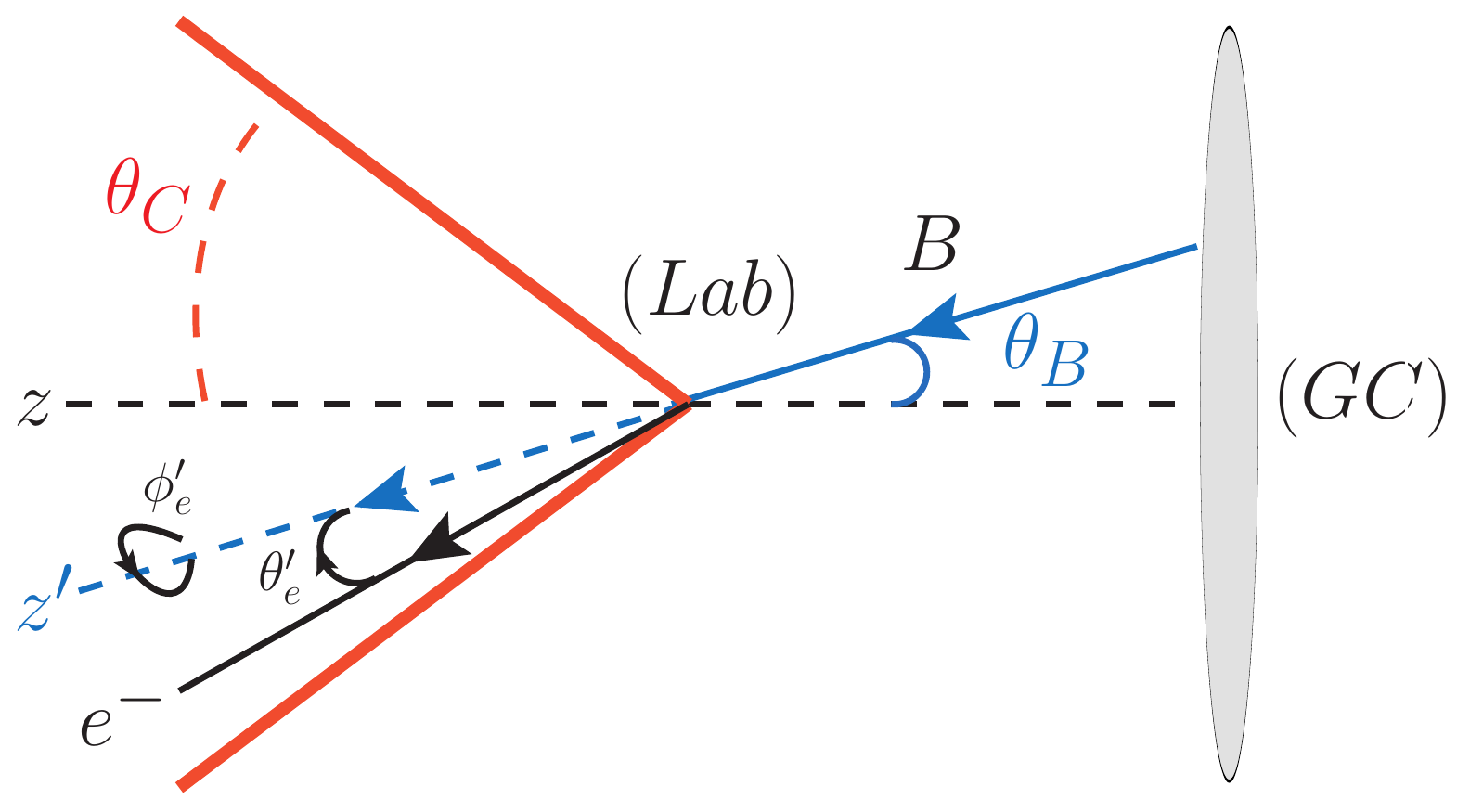} %
	\caption{Angles involved in boosted DM detection.  When a $\B$ particle arrives at an angle $\theta_B$ from the GC, it scatters to produce an electron at angle $\theta_e$ with respect to $z$ ($\theta_e'$ and $\phi'_e$ with respect to $z'$).  To better isolate the signal from the uniform atmospheric neutrino background, we impose a search cone of half-angle $\theta_C$.}
	\label{fig:angles}%
\end{figure}

There are a number of discriminants one could use to (statistically) separate our signal from the neutrino background.
\begin{itemize}
\item \textit{Angular restriction}:  Boosted $\B$ particles have a definite direction because they come from the GC.  In galactic coordinates, the atmospheric neutrino background has no preferred direction.  Therefore, one can impose that the detected electron falls within a cone of half-opening angle $\theta_C$ with respect to the GC.  As shown in \Fig{fig:angles}, there are two relevant axes to consider:  the $z$-axis connecting the earth to the GC and the $z'$-axis in the direction that the $\B$ travels along.  Through $\B e^- \to \B e^-$ scattering, a $\B$ particle coming from an angle $\theta_B$ ($\theta'_B = 0$) will yield a final state electron with scattering angle $(\theta'_e, \phi'_e)$, with 
\begin{align}
\cos \theta_e &= \cos \theta_B \cos \theta'_e - \sin \theta_B \sin \phi'_e \sin \theta'_e, \label{eq:defthetae}\\
\cos \theta'_e &= \frac{(m_A + m_e)}{\sqrt{m_A^2 - m_B^2} } \frac{\sqrt{ E_e - m_e} }{\sqrt{ E_e + m_e} }, \label{eq:thetaprime}
\end{align}
and $\phi'_e$ uniformly distributes between $0$ and $2\pi$.  To the extent that the electron energy is large and $m_A \gg m_B \gg m_e$, we have $\cos \theta_e \approx \cos \theta_B$.  As we will see in \Sec{subsec:significance}, the optimum angle $\theta_C$ to maximize the signal acceptance while minimizing the neutrino background is around $10^\circ$, assuming perfect angular resolution.

\item \textit{Energy restriction}:  Boosted $\B$ particles have a mono-energetic spectrum ($E_B = m_A$), compared to the continuous atmospheric neutrino energy spectrum.  This implies a correlation between the measured $E_e$ and $\cos \theta_e$.  That said, we suspect that the typical angular resolution of neutrino experiments is not fine enough to make use of this feature.  In fact, more important than energy resolution is to have a low energy threshold, since as shown in \Fig{fig:electronkinematics}, the signal cross section peaks at small $E_e$.
\item \textit{Absence of muon excess}:  The process $\B e^- \to \B e^-$ does not have a corresponding muon signature, whereas the neutrino CC process $\nu_e n \rightarrow e^- p$ is always accompanied by $\nu_\mu n \rightarrow \mu^- p$.  So an electron excess from boosted DM should \emph{not} have a correlated excess in muon events. One can also require fully-contained events to reduce the cosmic ray muon background.

\item \textit{Anti-neutrino discrimination:} The anti-neutrino background $\overline{\nu}_e p \rightarrow e^+ n$ is in principle reducible since it involves a final state positron instead of an electron. Super-K cannot perfectly distinguish $\nu_e$ from $\overline{\nu}_e$ events, but as of the SK-IV analyses, they have used likelihood methods to separate these two categories by studying the number of decay electrons in each process. The purity of the $\nu_e$ sample is $62.8 \%$ and that of $\overline{\nu}_e$ is $36.7\%$ \cite{LeeKaThesis,Dziomba}. We have not used this feature in our current analysis. Adding gadolinium to Super-K would help tagging neutrons from the $\overline{\nu}_e$ CC process, and thus might improve the purity of these samples \cite{Beacom:2003nk}.

\item \textit{Multi-ring veto}:  The process $\B e^- \to \B e^-$ leads to electron-like single-ring events only, without correlated multi-ring events. In contrast, neutrino CC process $\nu_e n \rightarrow e^- p$ can lead to multi-ring events when the outgoing proton energy is above Cherenkov threshold \cite{Fechner:2009aa}, or when the scattering is inelastic so that other charged hadronic states such as $\pi^\pm$ are produced. Argon-based detectors could improve the background discrimination since they can detect the hadronic final states from neutrino scattering better than water-based experiments.  We note that for some extreme parameters (increasing $g'$ or $\epsilon$), it is possible for $\B$ to interact twice (or more) within the detector, also creating a potential multi-ring signal (or a lightly-ionizing track in a scintillator detector).  That said, for such high cross sections, the signal would be heavily attenuated while traversing the earth (see \Sec{subsec:earth}).  Another potential disadvantage of a multi-ring veto is that we might miss out on interesting signals such as $\B$ scattering accompanied by $\gamma'$ bremsstrahlung ($\B e^- \to \B e^- \gamma' \to \B e^- e^+ e^-$).

\item \textit{Solar neutrino/muon veto}:  Solar neutrinos dominate the background under around 20 MeV \cite{Gaisser:2002jj}, though one can of course preform an analysis in solar coordinates and exclude events from the sun.  In addition, there is a background from muons that do not Cherenkov radiate but decay to neutrinos in the detector volume; these are relevant in the range of 30--50 MeV and can be mitigated through fiducial volume cuts \cite{Bays:2011si}.  To avoid both of these complications, we will use a cut of $E_e > 100~\MeV$ in our analysis below. Of course the threshold of 100 MeV in Super-K could be brought down as low as 50 MeV (where solar neutrino backgrounds start to dominate).  The potential advantage of looking in the 50-100 MeV range is that the backgrounds from atmospheric neutrinos are lower.  The main disadvantage is the degradation of the angular resolution of the detector \cite{Abe:2010hy}.  

\end{itemize}
The first two points favor detectors with excellent angular resolution and low energy thresholds on the outgoing electron.  The next three points mean that one could distinguish the boosted DM signal from neutrinos coming from WIMP DM annihilation in the GC; boosted DM only gives a single-ring electron signal whereas neutrinos from WIMPs would give equal contributions to an electron and muon signal, both single- and multi-ring events, and equal contributions to a neutrino and anti-neutrino signal.  The last point suggests the interesting possibility of looking for boosted DM from the sun due to DM solar capture, though in the particular model we study in this paper, the solar capture rate is too small to be visible, and any boosted DM particles from the sun would face considerable solar attenuation (see \Sec{subsec:earth} below).  The above criteria can be thought of as a general algorithm for background rejection, while specifics can be tailored to a particular experiment. For instance, ``multi-ring veto'' does not apply to PINGU where Cherenkov rings cannot be reconstructed and all non-$\mu$-like events are classified as ``cascade events''.

\subsection{Impact of Earth Attenuation} \label{subsec:earth}
 
As seen in \Eq{eq:xsectot}, the signal cross section $\sigma_{Be^- \rightarrow Be^-}$ is relatively high, so as they cross the earth, the $\B$ particles might get deflected and lose energy.   This is a particularly important effect for Northern Hemisphere experiments like Super-K, where a typical $\B$ would have to traverse through $\sim 10^5$ km (75\% of the earth's diameter).  The dominant cause for energy loss is (minimum) ionization of atoms.  While not relevant for detection, the main source of angular deflection is scattering off nuclei. In the following, we base our discussion on the standard analysis of particle propagation through matter as developed in the PDG \cite{Beringer:1900zz}.

First we estimate the $\B$'s energy loss.  Just as for a heavy charged particle traversing the earth (see, e.g., \Ref{Albuquerque:2003mi}), the main energy loss mechanism is through ionization.   For $\beta \gamma$ factors of 10--100, a muon loses $\approx$ 1 GeV of energy per meter of rock \cite{Beringer:1900zz}.  A muon scatters off nuclei via a $t$-channel $\gamma$ exchange, while a $\B$ scatters off nuclei via the exchange of a $\gamma'$.  We can approximate the length required for a $\B$ to lose 1 GeV by scaling the couplings and the propagator of the $\B$-$e^-$ scattering process to those of the $\mu$-$e^-$ scatterings:
\be
L_B \approx L_\mu \, \frac{e^2}{\epsilon^2 g'^2} \left(\frac{t- m_{\gamma'}^2}{t} \right)^2,
\ee
where $t = 2 m_e (m_e - E_e ) \approx -10^{-4}~\GeV^2$ for our key benchmark in \Eq{eq:keybenchmark}.  In this case, $\B$ loses $\approx 1~\GeV$ per $9 \times 10^8 ~\text{cm}$, giving rise to a total expected loss of 1 GeV per trip through the earth ($R_\oplus = 6.4 \times 10^8 ~\text{cm}$).  Since 1 GeV of energy loss is never more than $\simeq 10 \%$ of the $\B$'s initial energy in the parameter space of interest, we will assume the earth is transparent to $\B$'s for the rest of the analysis.  Accounting for the energy loss is approximately equivalent to shifting the plots in \Figs{fig:signalAB}{fig:significance} by the energy loss on the $m_A$ axis.  The parameter space of small $m_A$ is the most affected, but that region is already constrained by CMB bounds as shown in \Sec{sec:constraints}.

Turning to the angular distribution, the dominant source of deflection is from elastic scattering off of nuclei.   Note that  $\B$-$e^-$ scattering processes lead to very small angles of deflection because of the mass hierarchy $m_B \gg m_e$; indeed for the key benchmark in \Eq{eq:keybenchmark}, the maximum possible deflection per scatter is 0.14$^\circ$.  In contrast, Coulomb-like scattering of $\B$ particles of nuclei can give rise to a more substantial deflection (including full reversal).  The mean-square change in angle per collision process is
\begin{align}
\langle \theta_B^2 \rangle &\simeq 2 -  2 \, \langle \cos(\theta_B) \rangle, \nonumber \\ 
\langle \cos(\theta_B) \rangle &= \frac{1}{\sigma_{BN\rightarrow BN}} \int_{0}^{E_N^\text{max}} \cos (\theta_B (E_N)) \frac{d \sigma_{BN\rightarrow BN}}{d E_N} dE_N  \simeq \cos(0.2 ^\circ), \label{eq:cosav}
\end{align}
where $\sigma_{BN\rightarrow BN}$ is the scattering cross section of $\B$'s off a nucleus $N$ (see \Eq{eq:xsecfe}), and $E_N^\text{max}$ is defined similarly to \Eq{eq:emax}.  In the last step of \Eq{eq:cosav}, we have inserted the benchmark value from \Eq{eq:keybenchmark}.  Treating the deflection of $\B$ particles as a random walk through the earth, the total deflection is
\be
\langle \theta_{\rm total}^2 \rangle^{1/2} = \langle \theta_B^2 \rangle^{1/2} \sqrt{\frac{\ell_{BN\rightarrow BN}}{R_\oplus}},
\ee
where the quantity under the square root is the number of steps (interactions).  The mean free path to interact with a nucleus of charge number $Z$ and atomic number $A$ is
\be
\ell_{BN\rightarrow BN} = \frac{1}{n \, \sigma_{BN \rightarrow BN}  \left(\frac{Z}{26} \right)^2 \left(\frac{55.84}{A} \right)}  = 1.5 \times 10^7 ~\text{cm},
\ee
where $n$ is the number density.  Under the conservative assumption that the earth is entirely made of iron (the benchmark $A$ and $Z$ values above), and taking the mass density of earth to be $\rho = 5.5 \text{ g/cm}^3$, the number of scatters is $\approx 64$ for the benchmark in \Eq{eq:keybenchmark}, giving a total deflection of: 
\be
\langle \theta_{\rm total}^2 \rangle^{1/2} = 1.6^\circ.  \label{eq:totdeflection}
\ee
We checked that for different values of $m_A$, $m_B$, and $m_{\gamma'}$, the total deflection does not vary much compared to \Eq{eq:totdeflection}.  Since this deflection is small compared with the search cone of 10$^\circ$ that is used in \Sec{sec:signalrates}, we neglect the angular deflection of $\B$'s in our analysis.

Interestingly, if a signal of boosted DM is found, we could potentially use the earth attenuation to our advantage by correlating candidate signal events with the position of the GC with respect to the experiment. Indeed, with high enough statistics, the effect of earth shadowing would give rise to time-dependent rates, energies, and angles for $\B$ scattering.  As mentioned in footnote~\ref{footnote:suncapture}, solar attenuation would have an adverse effect on possible boosted DM signals from the sun.  Since the radius of the sun is 100 times larger than that of the earth, the $\B$ particles would lose a factor of 100 more energy, so we would need $m_A \gtrsim \mathcal{O} \text{(100 GeV-1 TeV)}$ for $\B$ particle to escape the sun. Alternatively, for a smaller scattering cross section of $\B$ particles with the SM, the sun might then be a viable source of signal \cite{Berger:2014sqa}.

\section{Detection Prospects for Present and Future Experiments}
\label{sec:experiments}

\begin{table}[t]
\begin{center}
\begin{tabular}{ccccc}
\hline\hline
Experiment & Volume (MTon) & $E_e^{\rm thresh}$ (GeV) & $\theta_e^{\rm res}$ (degree) & Refs. \\
\hline
Super-K & $2.24 \times 10^{-2}$ & 0.01 & $3^\circ$ &\cite{Ashie:2005ik} \\
Hyper-K & $0.56 $ &0.01 & $3^\circ $& \cite{Kearns:2013lea}\\
\hline
IceCube &$10^3$ & 100   & $30^\circ$ & \cite{Abbasi:2011eq, Aartsen:2013vca} \\
PINGU & $0.5$ & 1 &  $23^\circ (\text{at GeV scale})$ & \cite{Aartsen:2014oha}\\
MICA & $5$ & 0.01  & $30^\circ (\text{at 10 MeV scale})$ & \cite{MICA,MICA2} \\
\hline
\hline
\end{tabular}
\end{center}\caption{List of experiments studied in this paper, their angular resolutions $\theta_e^{\rm res}$ on the Cherenkov-emitted electron direction, and the typical minimum energy threshold $E_e^{\rm thresh}$ of the detected electron. We note here that IceCube has too high of an energy threshold for our analysis, but we are interested in its future low-energy extensions such as PINGU and MICA. For PINGU, we have scaled the nominal volume (1 MTon) down by a factor of 2 to estimate particle identification efficiency.  The MICA values are speculative at present, since there is not yet a technical design report.  } \label{tab:expt_summary1}
\end{table}

We now assess the detection prospects for boosted DM at present and future detectors for neutrinos and/or proton decay.  In \Tab{tab:expt_summary1}, we summarize the (approximate) capacities/sensitivities of some of the representative relevant experiments, given in terms of the detector volume $V_{\rm exp}$, electron energy threshold $E_e^{\rm thresh}$, and angular resolution $\theta_e^{\rm res}$.  From this table, we can already anticipate which experiments are going to be best suited for boosted DM detection.

Due to the relatively small flux of boosted DM, a larger volume detector, such as IceCube, KM3NeT, or ANTARES would be favored in order to catch more signal events.  However, the energy threshold for the original IceCube are much too high for our purposes (and similarly for KM3NeT and ANTARES), since the energy transferred to the outgoing electron is suppressed due to the $t$-channel $\gamma'$ (see \Fig{fig:electronkinematics}).  Even the $\simeq 1$ GeV threshold of PINGU is not ideal, though it will have some sensitivity.

So although Super-K/Hyper-K have smaller detector volumes, their low energy threshold is better matched to the boosted DM signal.  In addition, Super-K/Hyper-K have excellent angular resolution,\footnote{More accurately, the angular resolution of fully contained 1 ring Multi-GeV electrons is $1.5^\circ$ while that of fully contained 1 ring Sub-GeV electrons is less than $3.3^\circ$ as shown in \Ref{FDthesis}.  }
which makes it possible to optimize the $\theta_C$ search cone criteria.  Ultimately, MICA would offer better coverage in the energy range of interest.  It is also worth mentioning that the proposed experiments for proton decay based on large scale liquid Argon detectors \cite{Bueno:2007um,Badertscher:2010sy} can also be sensitive to boosted DM due to their low thresholds and large volume.  As mentioned in footnote~\ref{footnote:protonissue}, liquid Argon detectors may also have sensitivity to the proton scattering channel as well.

In the following subsections, we discuss event selection, signal/background rates, and expected signal significance in the above experiments.  For signal-only studies of the subdominant channels involving $\B$ scattering off protons/nuclei, see \App{app:detection}.

\subsection{Event Selection}

As discussed in \Secs{bDM_detect}{subsec:background}, the leading boosted DM signal comes from elastic scattering off electrons ($\B e^- \to \B e^-$) and the leading background is from atmospheric neutrinos (mostly $\nu_e n \rightarrow e^- p$).  In principle, one could use the full multivariate information about the kinematics of the outgoing electron to separate signal and background.  In order to keep the analysis simple, we will do a cut-and-count study to estimate the sensitivity.

To isolate events coming from the GC, we will use the search cone $\theta_C$ described in \Fig{fig:angles}.   The dominant background from CC $\nu_e$ scattering of atmospheric neutrinos is assumed to be uniform across the sky, so the background in a search cone of half-angle $\theta_C$ scales proportional to $\theta_C^2$.   Of course, one cannot take $\theta_C$ to be too small, otherwise the signal acceptance degrades.  To optimize for the signal significance in \Sec{subsec:significance}, we will convolve the angular dependence of halo $J$-factor and the angular dependence of the $\B e^- \to \B e^-$ cross section to figure out the optimum $\theta_C$.  Anticipating that result, we will find
\begin{equation}
\label{eq:searchconechoice}
\theta_C = \text{max} \{10^\circ, \theta_e^\text{res}\},
\end{equation}
where $10^\circ$ applies to the high resolution experiments (Super-K/Hyper-K), and the other experiments are limited by their angular resolutions.

From \Eqs{eq:emax}{eq:emin}, we have minimum and maximum electron energies $E_e^{\rm min}$ and $E_e^{\rm max}$ for the signal.  Ideally, one would adjust the energy selection for a given value of $m_A$ and $m_B$, and try to push the analysis threshold $E_e^{\rm min}$ to be as low as possible.  To be conservative, we will take the standard Super-K event categories for fully-contained single-ring  electron events (see e.g.~\Ref{Dziomba})
\begin{align}
\text{Sub-GeV: } & \{ 100~\MeV, 1.33~\GeV\},  \label{eq:subgev}\\
\text{Multi-GeV: } & \{ 1.33~\GeV, 100~\GeV\},
\end{align}
without attempting to do finer energy binning. For Super-K, Hyper-K, and MICA, we will use both categories as separate event selections; for the Sub-GeV category we will choose only zero-decay events.  PINGU has a higher energy threshold and cannot reconstruct Cherenkov rings nor efficiently separate $\mu$-like and $e$-like events near threshold, so we will only use the Multi-GeV category, while also adding in backgrounds from multi-ring events and $\mu$-like events; we will also scale the PINGU effective volume down by a factor of 2 to account for an estimated reconstruction efficiency of $\sim50\%$ \cite{Aartsen:2014oha}.  Note that the 100 MeV lower bound of the Sub-GeV category is above the nominal 10 MeV threshold of Super-K, so there is room for improved signal acceptance.  Similarly, when the 1.33 GeV upper bound of the Sub-GeV category is above $E_e^{\rm max}$, then we are overestimating the background.  

\subsection{Signal Rates}
\label{sec:signalrates}

Imposing the $\theta_C$ and energy range requirements, the number of signal electron events is:
\begin{eqnarray}
N^{\theta_C}_\text{signal}&=& \Delta T \, N_\text{target} \, (\Phi_\text{GC} \otimes \sigma_{B e^- \rightarrow B e^-})\bigr\rvert_{\theta_C} \nonumber \\
&=& \frac{1}{2} \Delta T \, \frac{10 \, \rho_\text{Water/Ice} V_\text{exp}}{m_{\rm H_2O}} \frac{r_\text{Sun}}{4 \pi} \left( \frac{\rho_\text{local}}{m_A}\right)^2 \langle \sigma_{A \overline{A} \rightarrow B \overline{B} } v\rangle_{v \rightarrow 0}  \\\nonumber
&&~\times  \int_0^{2 \pi} \frac{d \phi'_e}{ 2 \pi}  \int^{\theta'_{\rm max}}_{\theta'_{\rm min}} d \theta'_e \, \sin \theta'_e \, \frac{d \sigma_{B e^- \rightarrow B e^-}}{d \cos\theta'_e} \int_0^{\pi/2} d \theta_B \sin \theta_B  \, 2\pi J(\theta_{B}) \Theta(\theta_C-\theta_e), \label{eq:signalconvolve}
\end{eqnarray}
where $\Delta T$ is the time duration of the observation, $N_{\text{target}}$ is the number of target electrons, $\Phi_{\text{GC}}$ is the DM flux from the GC, and $\sigma_{B e^- \rightarrow B e^-}$ is the $\B$-electron scattering cross section (which depends on the energy integration range in \Eq{eq:diffBeBe}). The factor of 10 in the second line is the number of electrons per molecule of water. The DM flux and scattering cross section have to be convolved in order to isolate events that pass the $\theta_C$ requirement, and the angles in the last line are the same as in \Fig{fig:angles} with $\theta_e$ given in \Eq{eq:defthetae}.  The integration limits $\theta'_e \in \{\theta'_{\rm min}, \theta'_{\rm max}\}$ are given by \Eq{eq:thetaprime} by requiring $E_e \in \{E_e^{\rm max},E_e^{\rm min} \}$ (note the reversal of the limits, and that $\theta'_{\rm min} = 0$ if \Eq{eq:emax} is more restrictive than the energy categories above).

\begin{figure}[t]%
  \centering
    \subfloat[]{{\includegraphics[width=5cm]{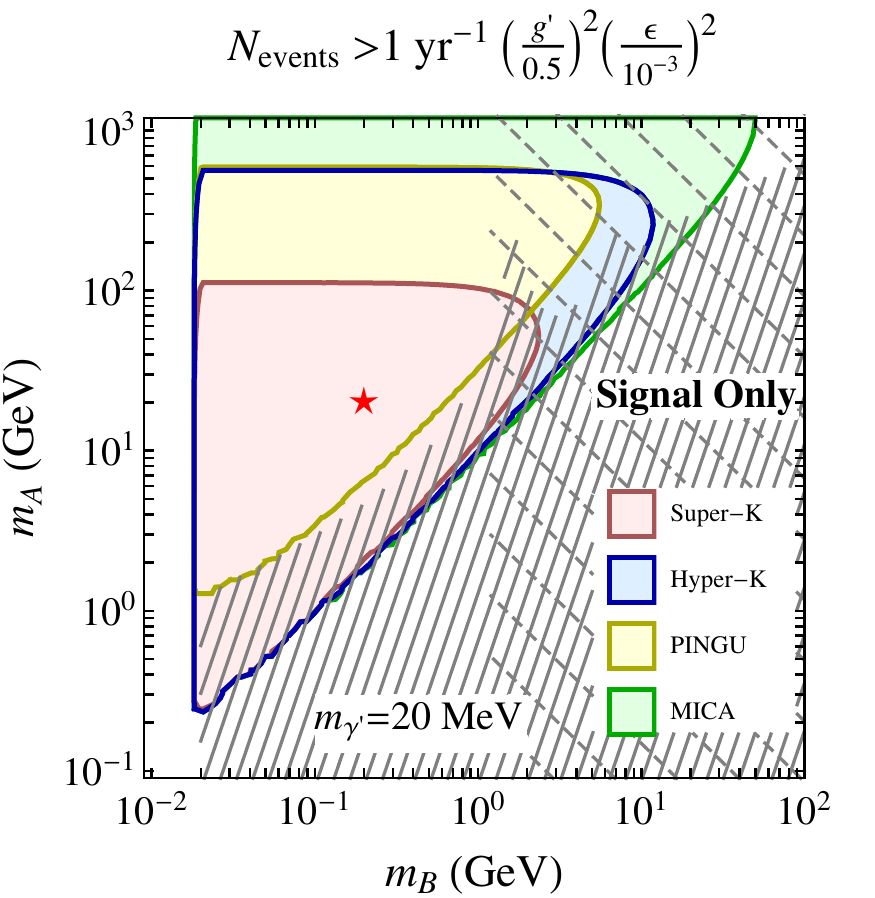} }}%
    \subfloat[]{{\includegraphics[width=5cm]{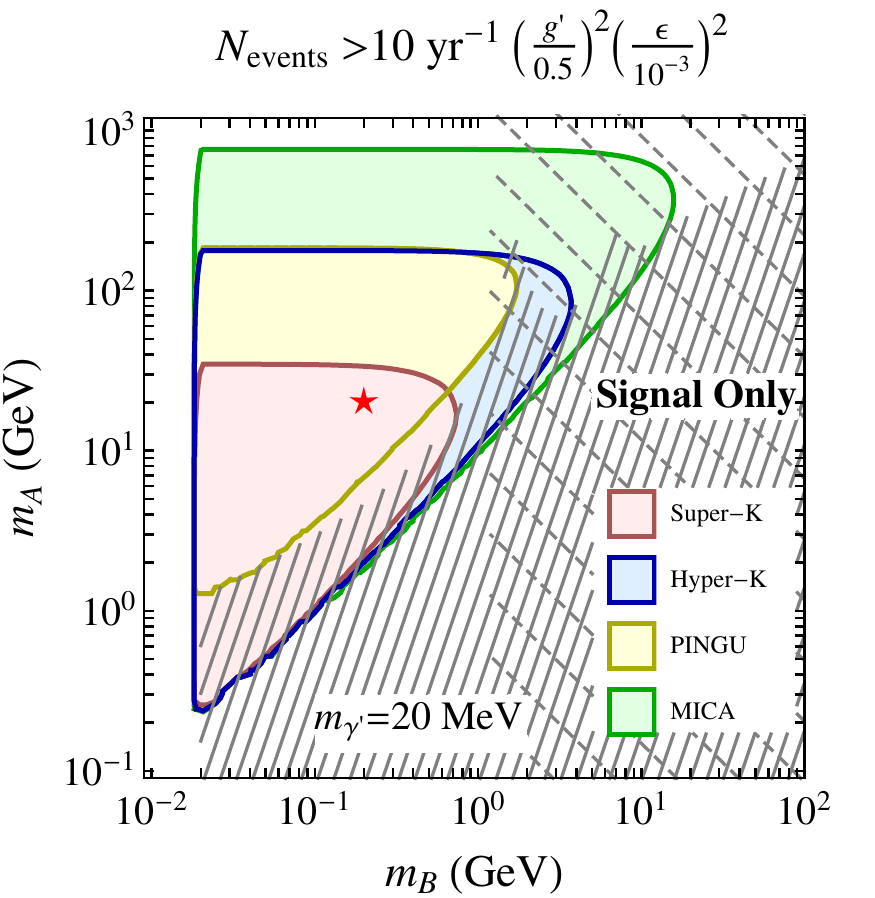} }}%
    \subfloat[]{{\includegraphics[width=5cm]{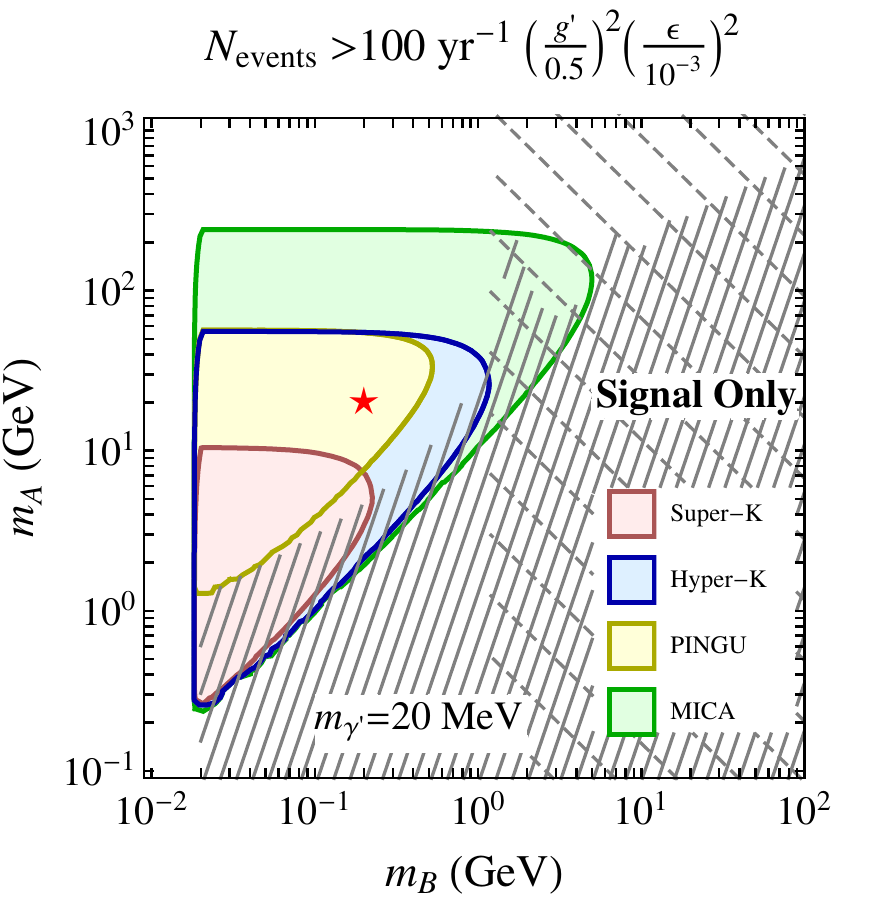} }}%
         \qquad
      \subfloat[]{{\includegraphics[width=5cm]{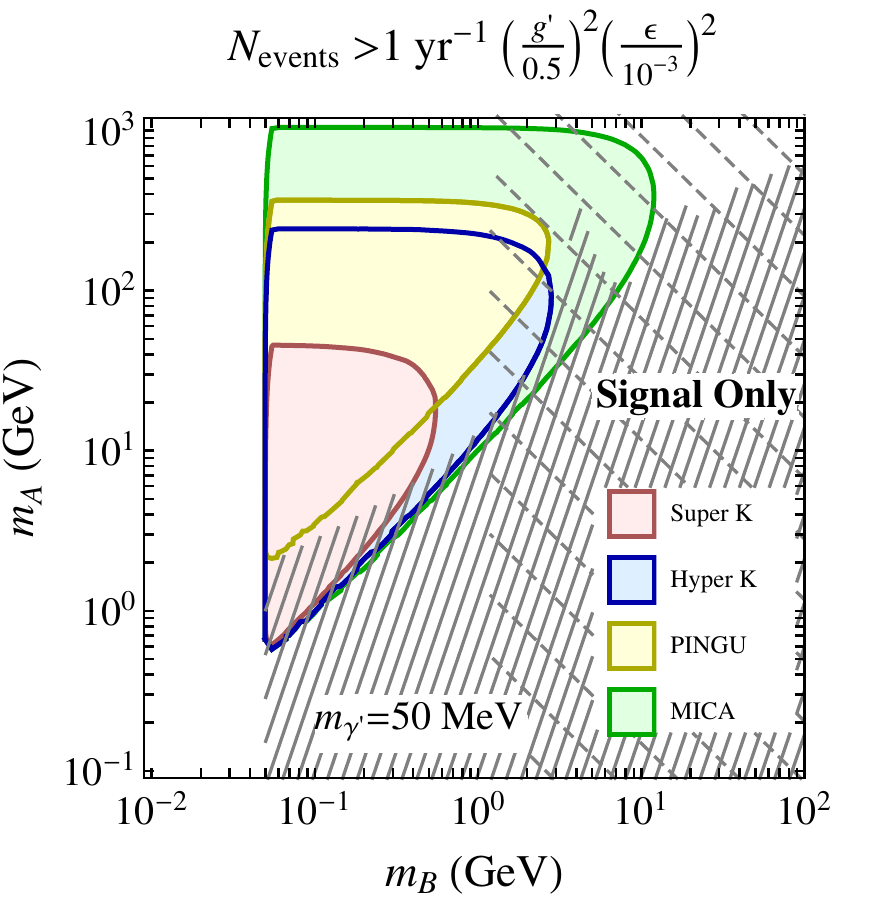} }}%
    \subfloat[]{{\includegraphics[width=5cm]{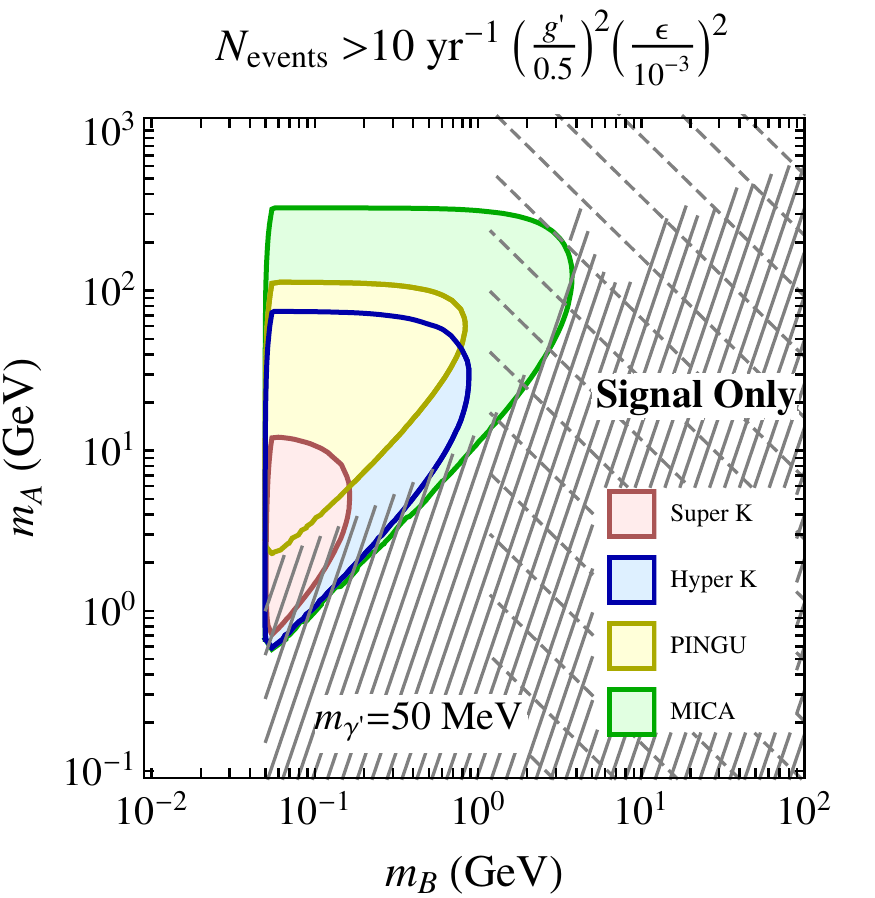} }}%
    \subfloat[]{{\includegraphics[width=5cm]{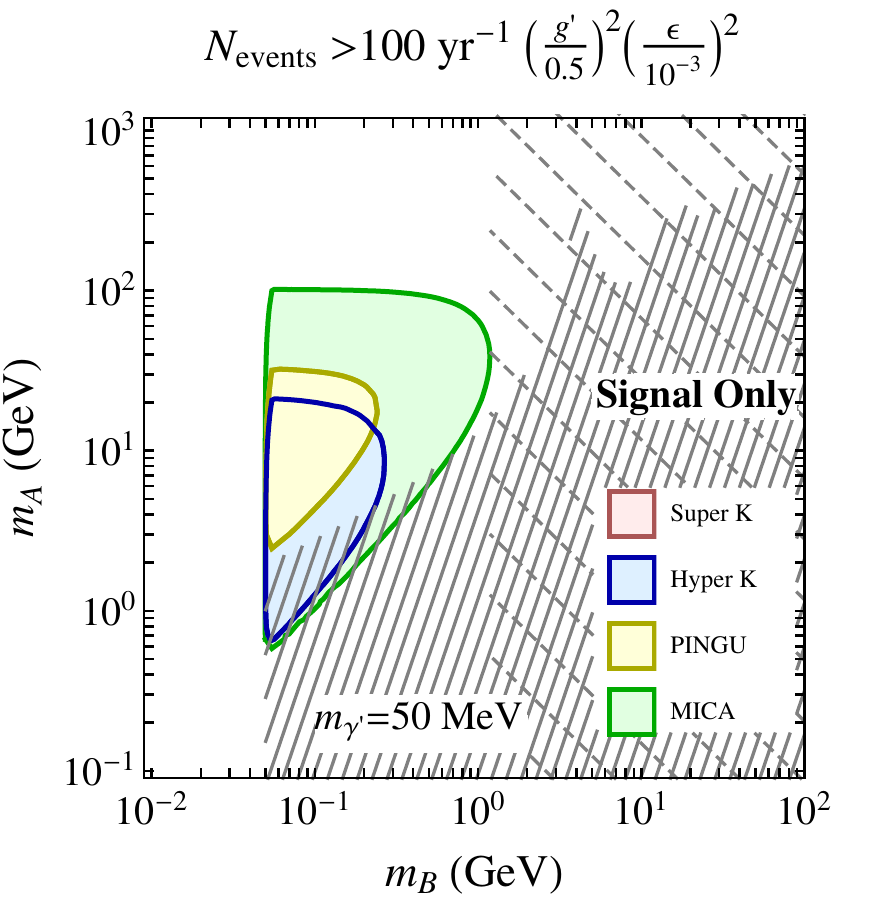} }}%
    \caption{Number of $\B e^- \rightarrow \B e^- $ signal events in Super-K, Hyper-K, PINGU, and MICA in the $m_A$/$m_B$ plane, for $m_{\gamma'} = 20~\MeV$ (top) and $m_{\gamma'} = 50~\MeV$ (bottom).  The indicated regions are for 1 (left), 10 (center), 100 (right) detected events in a one year period, normalized to the couplings $\epsilon= 10^{-3}$ and $g'=0.5$.  We have imposed the angular criteria of $\theta_C = 10^\circ$ and the electron energy range of $\{100~\MeV,100~\GeV\}$ ($\{1.33~\GeV, 100~\GeV\}$ for PINGU).  Also shown are model-dependent constraints on the relic $\B$ population from \Sec{sec:constraints}:  the solid gray lines are from CMB heating (shown only for $g' = 0.5$), and the dashed gray lines are from DAMIC direct detection (which are independent of $g'$, but can be eliminated by adding an inelastic splitting).  The red star indicates the benchmark in \Eq{eq:keybenchmark}.}
    \label{fig:signalAB}%
\end{figure}

To get a sense of the expected signal rate, we consider the number of signal events for $\theta_C = 10^\circ$ in the combined categories:  
\be \label{eq:nevents}
\frac{N^{10^\circ}_{\rm signal}}{\Delta T} = 25.1~\text{year}^{-1} \left(\frac{\vev{\sigma_{A \overline{A}  \rightarrow B \overline{B}}v}}{ 5 \times 10^{-26}~\text{cm}^3/\text{s}} \right) \left(\frac{20~ \GeV}{m_A} \right)^2 \left(\frac{ \sigma_{B e^- \rightarrow B e^-} }{1.2 \times 10^{-33}~\text{cm}^2 } \right) \left(\frac{V_{\rm exp}}{22.4\times 10^3~\text{m}^3} \right),
\ee
broken down by $21.1/\text{year}$ for Sub-GeV and $4.0/\text{year}$ for Multi-GeV, and the reference cross sections are based on the benchmark in \Eq{eq:keybenchmark}. In our analysis below, we will always assume that $\vev{\sigma_{A \overline{A}  \rightarrow B \overline{B}}v}$ takes on the thermal relic reference value.

Because $\sigma_{B e^- \rightarrow B e^-}$ scales homogeneously with $g'$ and $\epsilon$, the number of signal events does as well, so the only non-trivial dependence is on the mass parameters $m_A$, $m_B$, and $m_{\gamma'}$.  In \Fig{fig:signalAB}, we set two benchmark values $m_{\gamma'} = 20~\MeV$ and $m_{\gamma'} = 50~\MeV$, and show what part of the $m_A-m_B$ parameter space yields
\be
\frac{N^{10^\circ}_{\rm signal}}{\text{year}} = x \, \left( \frac{g'}{0.5} \right) ^2 \left(\frac{\epsilon}{10^{-3}}  \right)^2,
\ee
for $x = 1,10,100$.  These reference values for $m_{\gamma'}$ have been chosen such that the $t$-channel scattering processes are not overly suppressed by the dark photon mass, and the reference $\epsilon$ is close to the maximum allowed by dark photon constraints.  In the triangular regions in \Fig{fig:signalAB}, the top edge is set by $m_A$ which controls the DM number density (and therefore the annihilation rate), the left edge is set by the requirement that $m_B > m_{\gamma'}$, and the diagonal edge is set by the electron energy threshold.

In these figures, we have included model-dependent constraints from CMB heating and direct detection, discussed in the later \Sec{sec:constraints}.\footnote{The bump around $m_B = 10~\GeV$ in the CMB heating bound is due to a Sommerfeld resonance.}  It is worth emphasizing that both of these constraints are due to the thermal relic $\B$ population, and are independent of the boosted DM phenomenon.  Indeed, as discussed at the end of \Sec{sec:preliminaries}, we could give $\B$ a small Majorana mass splitting, which would eliminate the bound from (elastic) direct direction experiments while not affecting very much the kinematics of boosted $\B e^- \rightarrow \B e^-$ detection.  The CMB constraints are more robust since they mainly depend on $\B$ being in thermal contact with the SM via $\B\Bbar\rightarrow\gamma'\gamma'$, though the CMB constraints could potentially softened if $\gamma'$ somehow decays to neutrinos (or to non-SM states).

\subsection{Background Rates}
\label{sec:backgroundrates}

The atmospheric neutrino backgrounds have been measured by Super-K over a 10.7 year period, during runs SK-I (1489 days), SK-II (798 days), SK-III (518 days) and SK-IV (1096 days), and the final results are summarized in \Ref{Dziomba}.  In the Sub-GeV category, a total of 7755 fully-contained single-ring zero-decay electron events were seen the 100 MeV to 1.33 GeV energy range, giving a yearly background rate of
\be
\text{Sub-GeV:} \quad \frac{N_\text{bkgd}^{\text{all sky}}}{\Delta T} = 726~\text{year}^{-1} \left(\frac{V_{\rm exp}}{22.4\times 10^3~\text{m}^3} \right).
\ee
In the Multi-GeV category, 2105 fully-contained single-ring electron events were seen in the 1.33 GeV to 100 GeV energy range \cite{Dziomba,Wendell:2010md}, yielding
\be
\text{Multi-GeV:} \quad \frac{N_\text{bkgd}^{\text{all sky}}}{\Delta T} = 197~\text{year}^{-1} \left(\frac{V_{\rm exp}}{22.4\times 10^3~\text{m}^3} \right).
\ee
To estimate the background for PINGU (which lacks the ability to reconstruct Cherenkov rings), we add in multi-ring and $\mu$-like events in the Multi-GeV category, changing $197~\text{year}^{-1}$ to $634~\text{year}^{-1}$, which then has to be scaled by the effective PINGU detector volume.

For the boosted DM search, the background is reduced by considering only events where the electron lies in the search cone $\theta_C$.  We assume a uniform background distribution from the entire sky, so the background within a patch in the sky of angle $\theta_C$ is:
\begin{equation}
N_\text{bkgd}^{\theta_C} = \frac{1 - \cos \theta_C}{2} N_\text{bkgd}^{\text{all sky}},
\end{equation}
For $\theta_C =10^\circ$ relevant for Super-K, we have
\begin{align}
\text{Sub-GeV:} \quad \frac{N_\text{bkgd}^{10^\circ}}{\Delta T} &= 5.5~\text{year}^{-1}. \\
\text{Multi-GeV:}\quad \frac{N_\text{bkgd}^{10^\circ}}{\Delta T} &= 0.35~\text{year}^{-1}.
\end{align}
Ideally, we would use the full energy dependence of the background in order to optimize the signal/background separation, but given the rather low background rate, we will make the conservative choice to consider the whole Sub-GeV energy range.

Since one can estimate the background by looking at a side-band away from the GC, the background uncertainties in a $\theta_C$ cone should be dominated by Poisson fluctuations.  For the all sky background, we note that Super-K saw a $\simeq 10\%$ mismatch between the measured atmospheric background and the Monte Carlo estimate \cite{Wendell:2010md,Dziomba}, so there is in fact a bit of room beyond Poisson fluctuations to accommodate a boosted DM signal in the current Super-K data.\footnote{Associated with the published search for DM from the GC via upward going muons \cite{Desai:2004pq}, there is also unpublished electron data from SK-I, -II, and -III \cite{Mijakowski:2011zz,Mijakowski:slides}.  For $\cos \theta > 0.8$ ($\theta_C \simeq 37^\circ$), around 600 Sub-GeV fully-contained single-ring zero-decay electron events were observed in a 7.7 year period.  This number has subsequently been updated to around 850 events in the full 10.7 year data set \cite{Wendell:slides}.  In principle, these could be used to set a stronger bound than we show in this paper, since no statistically significant excess is seen.}

In order to have a fair comparison of the sensitivities at different experiments, for Hyper-K and MICA, we use the same event selection assuming the same exposure time, based on the available Super-K $\sim10$ year data set, and simply scale up the background rate proportional to the detector volume $V_\text{exp}$ (and adjust $\theta_C$ for MICA). As already mentioned, since PINGU has a higher energy threshold and an inability to reconstruct Cherenkov rings, we rescale the full Multi-GeV category (single-ring supplemented by the multi-ring and $\mu$-like events).

\subsection{Estimated Experiment Reach}
\label{subsec:significance}

\begin{figure}[t]%
	\centering
	\includegraphics[scale=0.5]{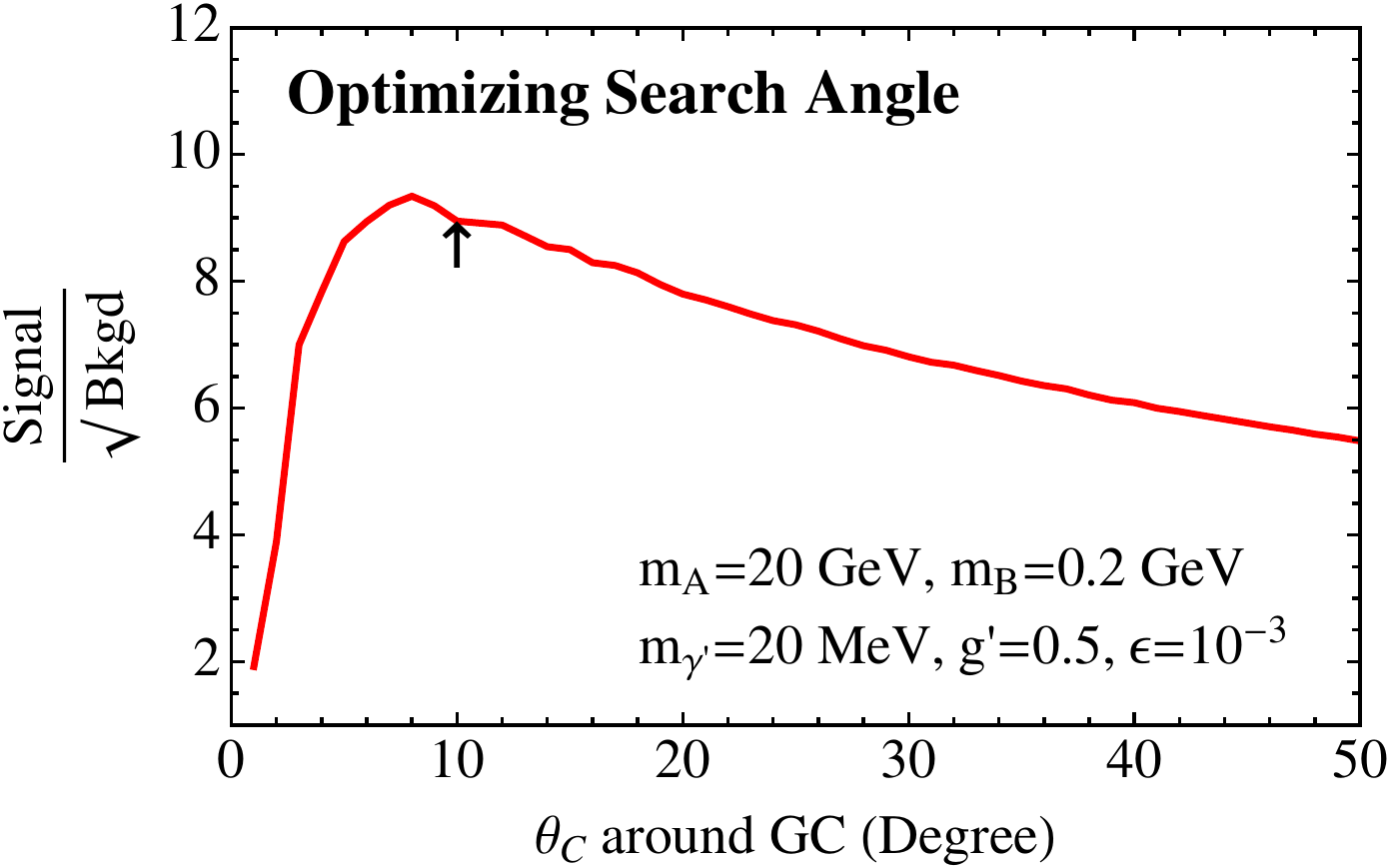} %
	\caption{\label{fig:optangle} Yearly signal significance in the Sub-GeV category for our benchmark in \Eq{eq:keybenchmark} as a function of the search cone angle $\theta_C$.  The peak around $10^\circ$ is seen for other parameter choices as well.} %
\end{figure} 

Given the signal and background rates above, we can find the optimal search cone $\theta_C$ to maximize the significance
\begin{equation}
\text{Sig}^{\theta_C} \equiv \frac{N^{\theta_C}_\text{signal}}{\sqrt{N_\text{bkgd}^{\theta_C}} }.
\end{equation}
In \Fig{fig:optangle}, we plot the significance as a function of search angle for our benchmark model in \Eq{eq:keybenchmark}; we checked that other parameter choices show similar behavior.  We see that the significance peaks at around $10^\circ$, and falls off somewhat slowly after that.  For Super-K/Hyper-K with $3^\circ$ resolution, we can effectively ignore experimental resolution effects and take $\theta_C$ at the optimal value.  For PINGU and MICA, we approximate the effect of the experimental resolution by taking $\theta_C = \theta_e^{\rm res}$; a more sophisticated treatment would be to apply Gaussian smearing to the electrons.  This is the logic behind \Eq{eq:searchconechoice} above.

\begin{figure}[t]%
    \centering
    \subfloat[]{{\includegraphics[scale=0.75]{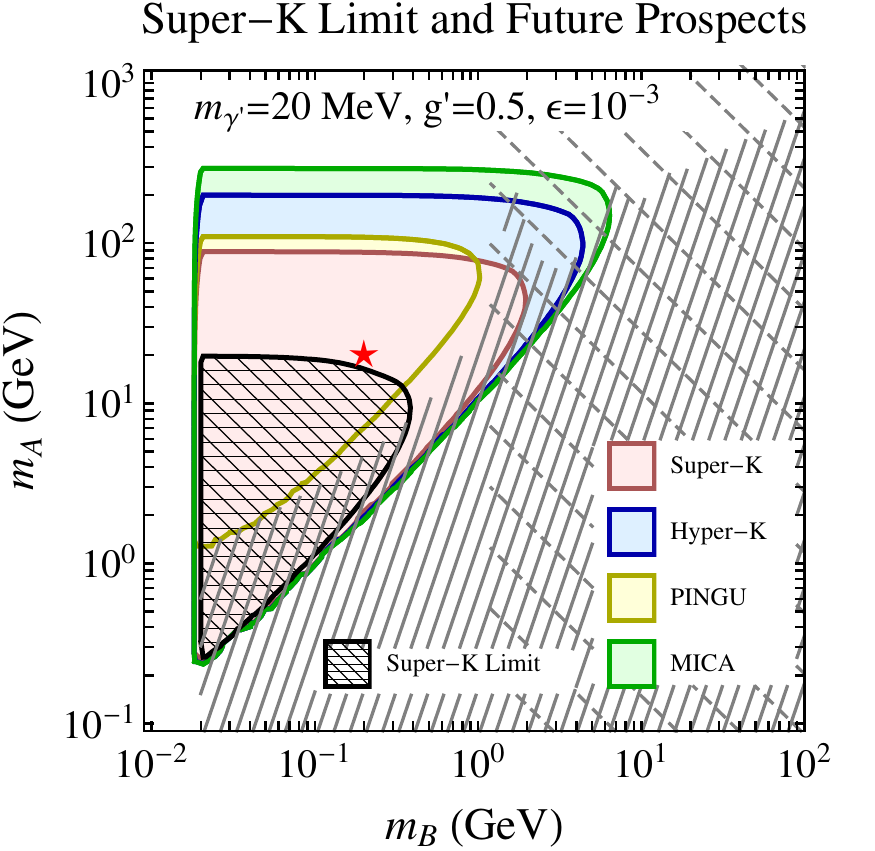} }}%
    \qquad
    \subfloat[]{{\includegraphics[scale=0.75]{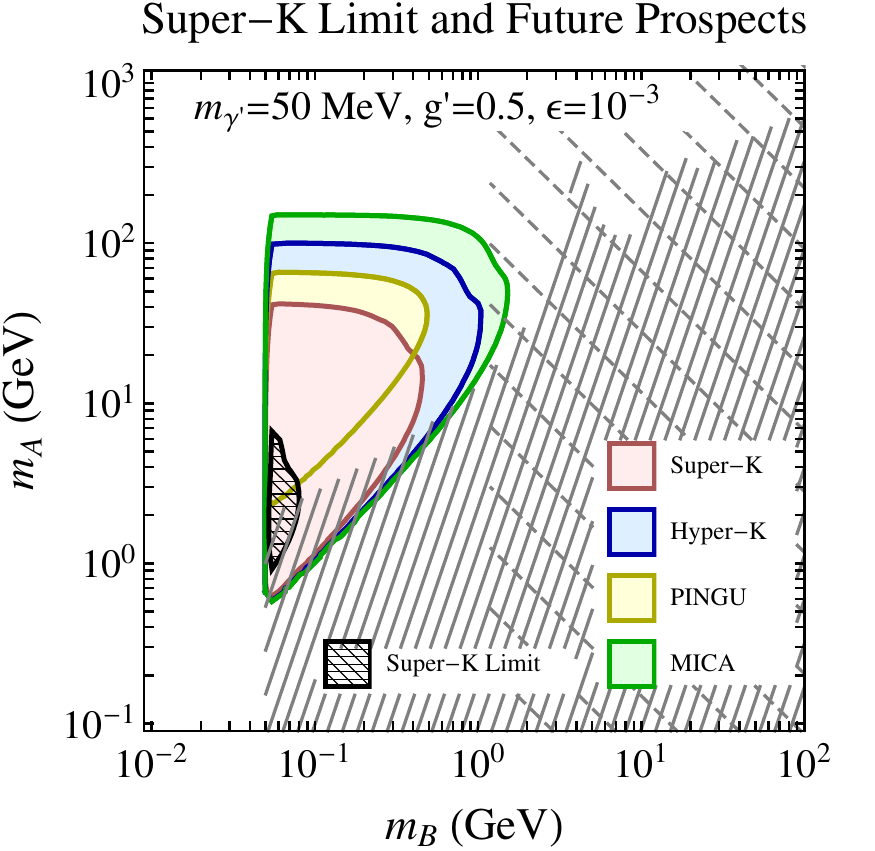} }}%
    \caption{Signal significance at Super-K, Hyper K, PINGU and MICA on the $m_A$/$m_B$ plane, for $m_{\gamma'} = 20~\MeV$ (left) and $m_{\gamma'} = 50~\MeV$ (right), fixing $\epsilon= 10^{-3}$ and $g'=0.5$.  Shown are the $2\sigma$ reaches with 10 years of data, taking $\theta_C = 10^\circ$ and adding the significances of the $E_e \in \{100~\MeV,1.33~\GeV\}$ and $E_e \in \{1.33~\GeV, 100~\GeV\}$ categories in quadrature (only the latter for PINGU).  Also shown is the current $2 \sigma$ exclusion using all-sky data from Super-K, where we assume a 10\% uncertainty on the background.  The grey model-dependent limits are the same as in \Fig{fig:signalAB}: the solid gray lines are constraints on $\B$ from CMB heating and the dashed gray lines are from DAMIC. The red star is the benchmark from \Eq{eq:keybenchmark}.}
    \label{fig:significance}%
\end{figure}

\begin{figure}[t]%
    \centering
    \subfloat[]{{\includegraphics[scale=0.75]{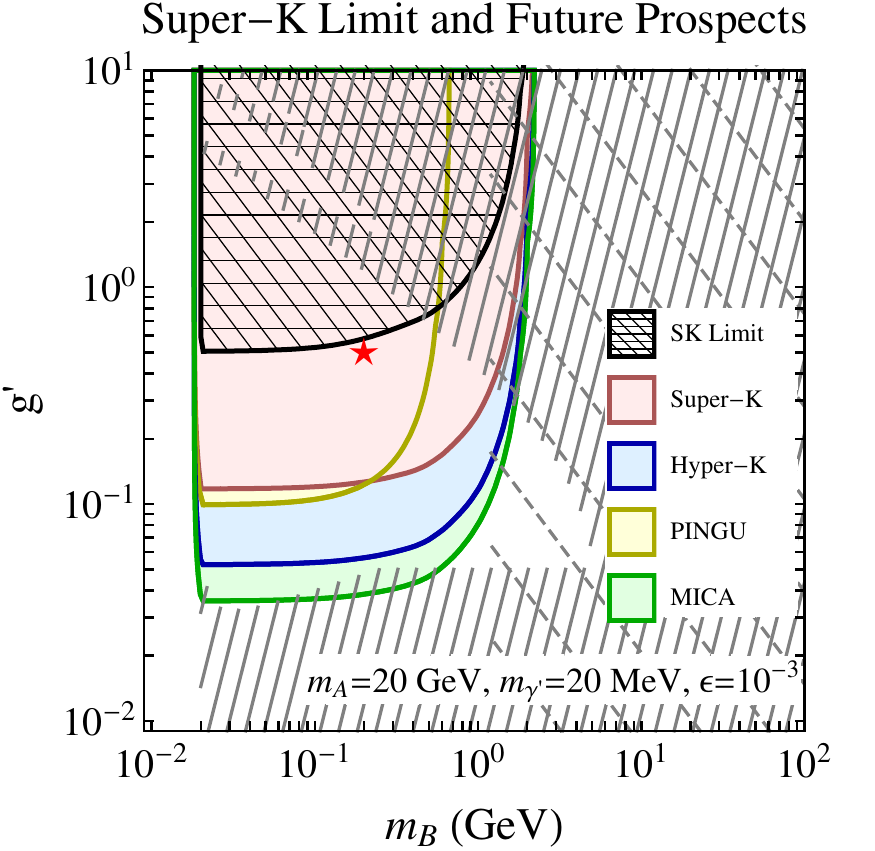} }}%
    \qquad
    \subfloat[]{{\includegraphics[scale=0.75]{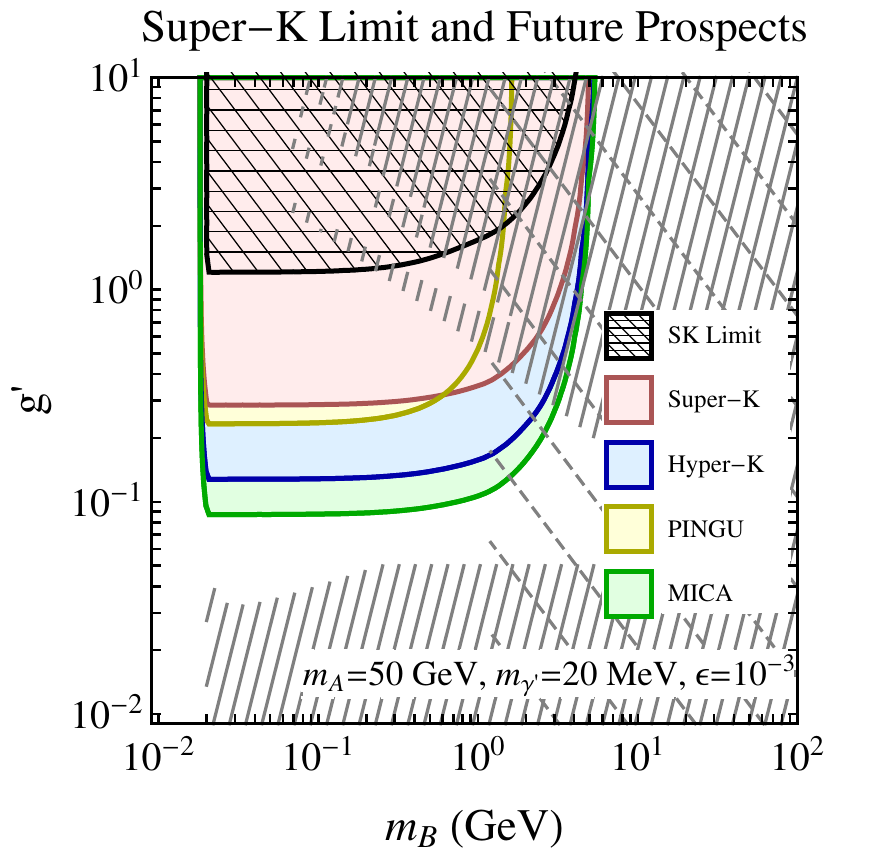} }}%
        \caption{Same as \Fig{fig:significance}, but on the $g'$/$m_B$ plane, for $m_A = 20~\GeV$ (left) and $m_A = 50~\GeV$ (right), fixing $\epsilon= 10^{-3}$ and $m_{\gamma'} = 20~\MeV$. The spikes in the CMB heating bounds (solid gray lines) are from Sommerfeld resonances.}
    \label{fig:significance2}%
\end{figure}

In \Figs{fig:significance}{fig:significance2}, we show the $2\sigma$ sensitively possible with the 10.7 years of Super-K data, using the optimal $\theta_C = 10^\circ$ selection criteria, as well as the estimated reach for Hyper K, PINGU, and MICA for the same period of time.  We treat the Sub-GeV and Multi-GeV categories separately and report the overall significance as the quadrature sum of the significances obtained from the two categories. We also show the current bounds from Super-K that one can place without the $\theta_C$ selection (i.e.~using the all-sky background), taking $\delta N_\text{bkgd} / N_\text{bkgd} = 10\%$ to account for systematic uncertainties in the all-sky background.  Here, we are only allowing for the two energy categories, and further improvements are possible if one adjusts the energy range as a function of $m_A$ and $m_B$. 

Taken together, these experiments have substantial reach for boosted DM.   The prospects for Super-K to find single-ring electron events from the GC are particularly promising, given that the data (with angular information) is already available \cite{Himmel:2013jva} and one simply needs to change from lab-coordinates to galactic coordinates (as in \Refs{Mijakowski:2011zz,Mijakowski:slides}).

\section{Summary of Existing Constraints}
\label{sec:constraints}

Apart from the measured neutrino fluxes discussed in \Sec{sec:backgroundrates}, we know of no model-independent constraints on the boosted DM phenomenon.  There are, however, constraints on the particular model described here, and we summarize those constraints in this section.  The most relevant bounds are due mainly to the relic $\B$ population left over from thermal freeze-out, which leads to bounds from ``Direct detection of non-relativistic $\B$'' and ``CMB constraints on $\B$ annihilation'' described below and seen in \Figs{fig:significance}{fig:significance2}.

\bi
\item \textit{Limits on the dark photon $\gamma'$.}  As discussed earlier, dark photon searches have set limits of $m_{\gamma'}\gtrsim \mathcal{O}(10~\MeV)$ and $\epsilon \lesssim 10^{-3}$, assuming the dominant decay mode is $\gamma' \to e^+ e^-$ \cite{Essig:2013lka}. For $m_{\gamma'} < \mathcal{O} (100 ~\MeV)$, beam dump experiments place a bound of roughly $\epsilon \gtrsim 10^{-5}$ \cite{Blumlein:2013cua}.   We have used $m_{\gamma'} = 20~ \MeV$ and $\epsilon = 10^{-3}$ as a benchmark in this paper, which yields a detectable boosted DM signal while satisfying the current dark photon bounds.  Our benchmark is also within the region of interest for explaining the muon $g-2$ anomaly \cite{Pospelov:2008zw,Fayet:2007ua}.

\item \textit{Direct detection of non-relativistic $\A$}.  Thermal relic $\A$ particles are subject to constraints from conventional DM direct detection experiments (e.g.~XENON, LUX, and CDMS) via their scattering off nuclei.  As discussed in more detail in \App{app:ADirectDetection}, the constraints on $\A$ are rather weak since $\A$ has no tree-level interactions with the SM.  That said, $\A$ can scatter off nuclear targets via a $\B$-loop.  Since we have approximated the $\A\Abar \B \Bbar$ interaction as a contact operator, this loop process is model-dependent.   In \Fig{fig:AApp} and \Eq{eq:Acompletion}, we give an example UV completion involving an extra scalar $\phi$ that allows us to estimate the $\A$-nucleon scattering cross section.   Due to the loop factor and the mass suppression from $m_\phi\gg m_A$, the limits on $\A$ are safe for most values of the parameter space, as shown in \Fig{fig:Adirectdetection}.  As already mentioned, one could introduce inelastic splitting within the $\A$ multiplet to  further soften direct detection constraints \cite{Finkbeiner:2009mi,Graham:2010ca,Pospelov:2013nea}.

\item  \textit{Direct detection of non-relativistic $\B$}.   Despite the small relic abundance of $\B$, it has a large $\B$-nucleon scattering cross section, as calculated in \App{app:elasticBp}.  
\be
\label{eq:directdirectionestimate}
\sigma_{Bp \rightarrow Bp} =  4.9 \times 10^{-31} \text{ cm}^2  \left(\frac{\epsilon}{10^{-3}} \right)^2 \left(\frac{g'}{0.5} \right)^2 \left(\frac{20~\MeV}{m_{\gamma'}} \right)^4 \left( \frac{m_B}{200~\MeV} \right)^2,
\ee 
where the scaling assumes $m_B \ll m_p$.  Thus, direct detection experiments essentially rule out any elastic $\B$-nucleon scattering above the detector threshold.  Of course, in the parameter space of our interest, the $\B$ mass is $\leq \mathcal{O}(1~\GeV)$, which is close to or below the threshold of LUX \cite{Akerib:2013tjd} and the low CDMS threshold analysis \cite{Agnese:2014aze}, and the most constraining limits come from CDMSLite \cite{PhysRevLett.112.041302} and DAMIC \cite{Barreto:2011zu}.  Because of this, light $\B$ particles can evade existing direct detection bounds.

In \Fig{fig:significance2}, we demonstrate the constraints on the $(g',m_B)$ plane from the DAMIC experiment (which has a lower threshold than CDMSLite), using the effective nuclear cross section\begin{equation}
\sigma_{Bp \rightarrow Bp}^\text{eff} =  \frac{\Omega_B}{\Omega_{\rm DM}} \sigma_{Bp \rightarrow Bp}.
\end{equation}
Essentially, the allowed parameter space is independent of $g'$ and $(m_{\gamma'})^{-4}$, since the expected $\B p \rightarrow \B p$ cross section is so large that any events above the energy threshold of the experiment would be seen.  There is also the fact that when $g'$ is $\mathcal{O}(10^{-2})$ and higher, the abundance scales as $g'^{-2}$ (see \Eq{eq:Babundanceestimate}), which cancels with the $g'^2$ scaling of $\sigma_{Bp \rightarrow Bp}$, yielding a $g'$-independent bound.\footnote{The $\B p \rightarrow \B p$ cross section scales as $(m_{\gamma'})^{-4}$, but we have checked that these bounds do not soften until $m_{\gamma'}$ is higher than $\mathcal{O}(1~\GeV)$, which is not the regime we are studying in this paper.}  Of course, as with $\A$, the direct detection bound on $\B$ could be alleviated by introducing inelastic splittings. 

It has been recently pointed out that sub-GeV DM might be better constrained by scattering off electrons rather than off nuclei \cite{Essig:2011nj}, as in recent XENON10 bounds \cite{Essig:2012yx}.  In our case, these bounds are subsumed by CMB heating bounds discussed below.  Note that for $\B e^- \rightarrow \B e^-$, the conventional direct detection process and the boosted DM detection process have very different kinematics, so one should not be surprised that the XENON10 bounds do not influence the boosted DM signal regions. 

\item \textit{Indirect detection of non-relativistic $\B$.}  The annihilation process $\B \Bbar \to \gamma' \gamma'$ and the subsequent $\gamma'$ decay to two $e^+/e^-$ pairs gives rise to a potential indirect detection signal in the positron and diffuse $\gamma$-ray channels.  The recent constraint on DM annihilation in positron channel from AMS-02 is demonstrated in \Refs{Bergstrom:2013jra,Ibarra:2013zia}, where the bound is strongest for 2-body final state and weaker when there are more particles in the final state like in our case. The suppressed relic abundance of $\Omega_B$ relative to $\Omega_{\rm DM}$ helps relieve the constraints on our model. In addition, at the sub-GeV mass which we are interested in, the background uncertainty of the above indirect detection limit is large due to solar modulation.  The CMB considerations below give stronger constraints for the parameter range of our interest.  The diffuse $\gamma$-ray signal from e.g.\ inverse Compton scattering of $e^\pm$ produced from $\B$ annihilation has a smaller cross section and also faces large background uncertainty in the sub-GeV region. In fact, the $\gamma$-ray search for DM at Fermi, for instance, has a lower energy cutoff at $\sim4$ GeV \cite{Ackermann:2012qk}. Indirect detection signals from $\A$ annihilation have to go through higher order or loop processes, and are much suppressed.

\item \textit{CMB constraints on $\B$ annihilation.}  With a light mass of $m_B \lesssim \mathcal{O}(1~\GeV)$, thermal $\B$ annihilation in the early universe may be subject to bounds from CMB heating \cite{Madhavacheril:2013cna}.  The CMB constrains the total power injected by DM into ionization, heating, and excitations.   For the dominant DM component with relic density $\Omega_{\rm DM}\approx 0.2$, the bound is directly imposed on the quantity:
\begin{equation}
 p_\text{ann,DM} = f_\text{eff} \frac{\vev{\sigma v}}{M_\chi},
\end{equation}
where $f_\text{eff}$ is the fraction of the annihilation power that goes into ionization, which depends on the annihilation channel and its energy scale.  Though $\B$ is a small fraction of total DM, it does annihilate into $\gamma'$ which subsequently decays via $\gamma' \to e^+ e^-$.  Therefore, the CMB spectrum constrains
\begin{equation}
p_{\rm ann, \B}= f_\text{eff}  \frac{\vev{\sigma_{B \overline B \rightarrow \gamma' \gamma'} v}}{m_B} \left(\frac{\Omega_B}{\Omega_{\rm DM}} \right)^2 \simeq f_\text{eff} \vev{\sigma_{A \overline{A} \rightarrow B \overline{B}}} \frac{m_B}{m_A^2},\label{CMB_power}
\end{equation}
where the last relation is obtained using \Eq{eq:Babundanceestimate} for $\OmegaB/\OmegaA$, which is valid for large values of $g'$ (typically for $g'\gtrsim0.1$) as explained in the \App{app:ABrelicstory}. These limits are illustrated in \Fig{fig:significance2} for $f_\text{eff}= 1$, which is a conservative assumption.  Due to the presence of a light $\gamma'$, there can be an extra Sommerfeld enhancement factor to the $\vev{\sigma_{B \overline B \rightarrow \gamma' \gamma'} v}$ in \Eq{CMB_power}. For the parameter space we consider, we expect that this enhancement saturates at CMB time, which leads to an extra factor of \cite{Slatyer:2009vg}
\begin{equation}
S = \frac{\pi }{\epsilon_v} \frac{\sinh \frac{12 \epsilon_v}{\pi \epsilon_\phi}}{\cosh \frac{12 \epsilon _v}{\pi \epsilon_\phi} - \cos \left(2 \pi \sqrt{\frac{6}{\pi^2 \epsilon_\phi} - \left(\frac{6}{\pi^2} \right)^2 \frac{\epsilon_v^2}{\epsilon_\phi^2}} \right)}, \qquad  \epsilon_v = \frac{4 \pi v}{g'^2 }, \qquad \epsilon_\phi = \frac{4 \pi m_{\gamma'}}{g'^2 m_B }.
\end{equation}
This enhancement contributes at low velocities, so we do not expect it to change the picture at freeze out, but it would be relevant in the CMB era where $v \approx 10^{-3}$.   For our current parameter space, $S \approx 1$ until high values of $g'=1$ where it becomes $\mathcal{O}(10)$. We incorporate the enhancement in the calculation of our CMB limits, as can be seen from the resonance peaks in \Fig{fig:significance2}.

\item \textit{BBN  constraints on $\B$ annihilation}.  The energy injection from $\B$ annihilation in the early universe can also alter standard BBN predictions \cite{Henning:2012rm, Berezhiani:2012ru}. The constraints from hadronic final states are the most stringent, comparable to or even somewhat stronger at $\mathcal{O}(1~\GeV)$  than those from the CMB heating as discussed above \cite{Henning:2012rm}.  However, as we focus on $m_{\gamma'}$ of $\mathcal{O}(10~\MeV)$, the production of hadronic final states ($n, p, \pi$) from the leading annihilation channel $\B\bar{\psi}_B\rightarrow \gamma'\gamma'$ followed by $\gamma'$ decay are not kinematically possible. The subleading channel $\B\Bbar \rightarrow q\bar{q}$ is $\epsilon^2$ suppressed.  Thus, the major energy injection to BBN is mostly electromagnetic from $\gamma'\rightarrow e^+e^-$, and the related constraint in this case are much weaker than the CMB bound we have considered above \cite{Henning:2012rm}.

\item \textit{Dark matter searches at colliders}. By crossing the Feynman diagrams in \Fig{fig:feynmandetection}, we see that $\B$ can be produced at colliders such as LEP, Tevatron, and the LHC. If $\B$ were to interact with SM electrons or quarks via a heavy mediator, then collider searches would provide a stronger bound than direct detection at these low DM masses.  However, this complementarity is lost when the interaction is due to a light mediator \cite{Goodman:2010ku, Fox:2011fx, Fox:2011pm}, which applies to our case where $\B$ interacts with SM states via an $\mathcal{O}(10~\MeV)$ dark photon. In addition, compared to the irreducible main background from electroweak processes, e.g. $e^+e^-\rightarrow Z^{(*)}\rightarrow \nu\bar{\nu}$, the production cross section of $\B$ is suppressed by $\epsilon^2\lesssim10^{-6}$, so the collider constraints on our model are rather weak.  

\ei

\section{Conclusions and Other Possibilities}
\label{sec:conclusions}

In this paper, we presented a novel DM scenario which incorporates the successful paradigm of WIMP thermal freeze-out, yet evades stringent constraints from direct and indirect detection experiments, and predicts a novel signal involving boosted DM.  The example model features two DM components, $\A$ and $\B$.   The heavier particle $\A$
(which is the dominant DM component)
experiences assisted freeze-out \cite{Belanger:2011ww} by annihilating into the lighter particle $\B$ (which is the subdominant DM component).
The whole dark sector is kept in thermal contact with the SM in the early universe via kinetic-mixing of a dark photon with the
SM photon.   Only $\B$ couples directly to the dark photon (and hence to the SM), so the dominant DM component $\A$ can largely evade current DM detection bounds.
If such a scenario were realized in nature, then the leading non-gravitational signal of DM would come from annihilating $\A$ particles in the galactic halo producing boosted $\B$ particles that could be detected on earth via neutral-current-like scattering via the dark photon.  
In large volume neutrino or proton-decay detectors, the smoking gun for this scenario would be an electron signal pointing toward the GC, with no corresponding excess in the muon channel.   Liquid argon detectors could potentially detect boosted DM through (quasi-)elastic proton scattering, as well as improve the rejection of the dominant neutrino CC backgrounds by vetoing on hadronic activity.  Future experiments that use LArTPC technology for tracing the particle paths \cite{Cennini:1999ih,Bromberg:2013fla} will provide both directionality and better background discrimination.

This phenomenon of boosted DM is generic in scenarios with multiple DM components.  In fact, models with a single component DM could also potentially give rise to the same signature.  If the stabilization symmetry is $\mathbb{Z}_3$, then the semi-annihilation process $\psi \psi \to \overline{\psi} \phi$ (where $\phi$ is a non-DM state) is allowed \cite{Belanger:2012zr,Ko:2014nha,Aoki:2014cja}.  For $m_\phi = 0$, the outgoing $\overline{\psi}$ would have energy $E_\psi = (5/4) m_\psi$.  In the limit $m_\psi \gg m_e$,  $\gamma_\psi = 1.25$ implies a maximum $\gamma_e^{\rm max} = 2 \gamma_\psi^2 - 1 = 2.125$, which is above the Cherenkov threshold in water (and ice).  
Of course, the $\mathbb{Z}_3$ symmetry is not consistent with $\psi$ being charged under a $U(1)'$, so additional model building would be necessary to get a sufficiently large scattering with the SM.  But this example shows why non-minimal dark sectors tend to have some production cross section for boosted DM.

It is intriguing to consider other scenarios where DM mostly annihilates to other stable states in the dark sector.  For example, if both $\A$ and $\B$ are charged under the $U(1)'$ and the mass hierarchy is
\be
m_A > m_{\gamma'} > m_B,
\ee
then the annihilation $\A \Abar \to \gamma' \gamma'$ would be followed by the decay $\gamma' \to \B \Bbar$, and the boosted $\B$ particles could again be detected via $t$-channel $\gamma'$ exchange with the SM.  Of course, now $\A$ itself has tree-level $\gamma'$ exchange diagrams with the SM, but if $\A$ has a Majorana mass splitting (allowing it to evade direct detection bounds), boosted DM would again be the dominant mode for DM discovery.\footnote{There would also be interesting signals for $\B$ in DM production/detection experiments \cite{Batell:2009di}.}

The above scenario is particularly interesting in light of the gamma ray excess recently seen in the GC \cite{Daylan:2014rsa}.  In the context of DM, this signal could be explained via cascade decays $\A \Abar \to \gamma' \gamma'$  followed by $\gamma' \to \text{SM} \, \text{SM}$ \cite{Boehm:2014bia,Abdullah:2014lla,Martin:2014sxa,Berlin:2014pya}.  Boosted DM could be produced in the same cascade process, since the dark photon could easily have comparable branching ratios for $\gamma' \to \text{SM} \, \text{SM}$ and $\gamma' \to \B \Bbar$ when $m_{\gamma'} > m_B$.  More generally, it is interesting to contemplate scenarios where $\A$ partially annihilates to boosted $\B$ and partially to SM states. For example, the bremsstrahlung process of $\A \A \rightarrow \B \B \gamma'$, where the $\gamma'$ decays to an electron-positron pair, can be a source of positrons that can be detected in experiments like AMS-02 \cite{Aguilar:2013qda} or indirectly in Gamma ray telescopes \cite{Essig:2013goa}. Of course, if the $\B$ states are not too depleted, then they could give indirect detection signals of their own. 

Finally, it is worth considering the broader experimental signatures possible in the paradigm of DM annihilating to stable dark sector states \cite{Ackerman:mha, Blennow:2012de, Fan:2013yva,Chu:2014lja, AB_CMB}, with simple extensions/variations based on our current model.  If $m_B \ll m_e$, then $\B$ acts effectively like dark radiation, which may leave signatures in CMB observables such as $N_{\rm eff}$ \cite{AB_CMB}.  If $\A$ has a non-negligible solar capture cross section, then boosted DM could emerge from the sun.  If $\B$ takes up sizable fraction of the total DM abundance (perhaps via a leading asymmetric component), then the fact that $\B$ has strong self-interactions may have implications for small scale structure of DM halos including the known anomalies such as cusp-core and too-big-to-fail problems \cite{deBlok:2009sp,BoylanKolchin:2011de}.  The potentially rich structure of the dark sector motivates a comprehensive approach to DM searches.

\acknowledgments{We thank Francesco D'Eramo for early discussions about boosted DM.  We also thank Brian Batell, Kfir Blum, Chris Kachulis, Ed Kearns, Kuver Sinha, Tracy Slatyer, Greg Sullivan, Zach Thomas, and Itay Yavin for helpful discussions. We thank our anonymous referee for detailed comments and insightful new detection options, and KC Kong and Jong-Chul Park for additional corrections.  K.A.\ and Y.C.\ are supported in part by NSF Grant No.\ PHY-1315155 and by the Maryland Center for Fundamental Physics.  L.N.\ and J.T.\ are supported by the U.S. Department of Energy (DOE) under cooperative research agreement DE-FG02-05ER-41360.  J.T.\ is also supported by the DOE Early Career research program DE-FG02-11ER-41741 and by a Sloan Research Fellowship from the Alfred P. Sloan Foundation.}

\appendix

\section{Analytic Approximations to Relic Abundances}
\label{app:ABrelicstory}

The coupled Boltzmann equations for the evolution of the $\A$/$\B$ abundances are
\begin{align}
 \frac{d n_A}{d t} + 3 H n_A &= - \frac{1}{2}\langle\sigma_{A\bar{A} \rightarrow B\bar{B}} v\rangle \left( n_A ^2 - \frac{(n_A ^{\rm eq})^2}{(n_B ^{\rm eq})^2} n_B^2 \right), \nonumber \\ 
  \frac{d n_B}{d t} + 3 H n_B &= - \frac{1}{2}\langle\sigma_{B\bar{B} \rightarrow \gamma' \gamma'} v\rangle \left( n_B ^2 - (n_B ^{\rm eq})^2 \right) - \frac{1}{2}\langle\sigma_{B\bar{B} \rightarrow A\bar{A}} v\rangle  \left( n_B^2 - \frac{(n_B ^{\rm eq})^2}{(n_A ^{\rm eq})^2} n_A^2 \right), \label{eq:Boltzmann}
\end{align}
where the factor of $\frac{1}{2}$ arises because $\A$ and $\B$ are Dirac fermions, and $n_A$ refers to the sum of the abundances for $\A$ and $\Abar$ (and similarly for $n_B$).  In terms of the comoving abundance $Y_i = n_i /s$, where $s$ is the entropy of the universe, and $x\equiv m_B/T$, we can rewrite the Boltzmann equations as
\begin{align}
 \frac{d Y_A}{dx} &= -\frac{\lambda_A}{x^2}  \left( Y_A ^2 - \frac{(Y_A ^{\rm eq})^2}{(Y_B ^{\rm eq})^2} Y_B^2 \right),  \label{dYdA} \\ 
  \frac{d Y_B}{dx} &= -\frac{\lambda_B}{x^2} \left( Y_B ^2 - (Y_B ^{\rm eq})^2 \right) + \frac{\lambda_A}{x^2} \left(Y_A^2-\frac{(Y_A^{\rm eq})^2}{(Y_B^{\rm eq})^2}Y_B^2\right), \label{eq:dYdB}
\end{align}
where we have introduced the shorthand notations:
\be
\lambda_A \equiv \frac{s x^3}{2 H(m_B)} \langle\sigma_{A\bar{A}\rightarrow B\bar{B}}v\rangle, \qquad \lambda_B \equiv \frac{s x^3}{2 H(m_B)} \langle\sigma_{B\bar{B} \rightarrow \gamma' \gamma'} v\rangle,
\ee
and used the fact the total DM number is not changed by the $\A\Abar \rightarrow \B\Bbar$ reaction, i.e.
\be
 - \langle\sigma_{B\bar{B} \rightarrow A\bar{A}} v\rangle \left(Y_B^2 - \frac{(Y_B ^{\rm eq})^2}{(Y_A ^{\rm eq})^2} Y_A^2 \right) = + \langle\sigma_{A\bar{A} \rightarrow B\bar{B}} v\rangle \left(Y_A^2-\frac{(Y_A^{\rm eq})^2}{(Y_B^{\rm eq})^2}Y_B^2\right).
\ee

Obtaining accurate solutions requires solving the above coupled equations numerically.  In much of the parameter space of interest, however, it is possible to obtain good analytic approximations based on two effectively decoupled equations.   When $m_B<m_A$ and $\lambda_B \gg \lambda_A$, $\B$ typically freezes out of equilibrium well after $\A$ does.  Therefore, the evolution of $Y_A$ in \Eq{dYdA} becomes the conventional Boltzmann equation for one species of DM by taking $Y_B\approx Y^{\rm eq}_B$ at least up until the $\A$ freeze-out time.\footnote{After $\B$ freezes out, $Y_B\approx Y^{\rm eq}_B$ is invalid, so the two equations formally ``re-couple''.  Since $Y_A$ has approached its asymptotic value by then, though, it is insensitive to late-time details.}   In the case of $s$-wave annihilation of our interest, the relic abundance of $\A$ can be well approximated by the familiar result \cite{Kolb:1990vq} (with an extra factor of 2 to account for both $\A$ and $\Abar$)
\be
Y_A(\infty) \simeq \frac{x_{f,A}}{\lambda_A} = \frac{7.6}{g_{*s}/g_*^{1/2}M_{pl} T_{f,A} \langle\sigma_{A\bar{A} \rightarrow B\bar{B}}v\rangle},\label{YAsol}
\ee
where $T_{f,A} = m_B / x_{f,A}$ is the freeze-out temperature for $\A$, and in the last step we used $s x^3/2 H(m_B)=0.132(g_{*s}/g_*^{1/2})M_{pl}m_B$. 

The solution for $Y_B$ is more subtle, but can also be greatly simplified when the freeze-out times of $\A$ and $\B$ are well separated.  If $x_{f,B} \gg x_{f,A}$, then we can drop terms suppressed by $(Y_A^{\rm eq}/Y_B^{\rm eq})^2$ in \Eq{eq:dYdB}, and we can treat the effect of $\A$ on $\B$ freeze-out by taking $Y_A(x_{f,B})\simeq Y_A(x_{f,A})\simeq Y_A(\infty)$.  Defining $\Delta\equiv Y_B-Y_B^{\rm eq}$, we rewrite \Eq{eq:dYdB} as:
\be
\frac{d \Delta}{d x}= - \frac{dY_B^{\rm eq}}{dx} -\lambda_Bx^{-2}\Delta(2Y_B^{\rm eq}+\Delta)+\lambda_Ax^{-2}Y_A^2(\infty)\label{YBsol}.
\ee
Focussing on the epoch when $\B$ starts to deviate from equilibrium, we can apply the ansatz $\Delta=c \, Y_B^{\rm eq}$, where $c$ is $\mathcal{O}(1)$. The equilibrium distribution for $x\gg1$ is
\begin{align}
Y_B^{\rm eq}(x) & \simeq +0.145\frac{g}{g_{*s}}x^{3/2}e^{-x},\\
\frac{dY_B^{\rm eq}}{dx} &\approx -0.145\frac{g}{g_{*s}}x^{3/2}e^{-x}=-Y_B^{\rm eq},
\end{align}
where we only keep the leading power term in $x$ in the second line. Combining all these, we can rewrite \Eq{YBsol} as a quadratic equation for $Y_B^{\rm eq}$,
\be
\lambda_Bc(2+c)(Y_B^{\rm eq})^2-x_f^2(c+1)Y_B^{\rm eq}-\lambda_AY_A^2(\infty)=0,
\ee
whose real positive solution is
\be
Y_B^{\rm eq}(x)=\frac{(c+1)x^2+\sqrt{(c+1)^2x^4+4\lambda_B\lambda_Ac(c+2)Y_A^2(\infty)}}{2\lambda_Bc(2+c)}.\label{YBresult}
\ee
We can then equate this equation with $Y_B^{\rm eq}(x) \simeq x^{3/2}e^{-x}$ to solve numerically for $x_{f,B}$.

We can see that by removing the contribution from $\A$ (i.e.~the term $\propto\lambda_AY_A^2(\infty)$) in \Eq{YBresult}, $\B$ freezes out in the standard way. In particular, we have the approximate relation $x_{f,B} \simeq \log \lambda_B - \frac{1}{2} \log x_{f,B}$ which yields
\be
Y_B(\infty) \simeq \frac{x_{f,B}}{\lambda_B}, \label{YBfinalsol}
\ee
in analogy with \Eq{YAsol}.  We also see that \Eq{YBresult} approaches the standard freeze-out solution when $\lambda_B$ decreases and approaches $\lambda_A$, such that $\B\Bbar\rightarrow \gamma' \gamma'$ freezes out at temperatures comparable to $\A\Abar\rightarrow \B\Bbar$; in that regime, the effect of $\A$ on the $\B$ evolution is subdominant since $Y_A^{\rm eq}<Y_B^{\rm eq}$ for $m_A>m_B$.  Standard freeze-out of $\B$ continues to hold when $\lambda_B\ll\lambda_A$, though the approximate solution \Eq{YBresult} would not be valid in that regime, since $\Omega_{B}>\Omega_{A}$, in contradiction to our ansatz that $\A$ constitutes the major DM component. 

More surprising is the case of large $\lambda_B$.  The $Y_A^2(\infty)$ term in \Eq{YBresult} dominates when
\be
\frac{\lambda_B}{\lambda_A} \left(\frac{m_B}{m_A}\right)^2 \gg  x_{f,B}^2,
\ee
where we have estimated $x_{f,A}/x_{f,B} \simeq m_A / m_B$.  Taking $Y_B^{\rm eq}(x_{f,B}) \simeq Y_B(\infty)$, \Eq{YBresult} reduces to
\be
\label{eq:balancedfreezeoutresult}
Y_B(\infty) = \sqrt{\frac{\lambda_A}{\lambda_B}} Y_A(\infty).
\ee
This behavior is very strange from the point of view of standard freeze-out, since the abundance of $\B$ scales like $1/\sqrt{\sigma_B}$ (instead of like the expected $1/\sigma_B$).  A naive quick way of understanding this behavior is by setting $d Y_B/d x \approx 0$ in \Eq{eq:dYdB} and dropping all $Y_i^{\rm eq}$ terms at late times, which immediately leads to \Eq{eq:balancedfreezeoutresult}.  We call this ``balanced freeze-out'', since the abundance of $\B$ is set by the balance between a depleting term ($\propto \lambda_B Y_B^2$) and a replenishing term ($\propto \lambda_A Y_A^2$).  Unlike in ordinary freeze-out where the expansion of the universe plays a key role in setting the abundance, in balanced freeze-out the main effect of the Hubble expansion is simply to drive $Y_i^{\rm eq}$ to zero at late times.

\section{Direct Detection of Non-Boosted DM}
\label{app:ADirectDetection}

In this paper, we have largely assumed that $\A$ has no couplings to the SM.  Given the contact interaction in \Eq{eq:AABBint}, though, $\A$ can interact with the dark photon via $\B$ loops.  In this appendix, we consider the direct detection bounds on $\A$ from these loop processes.  Of course, as with $\B$, one can relax direct detection limits by giving $\A$ an inelastic mass splitting.

\begin{figure}[t]
    \subfloat[\label{fig:AApp}]{{\includegraphics[scale=0.5, trim = 0 -2.2cm 0 0]{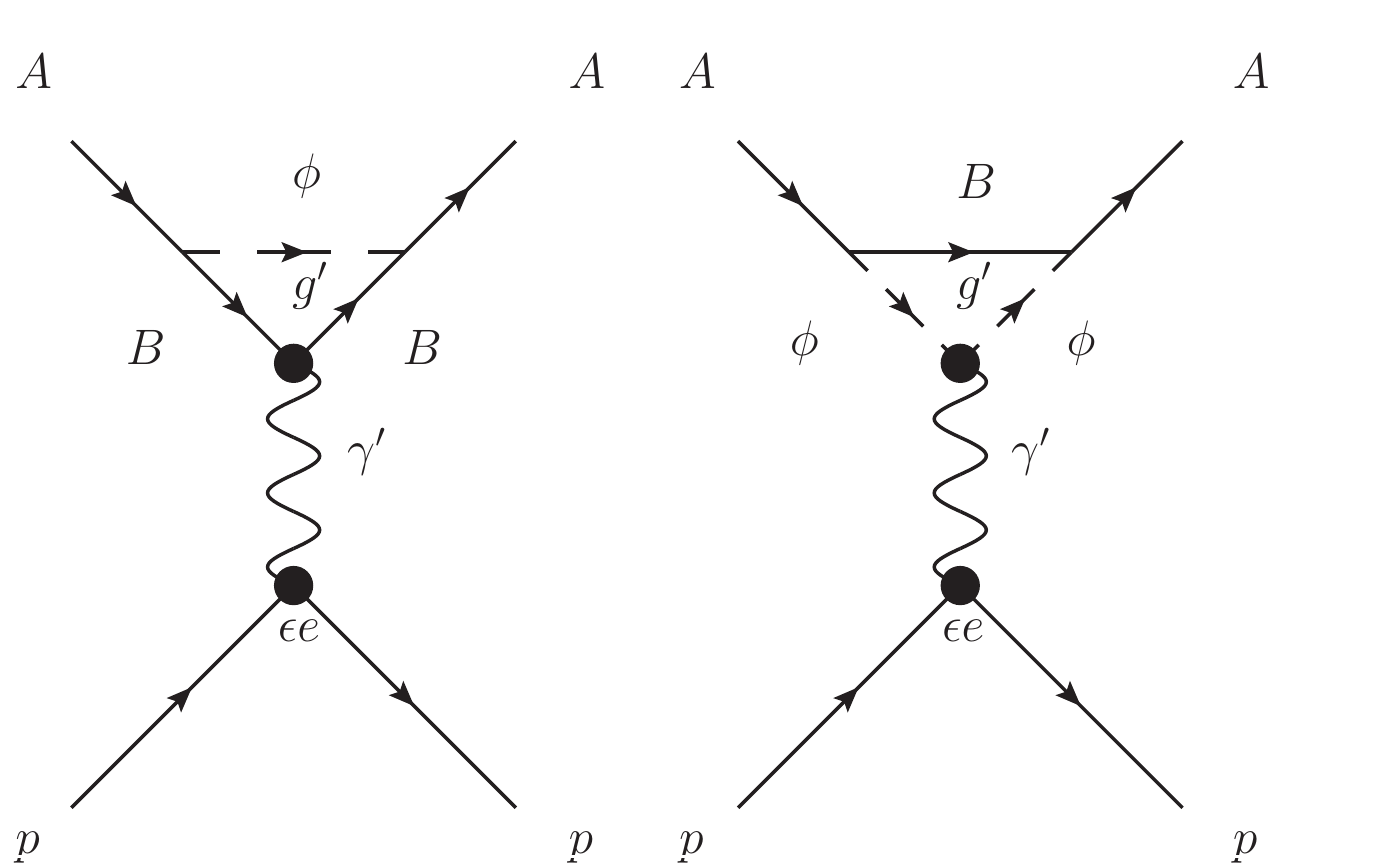} }}
    \qquad
    \subfloat[\label{fig:Adirectdetection}] {{
    \includegraphics[scale=0.6]{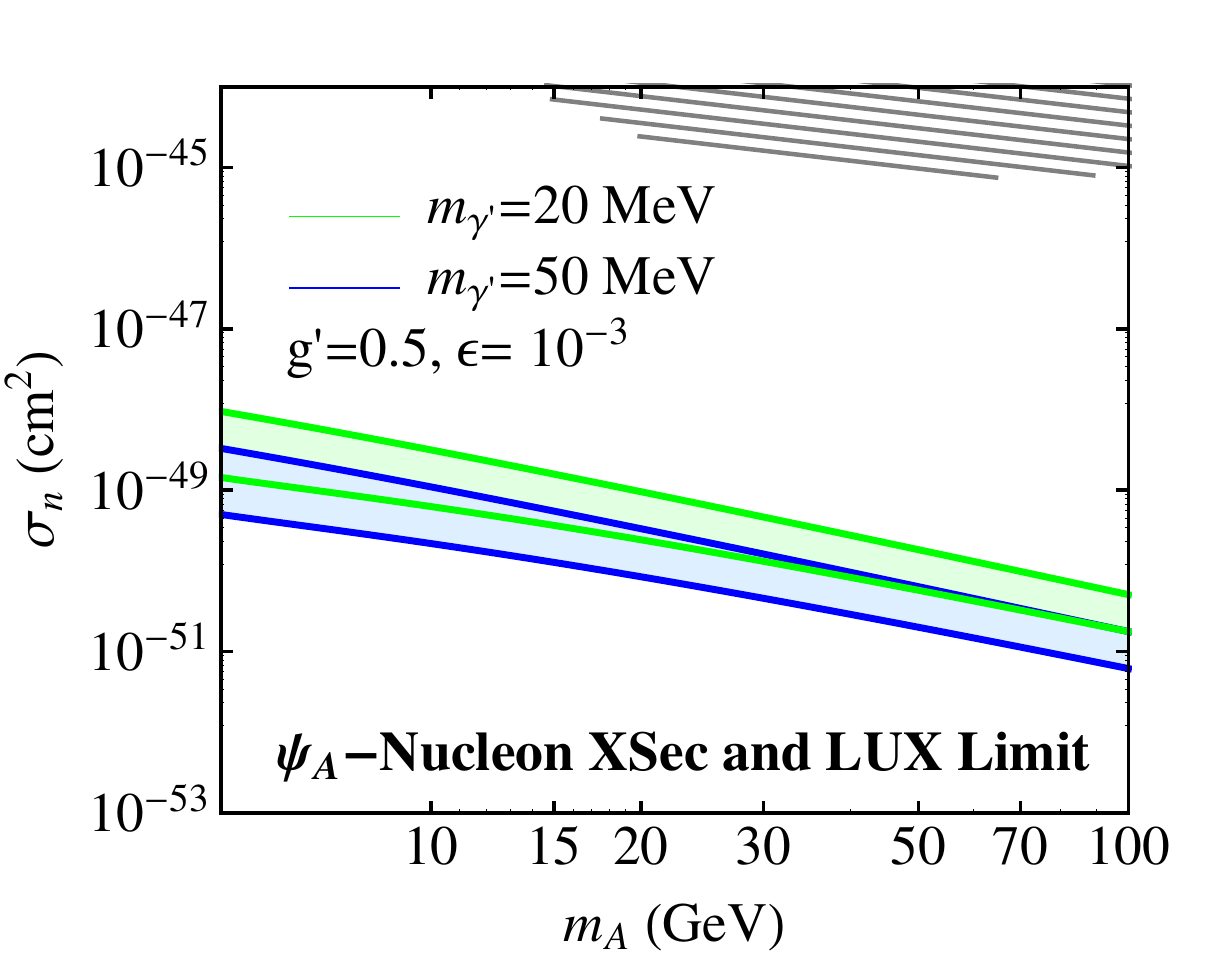}
     }}
   \caption{Left: Direct detection mechanism for $\A$ via a $\B$-$\phi$ loop.  Right: Scattering cross section of $\A$ on nucleons, sweeping $m_B = 0.1~\GeV$--$3~\GeV$ and fixing $g'=0.5$ and $\epsilon = 10^{-3}$.  Also shown are the current LUX limit (gray hashes).} \label{fig:AA}
\end{figure}

The loop-induced couplings of $\A$ to the dark photon depend on the UV completion of \Eq{eq:AABBint}, and we consider exchange of a complex scalar $\phi$ with $U(1)'$ charge as a concrete example.  The Lagrangian for this system is
\be
\label{eq:Acompletion}
\mathcal{L} \supset |D_\mu \phi|^2 - m_\phi^2 |\phi|^2  + \Bbar \slashed{D} \B  + (\lambda \Bbar \A \phi + \text{h.c.}),
\ee
where $D_\mu = \partial_\mu - i g' A'_\mu$.   Integrating out $\phi$ yields the contact interaction in \Eq{eq:AABBint} with
\be
\frac{1}{\Lambda^2} = \frac{\lambda^2}{m^2_{\phi}}.
\ee
Through $\B$-$\phi$ loops, $\A$ acquires a coupling to the dark photon.  In the limit $m_\phi \gg m_A \gg m_B$, the $\B$-$\phi$ loop generates the effective dimension six operator 
\be
\label{eq:effoperator}
\delta \mathcal{L} = \frac{ g' \lambda^2}{48 \pi^2}   \frac{ \log (m_B^2/m_\phi^2)}{m_\phi^2} \left( \Abar \gamma^\mu \partial^\nu \A F'_{\mu \nu} +\text{h.c.}\right),
\ee
which can lead to $\A$-nucleon scattering as in \Fig{fig:AApp}. As discussed in the appendix of \Ref{Agrawal:2011ze}, the standard dimension five dipole operator $\Abar \Sigma^{\mu \nu} \A F'_{\mu \nu}$ does not appear after integrating out $\B$ and $\phi$, because the interactions in \Eq{eq:Acompletion} respect a chiral symmetry acting on $\A$.

Similar to \Ref{Agrawal:2011ze} (but replacing the photon with a dark photon), the dominant effect of \Eq{eq:effoperator} is to give rise to a charge-charge interaction between DM and a nucleus $N$.  The spin-independent $\A N \rightarrow \A N$ cross section is
\be
\frac{d \sigma_{A N \to A N}}{d E_R} = \frac{m_N (Z \epsilon e)^2}{2 \pi v^2}  \frac{t^2}{(m_{\gamma'}^2 - t)^2} \left[\frac{ g' \lambda^2}{48 \pi^2}   \frac{ \log (m_B^2/m_\phi^2)}{m_\phi^2} \right]^2 F^2(E_R),
\ee
where $m_N$ is the nucleus mass, $E_R$ is the nucleus recoil energy, $t = -2 m_N E_R$ is the momentum-transfer-squared, $v$ is the DM velocity, $Z$ is the nucleus charge number, and $F^2(E_R)$ is the nucleus charge form factor.   The numerator in the expression above corresponds just to the lowest term in an expansion in $t$ (i.e.~small momentum transfer).  Spin-independent bounds on DM typically assume equal couplings to neutrons and protons, and can be expressed in terms of an effective nucleon cross section $\sigma_n$, with
\be
\frac{d \sigma_{A N \to A N}}{d E_R} =   \sigma_n \frac{m_N A^2}{2 \mu^2 v^2} F^2(E_R),
\ee
where $\mu$ is the DM-nucleon reduced mass, and $A$ is the nucleus mass number.  Thus, we have
\be
\sigma_n =\frac{ \mu^2 (Z \epsilon e)^2}{\pi A^2} \frac{t^2}{(m_{\gamma'}^2 - t)^2} \left[\frac{ g' \lambda^2}{48 \pi^2}   \frac{ \log (m_B^2/m_\phi^2)}{m_\phi^2} \right]^2.
\ee
Note that this cross section is momentum dependent, but for simplicity, we will take $E_R \simeq 10~\keV$ to determine the typical value of $t$.

Near the benchmark in \Eq{eq:keybenchmark}, $m_A$ is heavier than the proton so $\mu \simeq m_p$. The DM-nucleon cross section scales roughly as
\be
\sigma_{n}  \approx 4.5 \times10^{-49}~\text{cm}^2 \left(\frac{\epsilon}{10^{-3}} \right)^2  \left(\frac{g'}{0.5} \right)^2 \left(\frac{250~\GeV}{\Lambda} \right)^4 ,  \label{eq:Aloop}      
\ee  
where we have set $\lambda = 1$, ignored the logarithmic dependence on $m_\phi$ and $m_B$, and ignored the $m_{\gamma'}$ dependence since $m_{\gamma'}$  is comparable to the typical values of $t$.  Since we adjust $\Lambda$ (equivalently $m_\phi/\lambda$) to get the right abundance of DM, and since $\Lambda^4 \approx m_A^2 / \langle \sigma_{A \bar{A} \rightarrow B \bar{B}} v \rangle $ from \Eq{eq:thermav}, we can rewrite this dependence as:
\be
\sigma_{n}  \approx 4.5 \times10^{-49}~\text{cm}^2 \left(\frac{\epsilon}{10^{-3}} \right)^2  \left(\frac{g'}{0.5} \right)^2 \left( \frac{20 \text{ GeV}}{m_A} \right)^2   \left(\frac{\langle \sigma_{A \bar{A} \rightarrow B \bar{B}} v \rangle}{5 \times 10^{-26} \text{cm}^3/\text{sec}} \right). \label{eq:Aloop2}      
\ee
In \Fig{fig:Adirectdetection}, we show the limits of the LUX experiment \cite{Akerib:2013tjd} on the direct detection of $\A$ for different values of $(m_A, m_B, m_{\gamma'})$, and see that these constraints are easily satisfied, though future direct detection experiments would have sensitivity.

\section{Boosted DM Scattering Off Hadrons}
\label{app:detection}

In \Sec{bDM_detect}, we focused on the $\B e^- \rightarrow \B e^- $ mode for boosted DM detection.  Here, we summarize the  signal event rate for boosted DM scattering off protons or nuclei.  Since the number of signal events is rather small, we have not pursued a background study, though we remark that the angular pointing for hadronic scattering is rather poor at the low energies we are considering. As discussed in the main text, both event rate and angular resolution for scattering off proton can be improved at liquid Argon detectors.

\subsection{Elastic Scattering Off Hadrons}
\label{app:elasticBp}

The elastic scattering $\B N \rightarrow \B N$ has similar kinematics to electron scattering (with the obvious replacement $m_e \to m_N$), except we have to include the electromagnetic form factor. We will express the cross section as a function of the electric and magnetic Sachs form factors $G_E$ and $G_M$.  For protons, we can use the dipole approximation 
\be
G_E(q^2) = \frac{G_M(q^2)}{2.79} = \frac{1}{ \left( 1 + q^2/(0.71~\GeV^2)\right)^2}.
\label{eq:elasticprotonscattering}
\ee
To compute the cross section, we use the Rosenbluth formula in the lab frame as cited in \cite{Borie:2012tu}
\begin{align}
\label{eq:dsigmadO}
\frac{d \sigma }{ d \Omega}& = \frac{1}{(4\pi)^2)}\frac{(\epsilon e)^2 g'^2}{ (q^2 - m_{\gamma'}^2)^2 }  \frac{p'/p}{1 + (E_B - \frac{pE'_B}{p'} \cos \theta)/M} \nonumber \\ &~ \times \bigg(G_E^2 \frac{4 E_B E'_B + q^2}{1 - q^2/(4 M^2)}  + G_M^2 \left((4 E_B E'_B + q^2 )\left(1 - \frac{1}{1 - q^2/(4 M^2)} \right) + \frac{q^4}{2 M^2} + \frac{q^2 m_B^2}{M^2}\right) \bigg).
\end{align}
The energies and momenta are defined the same way as \Eq{eq:momentadefs}, $M$ is the mass of the proton, and $\theta$ is the scattering angle of $\B$.

The lowest momentum for a proton to Cherenkov radiate is $1.2~\GeV$, and for our benchmark in \Eq{eq:keybenchmark}, the proton cross section above this threshold is
\be
\sigma_{B p \to B p}^{\rm boost, Cher} = 1.4 \times 10^{-38} \text{ cm}^2 \left(\frac{\epsilon}{10^{-3}} \right)^2 \left( \frac{g'}{0.5} \right)^2  \label{eq:Bp_cherenkov},
\ee
yielding an all-sky event rate of
\be
\frac{N_{\rm events}}{\Delta T} =1.3 \times 10^{-3} \text{ year}^{-1}.
\ee
Due to the presence of the Cherenkov cutoff and the proton form factor, the elastic scattering rate in \Eq{eq:Bp_cherenkov} varies little within the mass range of interest, as given in \Eq{eq: mass scales}. When the transferred energy is above $2.5~\GeV$, the elastic scattering cross section is rather small, and protons instead typically produce secondary hadronic showers \cite{Fechner:2009mq}.  In that case, one should transition to the DIS calculation below.

As mentioned in footnote~\ref{footnote:protonissue}, large volume liquid Argon detectors are able to detect scattered protons with energies much below the Cherenkov threshold using ionization signals, where the total elastic scattering cross section off protons would be relevant.  We find the total elastic $\B \, p \to \B \, p$ scattering cross section for boosted $\B$ with $m_A\gtrsim1$ GeV to be
\be
\sigma_{B p \to B p}^{\rm boost, tot} = 1.8 \times 10^{-33} \text{ cm}^2 \left(\frac{\epsilon}{10^{-3}} \right)^2 \left( \frac{g'}{0.5} \right)^2\left(\frac{20\rm~MeV}{m_{\gamma'}}\right)^2  \label{eq:Bp_tot},
\ee
which is insensitive to $m_A$, $m_B$ in the boosted $\B$ regime due to the proton form factor.  We see that the total elastic scattering rate off proton is much larger than the one with a Cherenkov cutoff, so $\B \, p \to \B \, p$ could potentially be the leading signal detectable at a liquid Argon detector. 

Generalizing the previous calculation to a coherent nucleus of charge number Z:
\be \label{eq:xsecfe}
\sigma_{B N \rightarrow B N} = 1.2 \times 10^{-30} \text{ cm}^2  \left( \frac{Z}{26} \right)^2 \left(\frac{\epsilon}{10^{-3}} \right)^2 \left( \frac{g'}{0.5} \right)^2 \left(\frac{20\rm~MeV}{m_{\gamma'}}\right)^2. 
\ee

This same $\B \, p \to \B \, p$ calculation is relevant for direct detection of non-relativistic relic $\B$.  Taking the $q^2 \to 0$ limit and integrating over all angles, we have the cross section
\be
\sigma_{Bp \to Bp}^{v_B\rightarrow 0} = \frac{(\epsilon e)^2 g'^2}{\pi} \frac{ \mu_p^2}{m_{\gamma'}^4}.
\ee
where $\mu_p= m_p m_B /(m_p + m_B)$ is the reduced mass of the dark matter and the proton.  This is the basis for \Eq{eq:directdirectionestimate} shown earlier.

\subsection{Deep Inelastic Scattering Off Hadrons}

At sufficiently high energies, $\B$ scattering off hadrons will behave more like deep inelastic scattering (DIS), where the final state is a hadronic shower.  The DIS cross section is a convolution of the parton-level cross section with parton distribution functions (PDFs). The parton-level cross section $\hat{\sigma}$ is given by
\be
\frac{d \hat{\sigma}}{d \hat{t}} = \frac{1}{8 \pi} \frac{ (g' \epsilon \, Q_f)^2}{(\hat{t} - m_{\gamma'}^2)^2} \frac{ \left(\hat{s} - m_B^2 \right)^2  +  \left(\hat{u} - m_B^2 \right)^2 + 2m_B^2 \hat{t}   }{(\hat{s}- m_B^2)^2}. \label{eq:dsdt2}
\ee
For the $\B$-parton system: $\hat{s}+\hat{u} + \hat{t} = 2 m_B^2$, $\hat{t} = - Q^2$, and $\hat{s} = (1- x) m_B^2 + x s$.  We define $x$ by $p \equiv x P$ where $P$ is the 4-momentum of the initial proton at rest.  We define $y \equiv \frac{2 P\cdot q}{2 P\cdot k} = \frac{-\hat{t}}{\hat{s}- m_B^2}$, which characterizes the fraction of the energy transferred from $\B$ to the parton, since $y= \frac{q^0}{k^0}  = 1 - \frac{E'}{E}$ in the rest frame of the initial proton.

From these relations, we get the transferred momentum $Q^2 = x y (s-m_B^2)$, and $dx \, dQ^2 = \frac{dQ^2}{dy} dx \, dy =  x(s- m_B^2) \, dx \, dy$.
Including parton distribution functions, and using $x$/$y$ as variables, we obtain the resulting DIS cross section:
\be
\frac{d^2 \sigma}{dx \, dy} = \left( \sum_f x f_f(x,Q) Q_f^2 \right) \frac{(g' \epsilon)^2 }{8 \pi x} \frac{s (2x - 2 xy + xy^2) + m_B^2 (-2 x - xy^2 - 2y(1-x))}{(xy(s-m_B^2)+ m_{\gamma'}^2)^2},  \label{eq:d2sxy} 
\ee
where $f_f(x,Q)$ are PDFs with $_f$ indicating different flavor of fermion. For numerical evaluation, we use the MSTW2008 NNLO PDFs from \Ref{Martin:2009iq}. The integration limits of \Eq{eq:d2sxy} are $0 \leq x\leq1$ and $0 \leq y \leq y_{\rm max}$, where applying the condition $\cos\theta\leq1$ we obtain
\be
y_{\rm max}= \frac{4 (E_B^2 - m_B^2)(m_B^2 - s)x}{- 4 E_B^2 m_B^2 + 4 E_B^2 m_B^2 x - 4 E_B^2 s x - m_B^4 x^2 + 2 m_B^2 s x^2 - s^2 x^2 },
\ee
with $s= m_B^2 + M_p^2 + 2 M_p E_B$ and $E_B = m_A$.  Unlike the familiar case of DIS initiated by nearly massless incoming particles, for the massive $\B$ we consider here, $y_{\rm max}$ is not trivially $1$.  

Since the PDFs are only reliable for transferred energies over $\sim1$ GeV, we impose $Q^2\geq(1~\rm GeV)^2$ as a default cut for numerical integration. Analogous to the discussion for elastic scattering signals, for a particular experiment, there may be harder cut on phase space due to detector threshold energy.  For our benchmark in \Eq{eq:keybenchmark}, the DIS cross section above the $1~\GeV$ threshold is
\be
\sigma_{B p \rightarrow B X} = 1.42 \times 10^{-37} \text{ cm}^2 \left(\frac{\epsilon}{10^{-3}} \right)^2 \left( \frac{g'}{0.5} \right)^2,
\ee
yielding
\be
\frac{N_\text{events}}{\Delta T }  = 3.6 \times 10^{-2}\text{ year}^{-1}
\ee
for the all-sky event rate at Super-K.

\bibliography{biblio}{}

\providecommand{\href}[2]{#2}\begingroup\raggedright\begin{thebibliography}{100}

\bibitem{Zwicky:1933gu}
F.~Zwicky, {\it {Die Rotverschiebung von extragalaktischen Nebeln}},  {\em
  Helv.Phys.Acta} {\bf 6} (1933) 110--127.

\bibitem{Begeman:1991iy}
K.~Begeman, A.~Broeils, and R.~Sanders, {\it {Extended rotation curves of
  spiral galaxies: Dark haloes and modified dynamics}},  {\em
  Mon.Not.Roy.Astron.Soc.} {\bf 249} (1991) 523.

\bibitem{Bertone:2010zza}
G.~Bertone, J.~Silk, B.~Moore, J.~Diemand, J.~Bullock, et~al., {\em {Particle
  Dark Matter: Observations, Models and Searches}}.
\newblock Cambridge University Press, 2010.

\bibitem{Belanger:2011ww}
G.~Belanger and J.-C. Park, {\it {Assisted freeze-out}},  {\em JCAP} {\bf 1203}
  (2012) 038, [\href{http://xxx.lanl.gov/abs/1112.4491}{{\tt
  arXiv:1112.4491}}].

\bibitem{Fukuda:2002uc}
{\bf Super-Kamiokande} Collaboration, Y.~Fukuda et~al., {\it {The
  Super-Kamiokande detector}},  {\em Nucl.Instrum.Meth.} {\bf A501} (2003)
  418--462.

\bibitem{Abe:2011ts}
K.~Abe, T.~Abe, H.~Aihara, Y.~Fukuda, Y.~Hayato, et~al., {\it {Letter of
  Intent: The Hyper-Kamiokande Experiment --- Detector Design and Physics
  Potential ---}},  \href{http://xxx.lanl.gov/abs/1109.3262}{{\tt
  arXiv:1109.3262}}.

\bibitem{Ahrens:2002dv}
{\bf IceCube} Collaboration, J.~Ahrens et~al., {\it {Icecube - the next
  generation neutrino telescope at the south pole}},  {\em
  Nucl.Phys.Proc.Suppl.} {\bf 118} (2003) 388--395,
  [\href{http://xxx.lanl.gov/abs/astro-ph/0209556}{{\tt astro-ph/0209556}}].

\bibitem{Aartsen:2014oha}
{\bf IceCube-PINGU} Collaboration, M.~Aartsen et~al., {\it {Letter of Intent:
  The Precision IceCube Next Generation Upgrade (PINGU)}},
  \href{http://xxx.lanl.gov/abs/1401.2046}{{\tt arXiv:1401.2046}}.

\bibitem{MICA}
E.~Resconi, ``The stepping stones to proton decay: Icecube, pingu, mica.'' Talk
  at New Directions in Neutrino Physics, Aspen, Feb. 2013
  \url{http://indico.cern.ch/event/224351/contribution/33/material/slides/0.pdf}.

\bibitem{Katz:2006wv}
U.~F. Katz, {\it {KM3NeT: Towards a km**3 Mediterranean Neutrino Telescope}},
  {\em Nucl.Instrum.Meth.} {\bf A567} (2006) 457--461,
  [\href{http://xxx.lanl.gov/abs/astro-ph/0606068}{{\tt astro-ph/0606068}}].

\bibitem{Collaboration:2011nsa}
{\bf ANTARES} Collaboration, M.~Ageron et~al., {\it {ANTARES: the first
  undersea neutrino telescope}},  {\em Nucl.Instrum.Meth.} {\bf A656} (2011)
  11--38, [\href{http://xxx.lanl.gov/abs/1104.1607}{{\tt arXiv:1104.1607}}].

\bibitem{Bueno:2007um}
A.~Bueno, Z.~Dai, Y.~Ge, M.~Laffranchi, A.~Melgarejo, et~al., {\it {Nucleon
  decay searches with large liquid argon TPC detectors at shallow depths:
  Atmospheric neutrinos and cosmogenic backgrounds}},  {\em JHEP} {\bf 0704}
  (2007) 041, [\href{http://xxx.lanl.gov/abs/hep-ph/0701101}{{\tt
  hep-ph/0701101}}].

\bibitem{Badertscher:2010sy}
A.~Badertscher, A.~Curioni, U.~Degunda, L.~Epprecht, S.~Horikawa, et~al., {\it
  {Giant Liquid Argon Observatory for Proton Decay, Neutrino Astrophysics and
  CP-violation in the Lepton Sector (GLACIER)}},
  \href{http://xxx.lanl.gov/abs/1001.0076}{{\tt arXiv:1001.0076}}.

\bibitem{PhysRevD.88.013008}
Y.-F. Li, J.~Cao, Y.~Wang, and L.~Zhan, {\it Unambiguous determination of the
  neutrino mass hierarchy using reactor neutrinos},  {\em Phys. Rev. D} {\bf
  88} (Jul, 2013) 013008.

\bibitem{Li:2014qca}
Y.-F. Li, {\it {Overview of the Jiangmen Underground Neutrino Observatory
  (JUNO)}},  {\em Int.J.Mod.Phys.Conf.Ser.} {\bf 31} (2014) 1460300,
  [\href{http://xxx.lanl.gov/abs/1402.6143}{{\tt arXiv:1402.6143}}].

\bibitem{DEramo:2010ep}
F.~D'Eramo and J.~Thaler, {\it {Semi-annihilation of Dark Matter}},  {\em JHEP}
  {\bf 1006} (2010) 109, [\href{http://xxx.lanl.gov/abs/1003.5912}{{\tt
  arXiv:1003.5912}}].

\bibitem{SungCheon:2008ts}
H.~S. Cheon, S.~K. Kang, and C.~Kim, {\it {Doubly Coexisting Dark Matter
  Candidates in an Extended Seesaw Model}},  {\em Phys.Lett.} {\bf B675} (2009)
  203--209, [\href{http://xxx.lanl.gov/abs/0807.0981}{{\tt arXiv:0807.0981}}].

\bibitem{Hambye:2008bq}
T.~Hambye, {\it {Hidden vector dark matter}},  {\em JHEP} {\bf 0901} (2009)
  028, [\href{http://xxx.lanl.gov/abs/0811.0172}{{\tt arXiv:0811.0172}}].

\bibitem{Hambye:2009fg}
T.~Hambye and M.~H. Tytgat, {\it {Confined hidden vector dark matter}},  {\em
  Phys.Lett.} {\bf B683} (2010) 39--41,
  [\href{http://xxx.lanl.gov/abs/0907.1007}{{\tt arXiv:0907.1007}}].

\bibitem{Arina:2009uq}
C.~Arina, T.~Hambye, A.~Ibarra, and C.~Weniger, {\it {Intense Gamma-Ray Lines
  from Hidden Vector Dark Matter Decay}},  {\em JCAP} {\bf 1003} (2010) 024,
  [\href{http://xxx.lanl.gov/abs/0912.4496}{{\tt arXiv:0912.4496}}].

\bibitem{Belanger:2012vp}
G.~Belanger, K.~Kannike, A.~Pukhov, and M.~Raidal, {\it {Impact of
  semi-annihilations on dark matter phenomenology - an example of $Z_N$
  symmetric scalar dark matter}},  {\em JCAP} {\bf 1204} (2012) 010,
  [\href{http://xxx.lanl.gov/abs/1202.2962}{{\tt arXiv:1202.2962}}].

\bibitem{Carlson:1992fn}
E.~D. Carlson, M.~E. Machacek, and L.~J. Hall, {\it {Self-interacting dark
  matter}},  {\em Astrophys.J.} {\bf 398} (1992), no.~1 43--52.

\bibitem{deLaix:1995vi}
A.~A. de~Laix, R.~J. Scherrer, and R.~K. Schaefer, {\it {Constraints of
  selfinteracting dark matter}},  {\em Astrophys.J.} {\bf 452} (1995) 495,
  [\href{http://xxx.lanl.gov/abs/astro-ph/9502087}{{\tt astro-ph/9502087}}].

\bibitem{Hochberg:2014dra}
Y.~Hochberg, E.~Kuflik, T.~Volansky, and J.~G. Wacker, {\it {The SIMP
  Miracle}},  \href{http://xxx.lanl.gov/abs/1402.5143}{{\tt arXiv:1402.5143}}.

\bibitem{Berezhiani:1995am}
Z.~Berezhiani, A.~Dolgov, and R.~Mohapatra, {\it {Asymmetric inflationary
  reheating and the nature of mirror universe}},  {\em Phys.Lett.} {\bf B375}
  (1996) 26--36, [\href{http://xxx.lanl.gov/abs/hep-ph/9511221}{{\tt
  hep-ph/9511221}}].

\bibitem{Berezhiani:2000gw}
Z.~Berezhiani, D.~Comelli, and F.~L. Villante, {\it {The Early mirror universe:
  Inflation, baryogenesis, nucleosynthesis and dark matter}},  {\em Phys.Lett.}
  {\bf B503} (2001) 362--375,
  [\href{http://xxx.lanl.gov/abs/hep-ph/0008105}{{\tt hep-ph/0008105}}].

\bibitem{Ciarcelluti:2010zz}
P.~Ciarcelluti, {\it {Cosmology with mirror dark matter}},  {\em
  Int.J.Mod.Phys.} {\bf D19} (2010) 2151--2230,
  [\href{http://xxx.lanl.gov/abs/1102.5530}{{\tt arXiv:1102.5530}}].

\bibitem{Feng:2008mu}
J.~L. Feng, H.~Tu, and H.-B. Yu, {\it {Thermal Relics in Hidden Sectors}},
  {\em JCAP} {\bf 0810} (2008) 043,
  [\href{http://xxx.lanl.gov/abs/0808.2318}{{\tt arXiv:0808.2318}}].

\bibitem{Huang:2013xfa}
J.~Huang and Y.~Zhao, {\it {Dark Matter Induced Nucleon Decay: Model and
  Signatures}},  {\em JHEP} {\bf 1402} (2014) 077,
  [\href{http://xxx.lanl.gov/abs/1312.0011}{{\tt arXiv:1312.0011}}].

\bibitem{Pospelov:2007mp}
M.~Pospelov, A.~Ritz, and M.~B. Voloshin, {\it {Secluded WIMP Dark Matter}},
  {\em Phys.Lett.} {\bf B662} (2008) 53--61,
  [\href{http://xxx.lanl.gov/abs/0711.4866}{{\tt arXiv:0711.4866}}].

\bibitem{ArkaniHamed:2008qn}
N.~Arkani-Hamed, D.~P. Finkbeiner, T.~R. Slatyer, and N.~Weiner, {\it {A Theory
  of Dark Matter}},  {\em Phys.Rev.} {\bf D79} (2009) 015014,
  [\href{http://xxx.lanl.gov/abs/0810.0713}{{\tt arXiv:0810.0713}}].

\bibitem{Ackerman:mha}
L.~Ackerman, M.~R. Buckley, S.~M. Carroll, and M.~Kamionkowski, {\it {Dark
  Matter and Dark Radiation}},  {\em Phys.Rev.} {\bf D79} (2009) 023519,
  [\href{http://xxx.lanl.gov/abs/0810.5126}{{\tt arXiv:0810.5126}}].

\bibitem{Nomura:2008ru}
Y.~Nomura and J.~Thaler, {\it {Dark Matter through the Axion Portal}},  {\em
  Phys.Rev.} {\bf D79} (2009) 075008,
  [\href{http://xxx.lanl.gov/abs/0810.5397}{{\tt arXiv:0810.5397}}].

\bibitem{Mardon:2009gw}
J.~Mardon, Y.~Nomura, and J.~Thaler, {\it {Cosmic Signals from the Hidden
  Sector}},  {\em Phys.Rev.} {\bf D80} (2009) 035013,
  [\href{http://xxx.lanl.gov/abs/0905.3749}{{\tt arXiv:0905.3749}}].

\bibitem{AB_CMB}
Z.~Chacko, Y.~Cui, S.~Hong, and T.~Okui, ``Cosmological signals of a hidden
  dark matter sector.'' \!\!, work in preparation.

\bibitem{Bjorken:2009mm}
J.~D. Bjorken, R.~Essig, P.~Schuster, and N.~Toro, {\it {New Fixed-Target
  Experiments to Search for Dark Gauge Forces}},  {\em Phys.Rev.} {\bf D80}
  (2009) 075018, [\href{http://xxx.lanl.gov/abs/0906.0580}{{\tt
  arXiv:0906.0580}}].

\bibitem{Hodges:1993yb}
H.~Hodges, {\it {Mirror baryons as the dark matter}},  {\em Phys.Rev.} {\bf
  D47} (1993) 456--459.

\bibitem{Mohapatra:2000qx}
R.~N. Mohapatra and V.~L. Teplitz, {\it {Mirror dark matter and galaxy core
  densities of galaxies}},  {\em Phys.Rev.} {\bf D62} (2000) 063506,
  [\href{http://xxx.lanl.gov/abs/astro-ph/0001362}{{\tt astro-ph/0001362}}].

\bibitem{Foot:2001hc}
R.~Foot, {\it {Seven (and a half) reasons to believe in mirror matter: from
  neutrino puzzles to the inferred dark matter in the universe}},  {\em Acta
  Phys.Polon.} {\bf B32} (2001) 2253--2270,
  [\href{http://xxx.lanl.gov/abs/astro-ph/0102294}{{\tt astro-ph/0102294}}].

\bibitem{Fairbairn:2008fb}
M.~Fairbairn and J.~Zupan, {\it {Dark matter with a late decaying dark
  partner}},  {\em JCAP} {\bf 0907} (2009) 001,
  [\href{http://xxx.lanl.gov/abs/0810.4147}{{\tt arXiv:0810.4147}}].

\bibitem{Zurek:2008qg}
K.~M. Zurek, {\it {Multi-Component Dark Matter}},  {\em Phys.Rev.} {\bf D79}
  (2009) 115002, [\href{http://xxx.lanl.gov/abs/0811.4429}{{\tt
  arXiv:0811.4429}}].

\bibitem{Profumo:2009tb}
S.~Profumo, K.~Sigurdson, and L.~Ubaldi, {\it {Can we discover multi-component
  WIMP dark matter?}},  {\em JCAP} {\bf 0912} (2009) 016,
  [\href{http://xxx.lanl.gov/abs/0907.4374}{{\tt arXiv:0907.4374}}].

\bibitem{Fan:2013yva}
J.~Fan, A.~Katz, L.~Randall, and M.~Reece, {\it {Double-Disk Dark Matter}},
  {\em Phys.Dark Univ.} {\bf 2} (2013) 139--156,
  [\href{http://xxx.lanl.gov/abs/1303.1521}{{\tt arXiv:1303.1521}}].

\bibitem{Cui:2010ud}
Y.~Cui, J.~D. Mason, and L.~Randall, {\it {General Analysis of Antideuteron
  Searches for Dark Matter}},  {\em JHEP} {\bf 1011} (2010) 017,
  [\href{http://xxx.lanl.gov/abs/1006.0983}{{\tt arXiv:1006.0983}}].

\bibitem{Holdom:1985ag}
B.~Holdom, {\it {Two U(1)'s and Epsilon Charge Shifts}},  {\em Phys.Lett.} {\bf
  B166} (1986) 196.

\bibitem{Okun:1982xi}
L.~Okun, {\it {LIMITS OF ELECTRODYNAMICS: PARAPHOTONS?}},  {\em Sov.Phys.JETP}
  {\bf 56} (1982) 502.

\bibitem{Galison:1983pa}
P.~Galison and A.~Manohar, {\it {TWO Z's OR NOT TWO Z's?}},  {\em Phys.Lett.}
  {\bf B136} (1984) 279.

\bibitem{Stueckelberg:1900zz}
E.~Stueckelberg, {\it {Interaction energy in electrodynamics and in the field
  theory of nuclear forces}},  {\em Helv.Phys.Acta} {\bf 11} (1938) 225--244.

\bibitem{Kors:2005uz}
B.~Kors and P.~Nath, {\it {Aspects of the Stueckelberg extension}},  {\em JHEP}
  {\bf 0507} (2005) 069, [\href{http://xxx.lanl.gov/abs/hep-ph/0503208}{{\tt
  hep-ph/0503208}}].

\bibitem{TuckerSmith:2001hy}
D.~Tucker-Smith and N.~Weiner, {\it {Inelastic dark matter}},  {\em Phys.Rev.}
  {\bf D64} (2001) 043502, [\href{http://xxx.lanl.gov/abs/hep-ph/0101138}{{\tt
  hep-ph/0101138}}].

\bibitem{Cui:2009xq}
Y.~Cui, D.~E. Morrissey, D.~Poland, and L.~Randall, {\it {Candidates for
  Inelastic Dark Matter}},  {\em JHEP} {\bf 0905} (2009) 076,
  [\href{http://xxx.lanl.gov/abs/0901.0557}{{\tt arXiv:0901.0557}}].

\bibitem{Finkbeiner:2009mi}
D.~P. Finkbeiner, T.~R. Slatyer, N.~Weiner, and I.~Yavin, {\it {PAMELA, DAMA,
  INTEGRAL and Signatures of Metastable Excited WIMPs}},  {\em JCAP} {\bf 0909}
  (2009) 037, [\href{http://xxx.lanl.gov/abs/0903.1037}{{\tt
  arXiv:0903.1037}}].

\bibitem{Graham:2010ca}
P.~W. Graham, R.~Harnik, S.~Rajendran, and P.~Saraswat, {\it {Exothermic Dark
  Matter}},  {\em Phys.Rev.} {\bf D82} (2010) 063512,
  [\href{http://xxx.lanl.gov/abs/1004.0937}{{\tt arXiv:1004.0937}}].

\bibitem{Pospelov:2013nea}
M.~Pospelov, N.~Weiner, and I.~Yavin, {\it {Dark matter detection in two easy
  steps}},  \href{http://xxx.lanl.gov/abs/1312.1363}{{\tt arXiv:1312.1363}}.

\bibitem{Bhattacharya:2013hva}
S.~Bhattacharya, A.~Drozd, B.~Grzadkowski, and J.~Wudka, {\it {Two-Component
  Dark Matter}},  {\em JHEP} {\bf 1310} (2013) 158,
  [\href{http://xxx.lanl.gov/abs/1309.2986}{{\tt arXiv:1309.2986}}].

\bibitem{Modak:2013jya}
K.~P. Modak, D.~Majumdar, and S.~Rakshit, {\it {A Possible Explanation of Low
  Energy $\gamma$-ray Excess from Galactic Centre and Fermi Bubble by a Dark
  Matter Model with Two Real Scalars}},
  \href{http://xxx.lanl.gov/abs/1312.7488}{{\tt arXiv:1312.7488}}.

\bibitem{Steigman:2012nb}
G.~Steigman, B.~Dasgupta, and J.~F. Beacom, {\it {Precise Relic WIMP Abundance
  and its Impact on Searches for Dark Matter Annihilation}},  {\em Phys.Rev.}
  {\bf D86} (2012) 023506, [\href{http://xxx.lanl.gov/abs/1204.3622}{{\tt
  arXiv:1204.3622}}].

\bibitem{Fayet:2007ua}
P.~Fayet, {\it {U-boson production in e+ e- annihilations, psi and Upsilon
  decays, and Light Dark Matter}},  {\em Phys.Rev.} {\bf D75} (2007) 115017,
  [\href{http://xxx.lanl.gov/abs/hep-ph/0702176}{{\tt hep-ph/0702176}}].

\bibitem{Pospelov:2008zw}
M.~Pospelov, {\it {Secluded U(1) below the weak scale}},  {\em Phys.Rev.} {\bf
  D80} (2009) 095002, [\href{http://xxx.lanl.gov/abs/0811.1030}{{\tt
  arXiv:0811.1030}}].

\bibitem{Navarro:1995iw}
J.~F. Navarro, C.~S. Frenk, and S.~D. White, {\it {The Structure of cold dark
  matter halos}},  {\em Astrophys.J.} {\bf 462} (1996) 563--575,
  [\href{http://xxx.lanl.gov/abs/astro-ph/9508025}{{\tt astro-ph/9508025}}].

\bibitem{Cirelli:2010xx}
M.~Cirelli, G.~Corcella, A.~Hektor, G.~Hutsi, M.~Kadastik, et~al., {\it {PPPC 4
  DM ID: A Poor Particle Physicist Cookbook for Dark Matter Indirect
  Detection}},  {\em JCAP} {\bf 1103} (2011) 051,
  [\href{http://xxx.lanl.gov/abs/1012.4515}{{\tt arXiv:1012.4515}}].

\bibitem{Yuksel:2007ac}
H.~Yuksel, S.~Horiuchi, J.~F. Beacom, and S.~Ando, {\it {Neutrino Constraints
  on the Dark Matter Total Annihilation Cross Section}},  {\em Phys.Rev.} {\bf
  D76} (2007) 123506, [\href{http://xxx.lanl.gov/abs/0707.0196}{{\tt
  arXiv:0707.0196}}].

\bibitem{Mijakowski:2012dva}
{\bf Super-Kamiokande} Collaboration, P.~Mijakowski, {\it {Search for neutrinos
  from diffuse dark matter annihilation in Super-Kamiokande}},  {\em
  Nucl.Phys.Proc.Suppl.} {\bf 229-232} (2012) 546.

\bibitem{Beacom:2006tt}
J.~F. Beacom, N.~F. Bell, and G.~D. Mack, {\it {General Upper Bound on the Dark
  Matter Total Annihilation Cross Section}},  {\em Phys.Rev.Lett.} {\bf 99}
  (2007) 231301, [\href{http://xxx.lanl.gov/abs/astro-ph/0608090}{{\tt
  astro-ph/0608090}}].

\bibitem{Essig:2013lka}
R.~Essig, J.~A. Jaros, W.~Wester, P.~H. Adrian, S.~Andreas, et~al., {\it {Dark
  Sectors and New, Light, Weakly-Coupled Particles}},
  \href{http://xxx.lanl.gov/abs/1311.0029}{{\tt arXiv:1311.0029}}.

\bibitem{PhysRevD.67.093001}
J.~F. Beacom and S.~Palomares-Ruiz, {\it Neutral-current atmospheric neutrino
  flux measurement using neutrino-proton elastic scattering in
  super-kamiokande},  {\em Phys. Rev. D} {\bf 67} (May, 2003) 093001.

\bibitem{Fechner:2009mq}
M.~Fechner and C.~Walter, {\it {The physics impact of proton track
  identification in future megaton-scale water Cherenkov detectors}},  {\em
  JHEP} {\bf 0911} (2009) 040, [\href{http://xxx.lanl.gov/abs/0901.1950}{{\tt
  arXiv:0901.1950}}].

\bibitem{Fechner:2009aa}
{\bf Super-Kamiokande} Collaboration, M.~Fechner et~al., {\it {Kinematic
  reconstruction of atmospheric neutrino events in a large water Cherenkov
  detector with proton identification}},  {\em Phys.Rev.} {\bf D79} (2009)
  112010, [\href{http://xxx.lanl.gov/abs/0901.1645}{{\tt arXiv:0901.1645}}].

\bibitem{Ashie:2005ik}
{\bf Super-Kamiokande} Collaboration, Y.~Ashie et~al., {\it {A Measurement of
  atmospheric neutrino oscillation parameters by SUPER-KAMIOKANDE I}},  {\em
  Phys.Rev.} {\bf D71} (2005) 112005,
  [\href{http://xxx.lanl.gov/abs/hep-ex/0501064}{{\tt hep-ex/0501064}}].

\bibitem{Amenomori:2008zzd}
{\bf Tibet ASgamma} Collaboration, M.~Amenomori et~al., {\it {The cosmic-ray
  energy spectrum around the knee measured by the Tibet-III air-shower array}},
   {\em Nucl.Phys.Proc.Suppl.} {\bf 175-176} (2008) 318--321.

\bibitem{Formaggio:2013kya}
J.~Formaggio and G.~Zeller, {\it {From eV to EeV: Neutrino Cross Sections
  Across Energy Scales}},  {\em Rev.Mod.Phys.} {\bf 84} (2012) 1307,
  [\href{http://xxx.lanl.gov/abs/1305.7513}{{\tt arXiv:1305.7513}}].

\bibitem{LeeKaThesis}
L.~K. Pik, {\em Study of the neutrino mass hierarchy with the atmospheric
  neutrino data observed in Super-Kamiokande}.
\newblock PhD thesis, University of Tokyo, 2012.

\bibitem{Dziomba}
M.~Dziomba, {\em A Study of Neutrino Oscillation Models with Super-Kamiokande
  Atmospheric Neutrino Data}.
\newblock PhD thesis, University of Washington, 2012.

\bibitem{Beacom:2003nk}
J.~F. Beacom and M.~R. Vagins, {\it {GADZOOKS! Anti-neutrino spectroscopy with
  large water Cherenkov detectors}},  {\em Phys.Rev.Lett.} {\bf 93} (2004)
  171101, [\href{http://xxx.lanl.gov/abs/hep-ph/0309300}{{\tt
  hep-ph/0309300}}].

\bibitem{Gaisser:2002jj}
T.~Gaisser and M.~Honda, {\it {Flux of atmospheric neutrinos}},  {\em
  Ann.Rev.Nucl.Part.Sci.} {\bf 52} (2002) 153--199,
  [\href{http://xxx.lanl.gov/abs/hep-ph/0203272}{{\tt hep-ph/0203272}}].

\bibitem{Bays:2011si}
{\bf Super-Kamiokande} Collaboration, K.~Bays et~al., {\it {Supernova Relic
  Neutrino Search at Super-Kamiokande}},  {\em Phys.Rev.} {\bf D85} (2012)
  052007, [\href{http://xxx.lanl.gov/abs/1111.5031}{{\tt arXiv:1111.5031}}].

\bibitem{Abe:2010hy}
{\bf Super-Kamiokande} Collaboration, K.~Abe et~al., {\it {Solar neutrino
  results in Super-Kamiokande-III}},  {\em Phys.Rev.} {\bf D83} (2011) 052010,
  [\href{http://xxx.lanl.gov/abs/1010.0118}{{\tt arXiv:1010.0118}}].

\bibitem{Beringer:1900zz}
{\bf Particle Data Group} Collaboration, J.~Beringer et~al., {\it {Review of
  Particle Physics (RPP)}},  {\em Phys.Rev.} {\bf D86} (2012) 010001, Sec. 31
  (Passage of Particles Through Matter)
  \url{http://dx.doi.org/10.1103/PhysRevD.86.010001}.

\bibitem{Albuquerque:2003mi}
I.~Albuquerque, G.~Burdman, and Z.~Chacko, {\it {Neutrino telescopes as a
  direct probe of supersymmetry breaking}},  {\em Phys.Rev.Lett.} {\bf 92}
  (2004) 221802, [\href{http://xxx.lanl.gov/abs/hep-ph/0312197}{{\tt
  hep-ph/0312197}}].

\bibitem{Berger:2014sqa}
J.~Berger, Y.~Cui, and Y.~Zhao, {\it {Detecting Boosted Dark Matter from the
  Sun with Large Volume Neutrino Detectors}},
  \href{http://xxx.lanl.gov/abs/1410.2246}{{\tt arXiv:1410.2246}}.

\bibitem{Kearns:2013lea}
{\bf Hyper-Kamiokande Working Group} Collaboration, E.~Kearns et~al., {\it
  {Hyper-Kamiokande Physics Opportunities}},
  \href{http://xxx.lanl.gov/abs/1309.0184}{{\tt arXiv:1309.0184}}.

\bibitem{Abbasi:2011eq}
{\bf IceCube} Collaboration, R.~Abbasi et~al., {\it {Search for Dark Matter
  from the Galactic Halo with the IceCube Neutrino Observatory}},  {\em
  Phys.Rev.} {\bf D84} (2011) 022004,
  [\href{http://xxx.lanl.gov/abs/1101.3349}{{\tt arXiv:1101.3349}}].

\bibitem{Aartsen:2013vca}
{\bf IceCube} Collaboration, M.~Aartsen et~al., {\it {Search for
  neutrino-induced particle showers with IceCube-40}},
  \href{http://xxx.lanl.gov/abs/1312.0104}{{\tt arXiv:1312.0104}}.

\bibitem{MICA2}
D.~Cowen, ``Gev-scale physics in the south pole icecap.'' Talk at Baryon Number
  Violation Workshop, Argonne, Apr. 2013
  \url{https://indico.fnal.gov/getFile.py/access?contribId=118&sessionId=3&resId=0&materialId=slides&confId=6248}.

\bibitem{FDthesis}
F.~M. Dufour, {\em Precise study of the atmospheric neutrino oscillation
  pattern using Super-Kamiokande I and II}.
\newblock PhD thesis, Boston University, 2003.

\bibitem{Wendell:2010md}
{\bf Super-Kamiokande} Collaboration, R.~Wendell et~al., {\it {Atmospheric
  neutrino oscillation analysis with sub-leading effects in Super-Kamiokande I,
  II, and III}},  {\em Phys.Rev.} {\bf D81} (2010) 092004,
  [\href{http://xxx.lanl.gov/abs/1002.3471}{{\tt arXiv:1002.3471}}].

\bibitem{Desai:2004pq}
{\bf Super-Kamiokande} Collaboration, S.~Desai et~al., {\it {Search for dark
  matter WIMPs using upward through-going muons in Super-Kamiokande}},  {\em
  Phys.Rev.} {\bf D70} (2004) 083523,
  [\href{http://xxx.lanl.gov/abs/hep-ex/0404025}{{\tt hep-ex/0404025}}].

\bibitem{Mijakowski:2011zz}
P.~Mijakowski, {\em {Direct and Indirect Search for Dark Matter}}.
\newblock PhD thesis, University of Warsaw, 2011.

\bibitem{Mijakowski:slides}
P.~Mijakowski, ``{Indirect searches for dark matter particles at
  Super-Kamiokande}.''
  \url{http://moriond.in2p3.fr/J12/transparencies/11_Sunday_pm/mijakowski.pdf},
  2012.

\bibitem{Wendell:slides}
R.~Wendell, ``{Atmospheric Results from Super-Kamiokande}.''
  \url{https://indico.fnal.gov/getFile.py/access?contribId=260&sessionId=17&resId=0&materialId=slides&confId=8022},
  2014.

\bibitem{Himmel:2013jva}
{\bf Super-Kamiokande} Collaboration, A.~Himmel, {\it {Recent results from
  Super-Kamiokande}},  {\em AIP Conf.Proc.} {\bf 1604} (2014) 345--352,
  [\href{http://xxx.lanl.gov/abs/1310.6677}{{\tt arXiv:1310.6677}}].

\bibitem{Blumlein:2013cua}
J.~Bl{\"u}mlein and J.~Brunner, {\it {New Exclusion Limits on Dark Gauge Forces
  from Proton Bremsstrahlung in Beam-Dump Data}},  {\em Phys.Lett.} {\bf B731}
  (2014) 320--326, [\href{http://xxx.lanl.gov/abs/1311.3870}{{\tt
  arXiv:1311.3870}}].

\bibitem{Akerib:2013tjd}
{\bf LUX} Collaboration, D.~Akerib et~al., {\it {First results from the LUX
  dark matter experiment at the Sanford Underground Research Facility}},
  \href{http://xxx.lanl.gov/abs/1310.8214}{{\tt arXiv:1310.8214}}.

\bibitem{Agnese:2014aze}
{\bf SuperCDMS} Collaboration, R.~Agnese et~al., {\it {Search for Low-Mass
  WIMPs with SuperCDMS}},  \href{http://xxx.lanl.gov/abs/1402.7137}{{\tt
  arXiv:1402.7137}}.

\bibitem{PhysRevLett.112.041302}
{\bf SuperCDMS} Collaboration, R.~Agnese et~al., {\it Search for low-mass
  weakly interacting massive particles using voltage-assisted calorimetric
  ionization detection in the supercdms experiment},  {\em Phys. Rev. Lett.}
  {\bf 112} (Jan, 2014) 041302.

\bibitem{Barreto:2011zu}
{\bf DAMIC} Collaboration, J.~Barreto et~al., {\it {Direct Search for Low Mass
  Dark Matter Particles with CCDs}},  {\em Phys.Lett.} {\bf B711} (2012)
  264--269, [\href{http://xxx.lanl.gov/abs/1105.5191}{{\tt arXiv:1105.5191}}].

\bibitem{Essig:2011nj}
R.~Essig, J.~Mardon, and T.~Volansky, {\it {Direct Detection of Sub-GeV Dark
  Matter}},  {\em Phys.Rev.} {\bf D85} (2012) 076007,
  [\href{http://xxx.lanl.gov/abs/1108.5383}{{\tt arXiv:1108.5383}}].

\bibitem{Essig:2012yx}
R.~Essig, A.~Manalaysay, J.~Mardon, P.~Sorensen, and T.~Volansky, {\it {First
  Direct Detection Limits on sub-GeV Dark Matter from XENON10}},  {\em
  Phys.Rev.Lett.} {\bf 109} (2012) 021301,
  [\href{http://xxx.lanl.gov/abs/1206.2644}{{\tt arXiv:1206.2644}}].

\bibitem{Bergstrom:2013jra}
L.~Bergstrom, T.~Bringmann, I.~Cholis, D.~Hooper, and C.~Weniger, {\it {New
  limits on dark matter annihilation from AMS cosmic ray positron data}},  {\em
  Phys.Rev.Lett.} {\bf 111} (2013) 171101,
  [\href{http://xxx.lanl.gov/abs/1306.3983}{{\tt arXiv:1306.3983}}].

\bibitem{Ibarra:2013zia}
A.~Ibarra, A.~S. Lamperstorfer, and J.~Silk, {\it {Dark matter annihilations
  and decays after the AMS-02 positron measurements}},  {\em Phys.Rev.} {\bf
  D89} (2014) 063539, [\href{http://xxx.lanl.gov/abs/1309.2570}{{\tt
  arXiv:1309.2570}}].

\bibitem{Ackermann:2012qk}
{\bf LAT} Collaboration, M.~Ackermann et~al., {\it {Fermi LAT Search for Dark
  Matter in Gamma-ray Lines and the Inclusive Photon Spectrum}},  {\em
  Phys.Rev.} {\bf D86} (2012) 022002,
  [\href{http://xxx.lanl.gov/abs/1205.2739}{{\tt arXiv:1205.2739}}].

\bibitem{Madhavacheril:2013cna}
M.~S. Madhavacheril, N.~Sehgal, and T.~R. Slatyer, {\it {Current Dark Matter
  Annihilation Constraints from CMB and Low-Redshift Data}},
  \href{http://xxx.lanl.gov/abs/1310.3815}{{\tt arXiv:1310.3815}}.

\bibitem{Slatyer:2009vg}
T.~R. Slatyer, {\it {The Sommerfeld enhancement for dark matter with an excited
  state}},  {\em JCAP} {\bf 1002} (2010) 028,
  [\href{http://xxx.lanl.gov/abs/0910.5713}{{\tt arXiv:0910.5713}}].

\bibitem{Henning:2012rm}
B.~Henning and H.~Murayama, {\it {Constraints on Light Dark Matter from Big
  Bang Nucleosynthesis}},  \href{http://xxx.lanl.gov/abs/1205.6479}{{\tt
  arXiv:1205.6479}}.

\bibitem{Berezhiani:2012ru}
Z.~Berezhiani, A.~Dolgov, and I.~Tkachev, {\it {BBN with light dark matter}},
  {\em JCAP} {\bf 1302} (2013) 010,
  [\href{http://xxx.lanl.gov/abs/1211.4937}{{\tt arXiv:1211.4937}}].

\bibitem{Goodman:2010ku}
J.~Goodman, M.~Ibe, A.~Rajaraman, W.~Shepherd, T.~M. Tait, et~al., {\it
  {Constraints on Dark Matter from Colliders}},  {\em Phys.Rev.} {\bf D82}
  (2010) 116010, [\href{http://xxx.lanl.gov/abs/1008.1783}{{\tt
  arXiv:1008.1783}}].

\bibitem{Fox:2011fx}
P.~J. Fox, R.~Harnik, J.~Kopp, and Y.~Tsai, {\it {LEP Shines Light on Dark
  Matter}},  {\em Phys.Rev.} {\bf D84} (2011) 014028,
  [\href{http://xxx.lanl.gov/abs/1103.0240}{{\tt arXiv:1103.0240}}].

\bibitem{Fox:2011pm}
P.~J. Fox, R.~Harnik, J.~Kopp, and Y.~Tsai, {\it {Missing Energy Signatures of
  Dark Matter at the LHC}},  {\em Phys.Rev.} {\bf D85} (2012) 056011,
  [\href{http://xxx.lanl.gov/abs/1109.4398}{{\tt arXiv:1109.4398}}].

\bibitem{Cennini:1999ih}
P.~Cennini, J.~Revol, C.~Rubbia, F.~Sergiampietri, A.~Bueno, et~al., {\it
  {Detection of scintillation light in coincidence with ionizing tracks in a
  liquid argon time projection chamber}},  {\em Nucl.Instrum.Meth.} {\bf A432}
  (1999) 240--248.

\bibitem{Bromberg:2013fla}
C.~Bromberg, F.~Cavanna, T.~Junk, T.~Katori, K.~Lang, et~al., {\it {Liquid
  Argon Time Projection Chamber Research and Development in the United
  States}},  \href{http://xxx.lanl.gov/abs/1307.8166}{{\tt arXiv:1307.8166}}.

\bibitem{Belanger:2012zr}
G.~Belanger, K.~Kannike, A.~Pukhov, and M.~Raidal, {\it {$Z_3$ Scalar Singlet
  Dark Matter}},  {\em JCAP} {\bf 1301} (2013) 022,
  [\href{http://xxx.lanl.gov/abs/1211.1014}{{\tt arXiv:1211.1014}}].

\bibitem{Ko:2014nha}
P.~Ko and Y.~Tang, {\it {Self-interacting scalar dark matter with local $Z_{3}$
  symmetry}},  \href{http://xxx.lanl.gov/abs/1402.6449}{{\tt arXiv:1402.6449}}.

\bibitem{Aoki:2014cja}
M.~Aoki and T.~Toma, {\it {Impact of Semi-annihilation of Z3 Symmetric Dark
  Matter with Radiative Neutrino Masses}},
  \href{http://xxx.lanl.gov/abs/1405.5870}{{\tt arXiv:1405.5870}}.

\bibitem{Batell:2009di}
B.~Batell, M.~Pospelov, and A.~Ritz, {\it {Exploring Portals to a Hidden Sector
  Through Fixed Targets}},  {\em Phys.Rev.} {\bf D80} (2009) 095024,
  [\href{http://xxx.lanl.gov/abs/0906.5614}{{\tt arXiv:0906.5614}}].

\bibitem{Daylan:2014rsa}
T.~Daylan, D.~P. Finkbeiner, D.~Hooper, T.~Linden, S.~K.~N. Portillo, et~al.,
  {\it {The Characterization of the Gamma-Ray Signal from the Central Milky
  Way: A Compelling Case for Annihilating Dark Matter}},
  \href{http://xxx.lanl.gov/abs/1402.6703}{{\tt arXiv:1402.6703}}.

\bibitem{Boehm:2014bia}
C.~Boehm, M.~J. Dolan, and C.~McCabe, {\it {A weighty interpretation of the
  Galactic Centre excess}},  \href{http://xxx.lanl.gov/abs/1404.4977}{{\tt
  arXiv:1404.4977}}.

\bibitem{Abdullah:2014lla}
M.~Abdullah, A.~DiFranzo, A.~Rajaraman, T.~M.~P. Tait, P.~Tanedo, et~al., {\it
  {Hidden On-Shell Mediators for the Galactic Center Gamma-Ray Excess}},
  \href{http://xxx.lanl.gov/abs/1404.6528}{{\tt arXiv:1404.6528}}.

\bibitem{Martin:2014sxa}
A.~Martin, J.~Shelton, and J.~Unwin, {\it {Fitting the Galactic Center
  Gamma-Ray Excess with Cascade Annihilations}},
  \href{http://xxx.lanl.gov/abs/1405.0272}{{\tt arXiv:1405.0272}}.

\bibitem{Berlin:2014pya}
A.~Berlin, P.~Gratia, D.~Hooper, and S.~D. McDermott, {\it {Hidden Sector Dark
  Matter Models for the Galactic Center Gamma-Ray Excess}},
  \href{http://xxx.lanl.gov/abs/1405.5204}{{\tt arXiv:1405.5204}}.

\bibitem{Aguilar:2013qda}
{\bf AMS} Collaboration, M.~Aguilar et~al., {\it {First Result from the Alpha
  Magnetic Spectrometer on the International Space Station: Precision
  Measurement of the Positron Fraction in Primary Cosmic Rays of 0.5--350
  GeV}},  {\em Phys.Rev.Lett.} {\bf 110} (2013) 141102.

\bibitem{Essig:2013goa}
R.~Essig, E.~Kuflik, S.~D. McDermott, T.~Volansky, and K.~M. Zurek, {\it
  {Constraining Light Dark Matter with Diffuse X-Ray and Gamma-Ray
  Observations}},  {\em JHEP} {\bf 1311} (2013) 193,
  [\href{http://xxx.lanl.gov/abs/1309.4091}{{\tt arXiv:1309.4091}}].

\bibitem{Blennow:2012de}
M.~Blennow, E.~Fernandez-Martinez, O.~Mena, J.~Redondo, and P.~Serra, {\it
  {Asymmetric Dark Matter and Dark Radiation}},  {\em JCAP} {\bf 1207} (2012)
  022, [\href{http://xxx.lanl.gov/abs/1203.5803}{{\tt arXiv:1203.5803}}].

\bibitem{Chu:2014lja}
X.~Chu and B.~Dasgupta, {\it {A "Pas de Deux" - Dark Radiation Fattens and
  Puffs-up Dark Matter Halos}},  \href{http://xxx.lanl.gov/abs/1404.6127}{{\tt
  arXiv:1404.6127}}.

\bibitem{deBlok:2009sp}
W.~de~Blok, {\it {The Core-Cusp Problem}},  {\em Adv.Astron.} {\bf 2010} (2010)
  789293, [\href{http://xxx.lanl.gov/abs/0910.3538}{{\tt arXiv:0910.3538}}].

\bibitem{BoylanKolchin:2011de}
M.~Boylan-Kolchin, J.~S. Bullock, and M.~Kaplinghat, {\it {Too big to fail? The
  puzzling darkness of massive Milky Way subhaloes}},  {\em
  Mon.Not.Roy.Astron.Soc.} {\bf 415} (2011) L40,
  [\href{http://xxx.lanl.gov/abs/1103.0007}{{\tt arXiv:1103.0007}}].

\bibitem{Kolb:1990vq}
E.~W. Kolb and M.~S. Turner, {\it {The Early Universe}},  {\em Front.Phys.}
  {\bf 69} (1990) 1--547.

\bibitem{Agrawal:2011ze}
P.~Agrawal, S.~Blanchet, Z.~Chacko, and C.~Kilic, {\it {Flavored Dark Matter,
  and Its Implications for Direct Detection and Colliders}},  {\em Phys.Rev.}
  {\bf D86} (2012) 055002, [\href{http://xxx.lanl.gov/abs/1109.3516}{{\tt
  arXiv:1109.3516}}].

\bibitem{Borie:2012tu}
E.~Borie, {\it {Muon-proton Scattering}},
  \href{http://xxx.lanl.gov/abs/1207.6651}{{\tt arXiv:1207.6651}}.

\bibitem{Martin:2009iq}
A.~Martin, W.~Stirling, R.~Thorne, and G.~Watt, {\it {Parton distributions for
  the LHC}},  {\em Eur.Phys.J.} {\bf C63} (2009) 189--285,
  [\href{http://xxx.lanl.gov/abs/0901.0002}{{\tt arXiv:0901.0002}}].

\end{thebibliography}\endgroup
\bibliographystyle{jhep}

\end{document}